\documentclass[twocolumn,apj]{emulateapj} 
\usepackage{apjfonts}
\newcommand{\rotate}{}
\usepackage{lscape}

\usepackage[pagebackref=false,letterpaper=true,colorlinks=true,citecolor=black,linkcolor=blue, breaklinks=true,bookmarks=true]{hyperref}

\shorttitle{Quasar Extended Emission-Line Regions} 
\shortauthors{Fu and Stockton}
\journalinfo{accepted by the Astrophys. J.}
\submitted{Received 2008 June 10; accepted 2008 September 5}

\newcommand{\eg}{e.g.,}
\newcommand{\ie}{i.e.,}

\newcommand{\kms}{{km~s$^{-1}$}}
\newcommand{\cc}{{cm$^{-3}$}}
\newcommand{\ergs}{{erg~s$^{-1}$}}
\newcommand{\msun}{$M_{\odot}$}
\newcommand{\zsun}{$Z_{\odot}$}
\newcommand{\lsun}{$L_{\odot}$}

\newcommand{\leothree}{$L_{\rm E\,[O\,III]}$}
\newcommand{\lnothree}{$L_{\rm N\,[O\,III]}$}
\newcommand{\hst}{{\it HST}}

\newcommand{\othree}{{[O\,{\sc iii}]}}
\newcommand{\otwo}{{[O\,{\sc ii}]}}
\newcommand{\oone}{{[O\,{\sc i}]}}
\newcommand{\hetwo}{He\,{\sc ii}}

\newcommand{\ntwo}{[N\,{\sc ii}]}
\newcommand{\nfive}{N\,{\sc v}}
\newcommand{\cfour}{C\,{\sc iv}}
\newcommand{\stwo}{[S\,{\sc ii}]}
\newcommand{\fetwo}{Fe\,{\sc ii}}
\newcommand{\nev}{[Ne\,{\sc v}]}
\newcommand{\hb}{H$\beta$}

\newcommand{\oiii}{{[O\,{\sc iii}]}}

\newcommand{\sii}{[S\,{\sc ii}]}

\newcommand{\oi}{{[O\,{\sc i}]}}

\begin{document}

\title{Extended Emission-Line Regions: Remnants of Quasar Superwinds?\altaffilmark{1}}
\author{Hai Fu\altaffilmark{2} and Alan Stockton}
\affil{Institute for Astronomy, University of Hawaii, Honolulu, HI 96822}

\altaffiltext{1}{
Based in part on observations obtained at the Gemini Observatory,
which is operated by the Association of Universities for Research in
Astronomy, Inc., under a cooperative agreement with the NSF on behalf of
the Gemini partnership: the National Science Foundation (United States),
the Particle Physics and Astronomy Research Council (United Kingdom),
the National Research Council (Canada), CONICYT (Chile), the Australian
Research Council (Australia), CNPq (Brazil) and CONICET (Argentina).
Gemini Program ID: GN-2007A-Q-43 and GN-2007B-Q-12.}
\altaffiltext{2}{
Current address: Department of Astronomy, California Institute of Technology, MS 105-24, Pasadena, CA 91125; fu@astro.caltech.edu
}

\begin{abstract}
We give an overview of our recent integral-field-unit spectroscopy of luminous extended emission-line regions (EELRs) around low-redshift quasars, including new observations of 5 fields. Previous work has shown that the most luminous EELRs are found almost exclusively around steep-spectrum radio-loud quasars, with apparently disordered global velocity fields, and little, if any, morphological correlation with either the host-galaxy or the radio structure. Our new observations confirm and expand these results. The EELRs often show some clouds with velocities exceeding 500 \kms, ranging up to 1100 \kms, but the velocity dispersions, with few exceptions, are in the 30--100 \kms\ range. Emission-line ratios show that the EELRs are clearly photoionized by the quasars. Masses of the EELRs range up to $>10^{10}$ \msun.  Essentially all of the EELRs show relatively low metallicities, and they are associated with quasars that, in contrast to most, show similarly low metallicities in their broad-line regions. The two objects in our sample that do not have classical double-lobed radio morphologies (3C\,48, with a compact-steep-spectrum source; Mrk\,1014, radio-quiet, but with a weak compact-steep-spectrum source) are the only ones that appear to have recent star formation. While some of the less-luminous EELRs may have other origins, the most likely explanation for the ones in our sample is that they are examples of gas swept out of the host galaxy by a large-solid-angle blast wave accompanying the production of the radio jets. The triggering of the quasar activity is almost certainly the result of the merger of a gas-rich galaxy with a massive, gas-poor galaxy hosting the supermassive black hole.
 \end{abstract}


\section{Introduction}

As soon as quasars were discovered, it was noticed that some of them showed detectable nebulosity surrounding their star-like nuclei \citep[\eg][]{Mat63}. In an attempt to understand the nature of the fuzz around 3C\,48 \citep[][]{Mat63,San66}, \citet{Wam75} serendipitously discovered extended emission-line gas in a spectrum taken at a region $\sim$4\arcsec\ north of the quasar, although this ``fuzz" they intended to look into turned out to be mostly the unusually large and disturbed host galaxy of 3C\,48 \citep{Bor82}. Shortly afterward similarly extended nebulae were identified in off-nuclear spectra of two other quasars: 4C\,37.43 \citep[][]{Sto76} and 3C\,249.1 \citep[][]{Ric77}. In spite of the limited spatial sampling of these observations, it was already clear that such nebulae extend well beyond the classical narrow-line region (NLR) which is typically confined within $\sim1$ kpc from the nucleus. These ``extended emission-line regions'' (EELRs; following \citealt{Sto83}), are clearly physically associated with the quasars:  the ionized gas shows almost the same redshift as the nearby quasar, and the spectrum of the gas suggests the quasar as a more likely photoionization source than massive young stars. 

Slit spectroscopy does not allow an accurate determination of the morphology of the extended nebulae. As the redshift of 3C\,249.1 conveniently places the \othree\ $\lambda$5007 line into the commonly available H$\alpha$ filter, its EELR was the first to be imaged through a narrow-band filter centered on this strongest optical nebulae line \citep{Sto83}. The gas was found to form two extensive ``tidal-tails" to the NW and SE of the nucleus. 
As more narrow-band filters were acquired, \citeauthor{Sto83} continued the imaging survey on a sample of 58 quasars at $z < 0.5$. By the end of the survey, 47 quasars were imaged in \othree, and extended emission was detected in 15 of them \citep{Sto87}. These images reveal that the morphologies of EELRs are often fascinatingly complex and clearly betray a non-equilibrium situation showing knots and filaments that sometimes extend for tens of kpc. 
Structures reminiscent of tidal tails such as the EELR of 3C\,249.1 turn out to be rare---Ton\,202 is the only other similar case known. Surprisingly, the distribution of the ionized gas generally bears {\it no} close morphological relationships either with the host galaxies or with the extended radio structures, if present.

The origin and nature of these impressive and often beautiful structures have proved to be elusive. In this paper, we present new imaging spectroscopic observations of 5 quasars with luminous EELRs, discussing these in the context of our more detailed investigations of 3 additional EELRs \citep{Fu06,Fu07b,Fu08} and the surprising correlation found between quasar broad-line metallicity and the presence of luminous EELRs \citep{Fu07a}. Section \ref{backgroundsec} gives an outline of previous work on EELRs, \S~\ref{obssec} describes our observations and data-reduction procedures, \S~\ref{resultssec} gives our results for the 5 quasars with new observations, \S~\ref{sec:3c48metal} discusses the ionization mechanism and the metallicities for these EELRs, and \S~\ref{chap:sum} gives an over-all summary, including a general scenario that we believe can account for the observed properties of luminous EELRs. 
Throughout the paper, we assume a concordance cosmological model with
$H_0=70$ km s$^{-1}$ Mpc$^{-1}$, $\Omega_m=0.3$, and
$\Omega_{\Lambda}= 0.7$. Unless stated otherwise, solar abundances
are defined by the results in \citet{And89} [12+log(O/H)$_{\odot}$ =
8.93 and \zsun\ = 0.02]. Even though these are not the most recent determinations of solar abundances, they have the advantage that they allow easy comparison with many previous studies that have used this metallicity scale.

\section{Background}\label{backgroundsec}
\subsection{Correlations between Extended Emission and Quasar Properties}\label{introsec:corr}

\citet{Bor84} and \citet{Bor85} carried out an important spectroscopic survey to obtain off-nuclear spectra of 9 radio-loud quasars and 3 QSOs\footnote{Here we consider only their high-luminosity sample to be consistent with the sample of \citet{Sto87} which we will discuss in the following paragraphs.}. In addition to finding more quasars with extended emission, an important discovery of this program was the strong correlation between luminous optical extended emission and radio morphology (or radio spectral index). \citeauthor{Bor85} found that all of the six quasars with strong extended emission lines are steep-spectrum radio sources, and the other six objects\footnote{Contrary to these original papers, 3C\,48, a compact steep-spectrum radio source with a powerful radio jet extending 0\farcs7 to the north, is in fact surrounded by a luminous EELR, as was later shown by \citet{Sto87}.} where such extended emission is weak or absent are either radio-quiet or flat-spectrum radio-loud sources. There is a strong correlation between radio spectral index and radio morphology, since extended radio structures such as radio lobes are dominated by optically thin synchrotron radiation, which produces a steep power-law spectrum, while unresolved radio cores generally shows a flat spectrum due to synchrotron self-absorption (or thermal bremsstrahlung radiation from a disk wind, see \citealt{Blu07}). This extended emission$-$spectral index correlation can thus be translated into an extended emission$-$radio morphology correlation---luminous extended emission was found exclusively in quasars having extended radio structures, most of which show Fanaroff-Riley type II (FR II) twin-jet morphologies \citep{Fan74}. 

\begin{figure*}[!t]
\epsscale{0.55}
\plotone{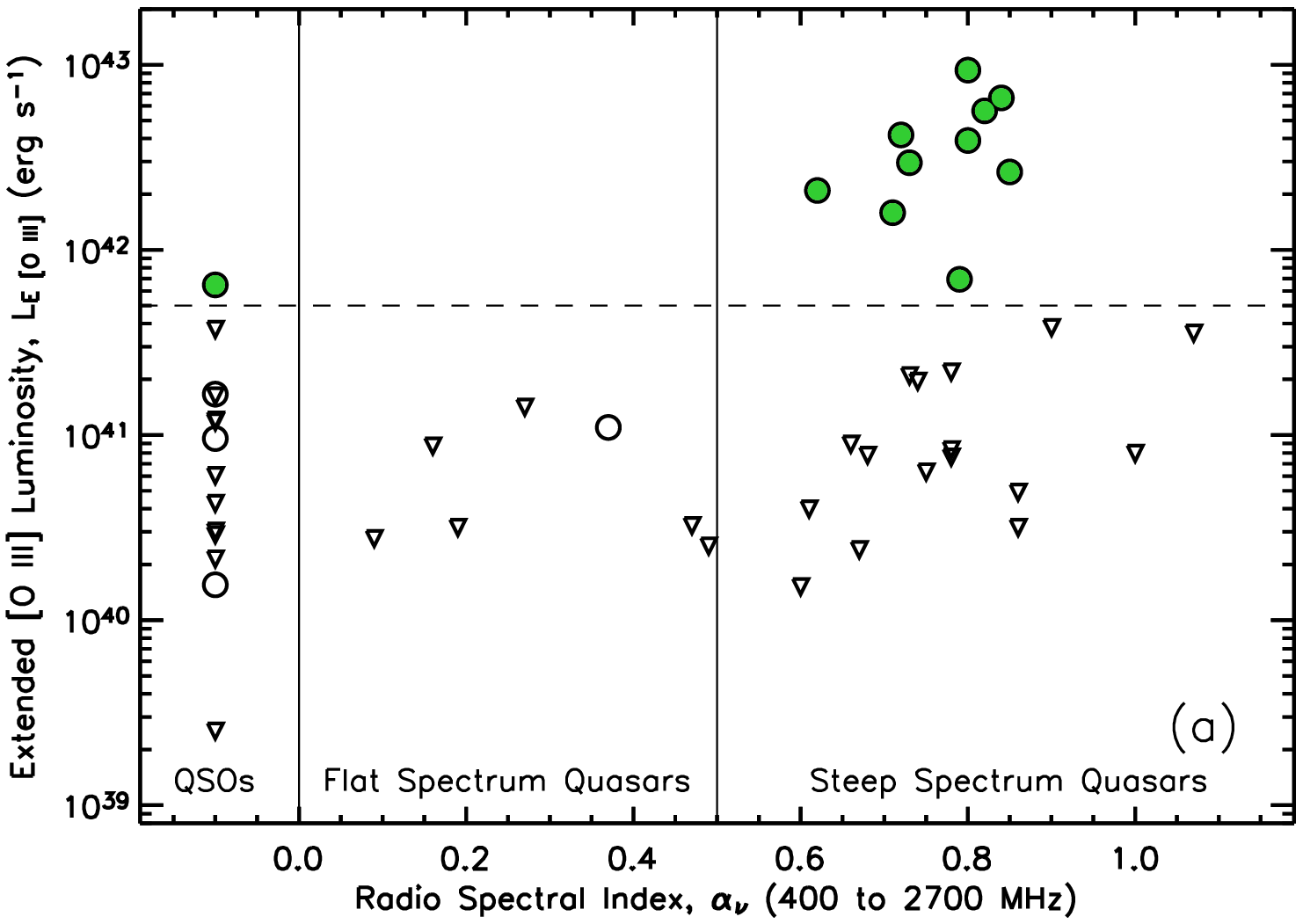}
\plotone{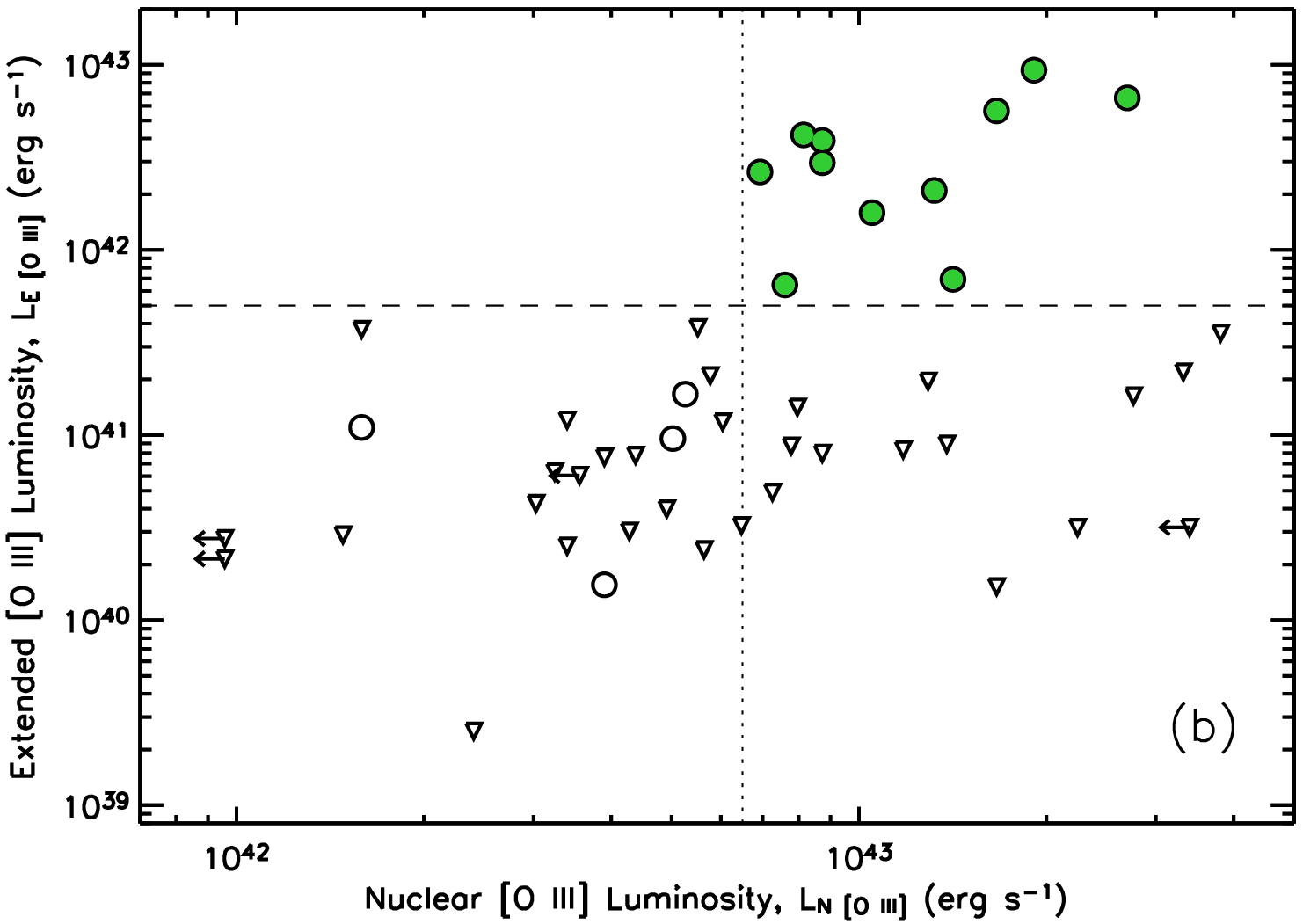}
\vskip 0.1in
\plotone{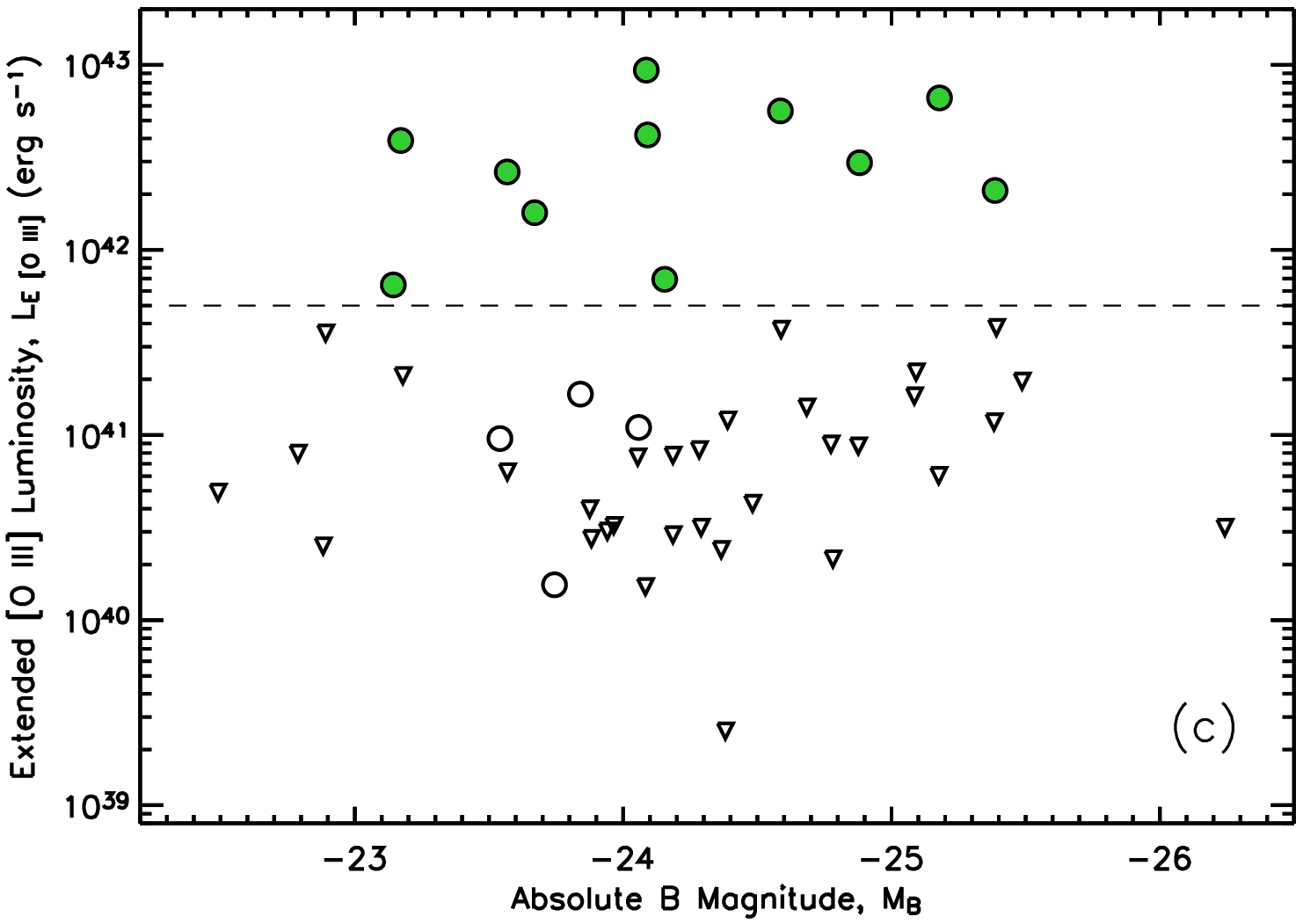}
\plotone{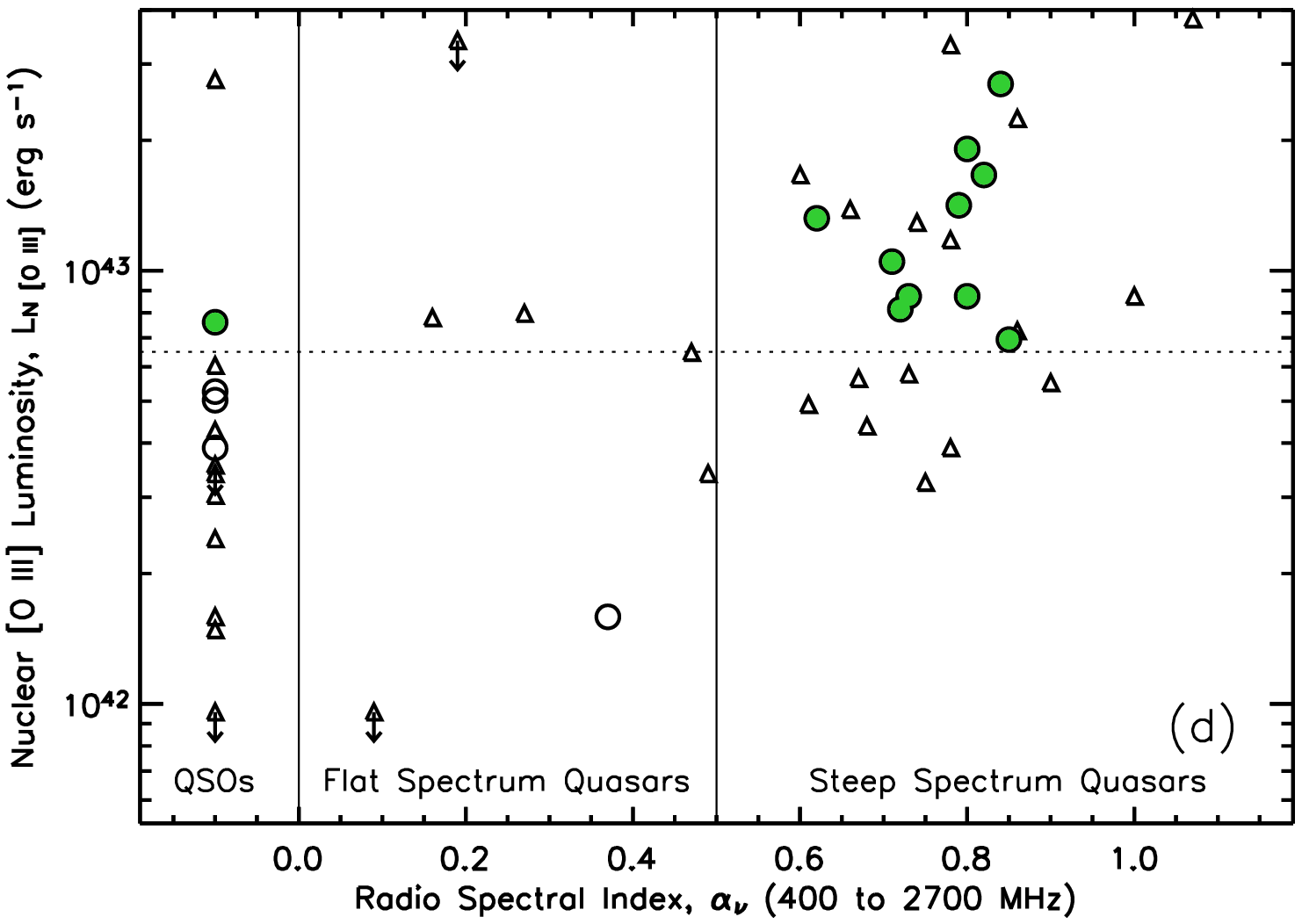}
\caption[Correlations between extended \othree\ emission and quasar properties]{
({\it a}) Correlation between extended \othree\ luminosity (\leothree) and radio spectral index ($\alpha_{\nu}$), which correlates strongly with radio morphology. 
({\it b}) Correlation between \leothree\ and nuclear \othree\ luminosity (\lnothree).
({\it c}) \leothree\ versus quasar absolute {\it B} magnitude.
({\it d}) \lnothree\ versus $\alpha_{\nu}$.
Open and filled circles represent objects with solid detections of extended \othree\ emission. Open triangles show objects where extended emission was undetected or ambiguous, they therefore represent upper limits of \leothree. Upper limits of \lnothree\ are drawn by arrows. 
{\it Solid lines} separate radio-quiet quasars (QSOs), flat-spectrum core-dominated quasars, and steep-spectrum lobe-dominated quasars. 
{\it Dashed lines} mark $L_{\rm cut} = 5\times10^{41}$ \ergs, which is our minimum \othree\ luminosity for any extended emission to be called an EELR. So the green filled circles highlight the {\it EELR quasars}, while the open circles and triangles together are considered as {\it non-EELR quasars} (even though the open circles indicate detection of weak extended emission). The spectral indices of radio quiet objects (\ie\ QSOs) have been arbitrarily set to be $-0.1$ for illustration purposes. 
{\it Dotted lines} indicate \lnothree\ = $6.5\times10^{42}$ \ergs, below which apparently no EELRs were found. Original data came from \citet{Sto87} and \citet{Ver06} and all except $\alpha_{\nu}$ have been converted to our concordance cosmology.
} 
\label{introfig:corr} 
\end{figure*}

This extended emission$-$radio morphology correlation was confirmed by the \othree\ narrow-band imaging survey of \citet{Sto87}, but it evolved into a more complicated picture as a result of their $4\times$ larger sample and their seamless spatial sampling. The authors found that (1) not all steep-spectrum ($\alpha_{\nu} > 0.5$) lobe-dominated quasars are surrounded by luminous EELRs---only 10 out of the 26 ($38\pm12$\%) have EELRs with \leothree\ $ \geq L_{\rm cut} = 5\times10^{41}$ \ergs (hereafter EELR quasars), where \leothree\ is the total \othree\ $\lambda5007$ luminosity within an annulus of inner radius 11.2 kpc and outer radius 44.9 kpc centered on the nucleus and $L_{\rm cut}$ is determined by the highest \leothree\ upper limits of the objects where extended emission is not detected (see Fig.~\ref{introfig:corr}$a$); and (2) none of the seven flat-spectrum ($\alpha_{\nu} < 0.5$) core-dominated quasars and only one of the 14 ($7\pm7$\%) optically selected QSOs show an EELR with \leothree\ $> L_{\rm cut}$. To illustrate this correlation more clearly, we have created a new figure (Fig.~\ref{introfig:corr}$a$) based on the original data of \citet{Sto87}. 

The sample observed by \citet{Sto87} included almost 2/3 of all optically luminous QSOs known at the time with $z\le0.45$ and $\delta > -25\arcdeg$. It is still nearly complete for radio-loud quasars within these limits but clearly incomplete for radio-quiet QSOs, which typically comprise about 90\%\ of the total population. Nevertheless, the radio-quiet objects in the sample are likely typical of the population of luminous, low-redshift, UV-selected QSOs.

\citet{Bor85} also noticed that the presence of strong extended emission was frequently accompanied by broader and bumpier nuclear Balmer lines, weaker nuclear \fetwo\ emission, and stronger nuclear narrow lines. The first two correlations are often regarded as by-products of the extended emission$-$radio morphology correlation, as there exist well-established correlations between radio morphology and nuclear broad-line profile and \fetwo\ emission \citep{Mil79,Ste81}. The last correlation seems to be an intrinsic one, and it was confirmed by \citet{Sto87}, who found that strong extended \othree\ emission is present only for objects with strong nuclear narrow-line emission. Specifically, all of the 11 EELR quasars in \citet{Sto87} have nuclear \othree\ $\lambda5007$ luminosity (\lnothree) of at least $6.5\times10^{42}$ \ergs, which is clearly seen in Fig.~\ref{introfig:corr}$b$. 

The nuclear \othree\ $\lambda5007$ luminosity has frequently been used as a proxy for the bolometric luminosity or accretion rate of AGN \citep[\eg][]{Hec04}, since it correlates with numerous multiwavelength broad-band luminosities \citep[\eg][]{Mul94}. However, since (1) the scatter of the \lnothree$-$continuum luminosity correlation is quite large \citep[see \eg][their Fig.~14]{Zak03}, and (2) the \citet{Sto87} quasar sample spans a limited range in \lnothree\ ($3\times10^8$ to $10^{10}$ \lsun), the existence of the apparent minimum \lnothree\ seems not to reflect a requirement of a minimum quasar accretion rate, instead, it may merely be a requirement on the availability of gaseous material in the nuclear region---in order to form an EELR, a quasar may have to have a fair amount of gas in the nuclear NLR (given a density of $10^3$ \cc\ and \othree\ $\lambda5007$/H$\beta$ = 10,  \lnothree\ = $6.5\times10^{42}$ \ergs\ corresponds to an ionized mass of $4.4\times10^6$ \msun). In fact, if we replace \lnothree\ with the absolute {\it B} magnitude of the quasar ($M_B$; compiled from \citealt{Ver06}) or continuum luminosity at a certain rest-frame wavelength, then this correlation immediately disappears (Fig.~\ref{introfig:corr}$c$), confirming that quasar accretion rate is not a dominant factor in the formation of an EELR.

Combining the two correlations together, one would expect to achieve a high success rate in identifying EELR objects. In Figure~\ref{introfig:corr}$d$, nuclear \othree\ luminosity is plotted against radio spectral index for the same 47 objects for which \citet{Sto87} obtained narrow-band \othree\ images. If one pre-selects only quasars with both $\alpha_{\nu} > 0.5$ and $L_{\rm N\,[O\,III]} > 6.5\times10^{42}$ \ergs, then roughly half (10/19) of the sample would turn out to be EELR quasars, as opposed to $<23\%$ (11/47)\footnote{When radio-quiet quasars are included in this kind of calculation, the derived percentage of EELR quasars should be treated as upper limits since \citet{Sto87}'s survey included a disproportionally small number of radio-quiet quasars compared to the radio-loud ones. Note that there are a total of 1876 quasars known at $z \leq 0.45$ and $\delta > -25^{\circ}$, among which less than 10\% are radio-loud \citep{Ver06}.}, 38\% (10/26), or $<46\%$ (11/24) if one performs the pre-selection on the basis of no {\it a priori} assumption, $\alpha_{\nu}$, or $L_{\rm N\,[O\,III]}$, respectively. 

Although the identification rate of EELRs can be improved significantly through this kind of exercise, one inevitably encounters this outstanding puzzle---why do only half of the otherwise identical quasars end up having luminous EELRs while the other half do not? The EELR quasars and the non-EELR quasars not only have similar radio morphology and nuclear narrow-line luminosity (given that they have been preselected with both criteria), but they also look remarkably similar in terms of the redshift range, the broad-band quasar luminosity, the luminosity of the host galaxy and the central black hole mass. It is natural to suspect that there is at least one concealed parameter at work, the difference in which results in the two distinct groups. 
This parameter has recently been identified. Quite unexpectedly, it turned out to be the gas metallicity of quasar broad-line regions (BLRs): quasars showing luminous EELRs have significantly lower metallicity BLRs ($Z \lesssim 0.6$ \zsun) than other quasars ($Z >$ \zsun) \citep{Fu07a}. 

\subsection{The Origin of the Extended Gas}\label{introsec:orig}

The suggestion that the extended ionized gas is organized into a rotating envelope or disk \citep{Wam75} was soon ruled out by the a counter example of 4C\,37.43, where blue-shifted clouds were detected on both sides of the nucleus \citep{Sto76}; instead, \citeauthor{Sto76} proposed that the ionized gas was ejected from the quasar itself, as strong outflows had been observed in a number of high-redshift quasars as blue-shifted absorption lines. But this scenario did not draw much attention. As 3C\,249.1 and Ton\,202 happened to be the first targets to be imaged in their redshifted \othree\ line, and both showed structures similar to the galactic tidal tails in the simulations of \citet{Too72}, \citet{Sto83} argued that the extended gas is tidal debris from a merger of two gas-rich galaxies and the merger is also responsible for triggering the quasar activity. However, a problem with this scenario was noticed soon afterward. 

The density of the ionized gas was known to range from tens to a few hundreds \cc, as indicated by both the density-sensitive \otwo\ $\lambda3727$ doublet profile \citep{Sto76} and ionization-state-sensitive \otwo\ $\lambda3727$/\othree\ $\lambda5007$ ratio in combination with the projected distance to the nucleus \citep{Fab87,Cra88}. Such a high density implies both a small size of the ionized cloud and a high pressure, which is unlikely to be balanced by that of the interstellar medium of a normal galaxy. An ionized cloud of this density would dissipate within a sound-crossing time of $\sim10^4$ years unless it were confined gravitationally or by the pressure of a hot surrounding medium. As there are not enough stars at the locations of the gaseous ``tidal tails" to gravitationally confine the gas, \citet{Fab87} suggested that the EELRs are embedded within a hot high-pressure medium. In this scenario, the warm ionized gas comprising the EELRs condenses from a hot halo sometimes seen even in poor clusters. The most direct proof for such a cooling flow would be a detection of diffuse X-ray$-$emitting gas. The X-ray luminosity can be used to estimate the density of the hot medium at the locations of the EELRs ($\sim10$ kpc), assuming a typical halo density profile and a typical temperature for the X-ray$-$emitting gas (\eg\ $kT$ = 1 keV, or $T$ = $1.2\times10^7$ K). Unfortunately, since quasars themselves are luminous X-ray sources and known EELR quasars typically have a redshift of $\sim$0.3, such an observation demands (1) a large-aperture X-ray telescope, (2) an arcsec-scale angular-resolution to confine most of the quasar emission within $\sim$10 kpc, and (3) an accurate knowledge of the instrument point-spread-function to remove the quasar spill-over; therefore it had to wait until {\it Chandra} X-ray observatory was launched near the turn of the millenium.

Contrary to the expectation, deep {\it Chandra} X-ray images of three EELR quasars failed to detect the hot X-ray halo from which the warm gas is supposed to condense, and the upper limits on the density of the hot gas were too low to allow sufficient cooling \citep{Sto06b}. In fact, the pressure of the hot gas, as implied by the density upper limits, is comparable to that of the ISM in our Galaxy ($n T \approx 10^4$ K \cc). To reiterate, without a confinement of a high-pressure surrounding medium, the warm ionized gas in the EELRs would dissipate quickly if its density is significantly greater than 1 \cc.

A somewhat related possibility that avoids these problems is that of a cold accretion flow along Mpc-scale filamentary structures \citep{Ker05}.  We postpone commenting on this option until our discussion in \S~\ref{sumsec:toy}.

Following the approach pioneered by \citet{Vie92} and \citet{Bin96}, \citet{Rob00} found that a photoionization model of a mixed medium with both optically thin and thick components not only overcomes the outstanding difficulties in fitting certain line ratios  suffered by previous single-phase models in the extended ionized gas in the radio galaxy 3C\,321, but it also results in an ionizing power more consistent with far-infrared (FIR) and radio observations of the hidden central engine where there is less extinction. \citet{Sto02} showed that the EELR of quasar 4C\,37.43 is also better modeled as a two-phase medium, consisting of a matter-bounded diffuse component with a unity filling factor ($n \sim 2$ \cc, $T \sim15,000$ K), in which are embedded small, dense clouds ($n \sim 500$ \cc, $T \sim 10,000$ K). Without assuming that most of the mass resides in a high-density medium, which raises the confinement crisis, this two-phase model can explain the high densities inferred from both the \otwo/\othree\ ratio and the \otwo\ doublet ratio, as the \otwo\ emission is almost entirely produced by the dense clouds even though they account for less than 1\% of the total mass. 

\citet{Sto02} also obtained a global \othree\ velocity map of the 4C\,37.43 EELR from spectroscopy with an image slicer. The velocity field appears largely disordered, consistent with the results from earlier integral field spectroscopy of similar objects \citep{Dur94,Cra97,Cra00}. Some faint clouds show velocities as high as 700 \kms, which are unlikely to be solely gravitational in origin. As the embedded dense clouds would have to be constantly resupplied because of their short lifetimes, \citeauthor{Sto02} suggested that shocks were the most likely mechanism for regenerating the small dense clouds. It was also known that two of the EELR quasars have FIR colors similar to those of ultraluminous infrared galaxies (ULIRGs), which are virtually all starburst galaxies triggered by major mergers \citep{San96}. The host galaxies of both quasars had been confirmed to have starburst or recent poststarburst stellar populations (3C\,48 and Mrk\,1014, \citealt{Can00a,Can00b}). Combining all these pieces, \citet{Sto02} suggested that the extended emission-line gas has been expelled during a merger and is being shocked by a galactic superwind driven by a starburst triggered by the merger. 

So three decades after EELRs were first recognized, the origin of the ionized gas and the physical forces that controlled its distribution remained uncertain: (1) the cooling flow model seemed discredited following the {\it XMM-Newton} high-resolution spectroscopy \citep{Pet03} and the {\it Chandra} imaging program \citep{Sto06b}, 
but a new possibility had arisen---that the gas could have originated in a cold accretion flow funneled along intergalactic filaments \citep{Ker05}, 
(2) the tidal debris model had to incorporate a new physical mechanism (\eg\ shocks from a starburst superwind) to explain the existence of small dense clouds embedded in the otherwise low-density photoionized medium, and (3) the old quasar ejection model \citep{Wam75,Sto76} had been rejuvenated because of the accumulating circumstantial evidence of shocks propagating in the immediate environments of EELR quasars. This evidence includes the two-phase photoionization model of the brightest condensation in the EELR of 4C\,37.43 \citep{Sto02} and the discrete X-ray sources detected at distances within 40 kpc from two EELR quasars \citep[3C\,249.1 and 4C\,37.43,][]{Sto06b}. There was thus some evidence for an outflow, but it was unclear what was driving it---a starburst, or the quasar itself, or both? 

\begin{figure*}[!t]
\epsscale{0.285}
\plotone{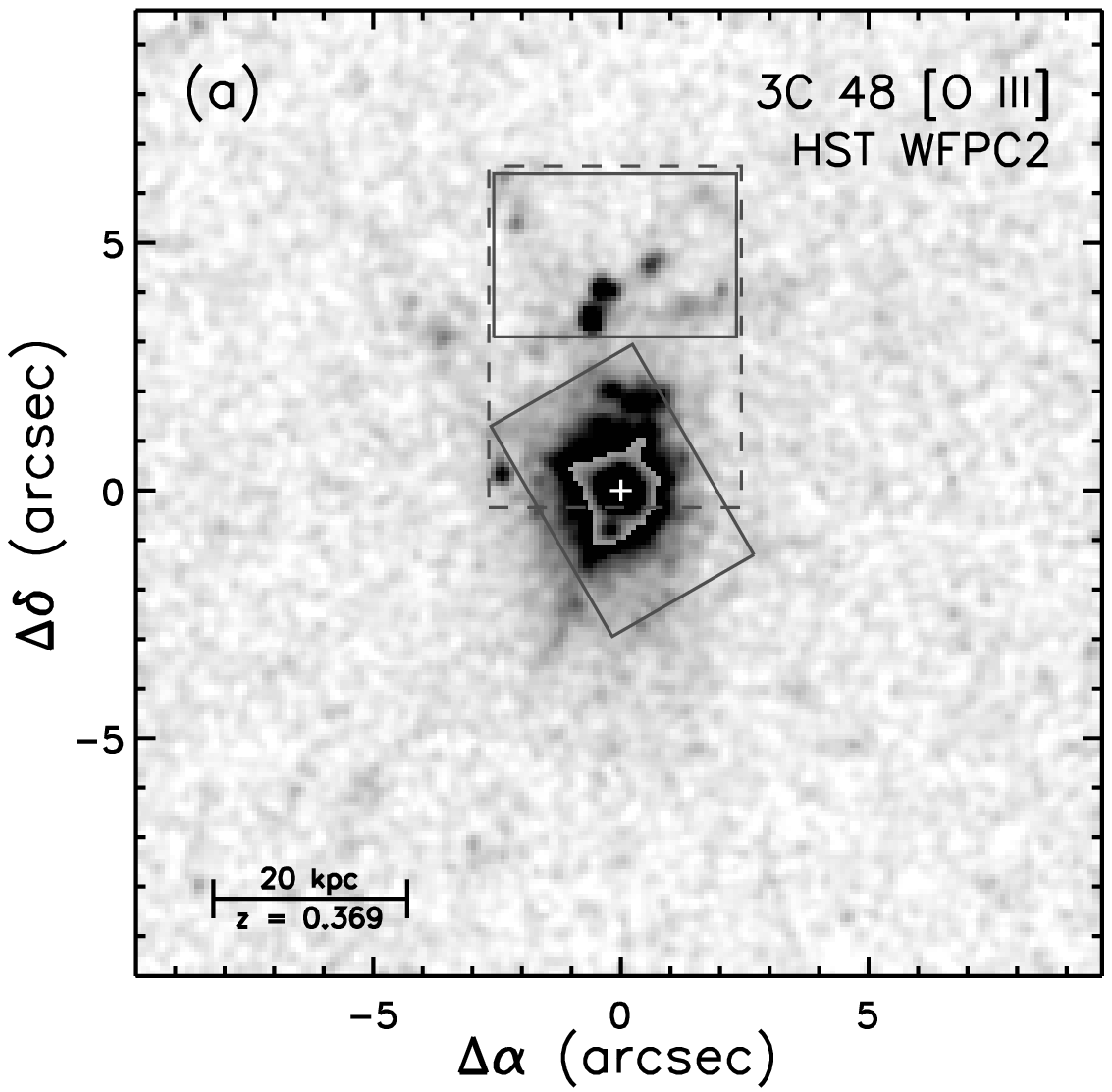}
\plotone{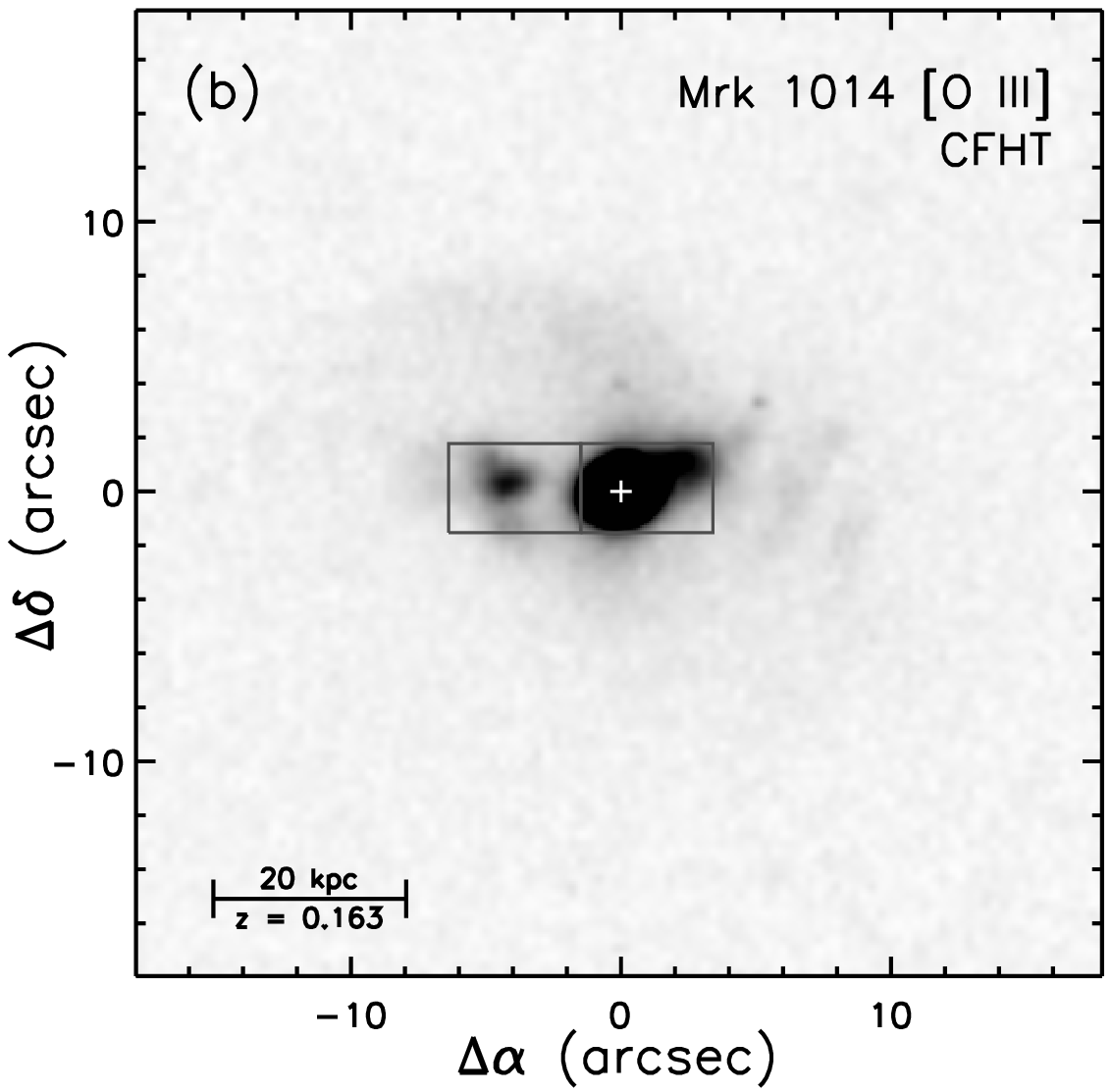}
\plotone{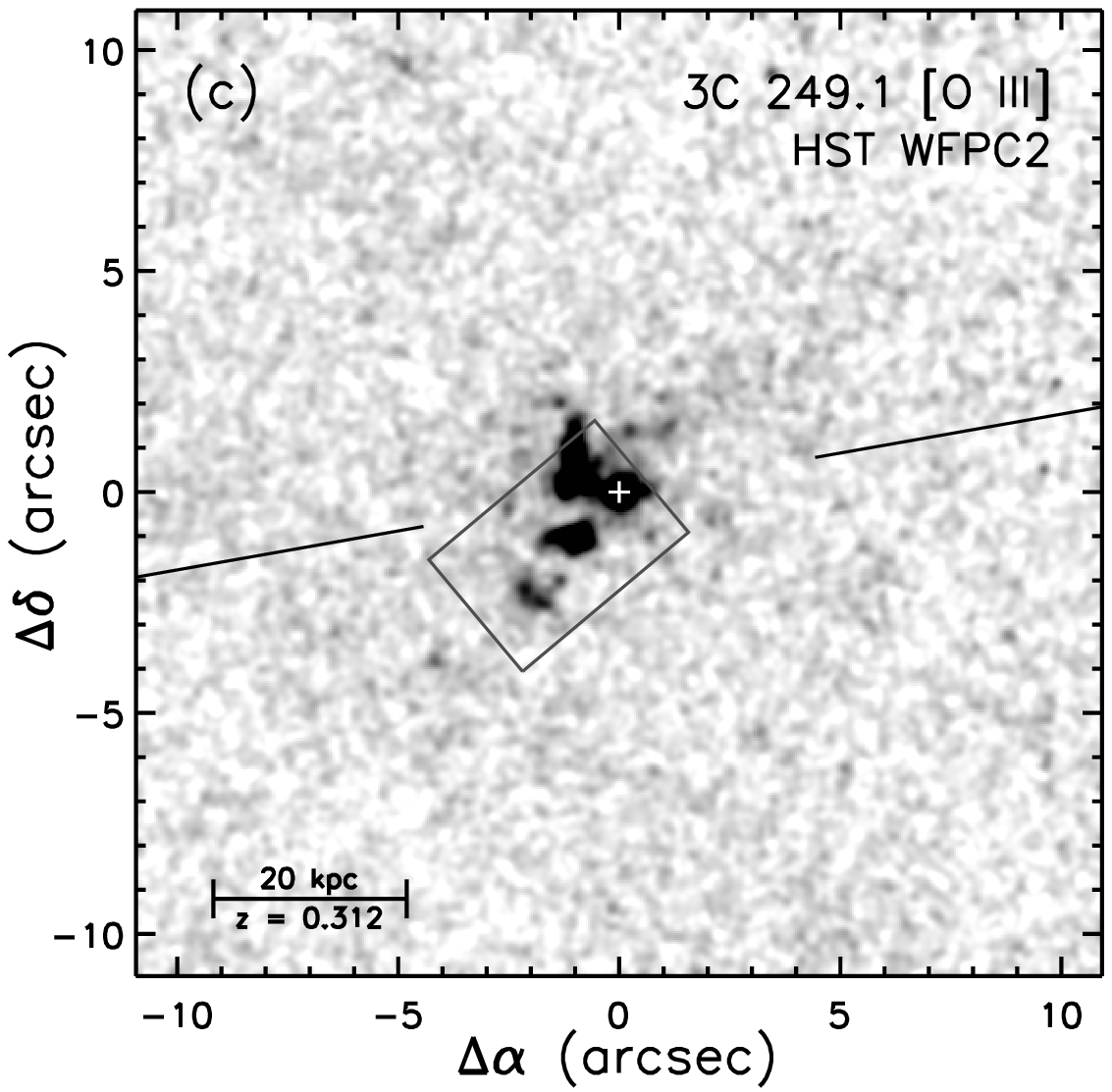}
\plotone{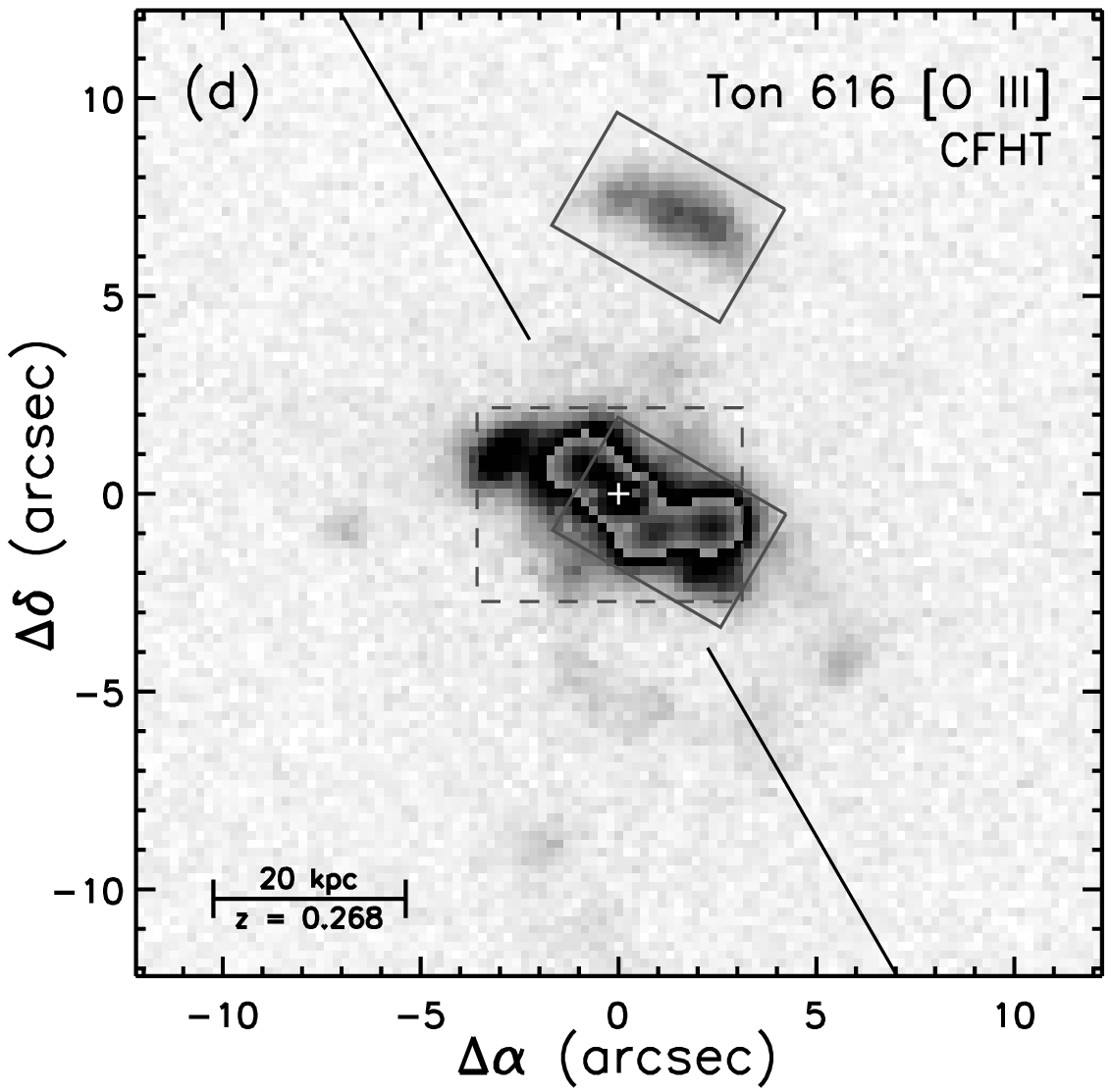}
\vskip 0.1in
\plotone{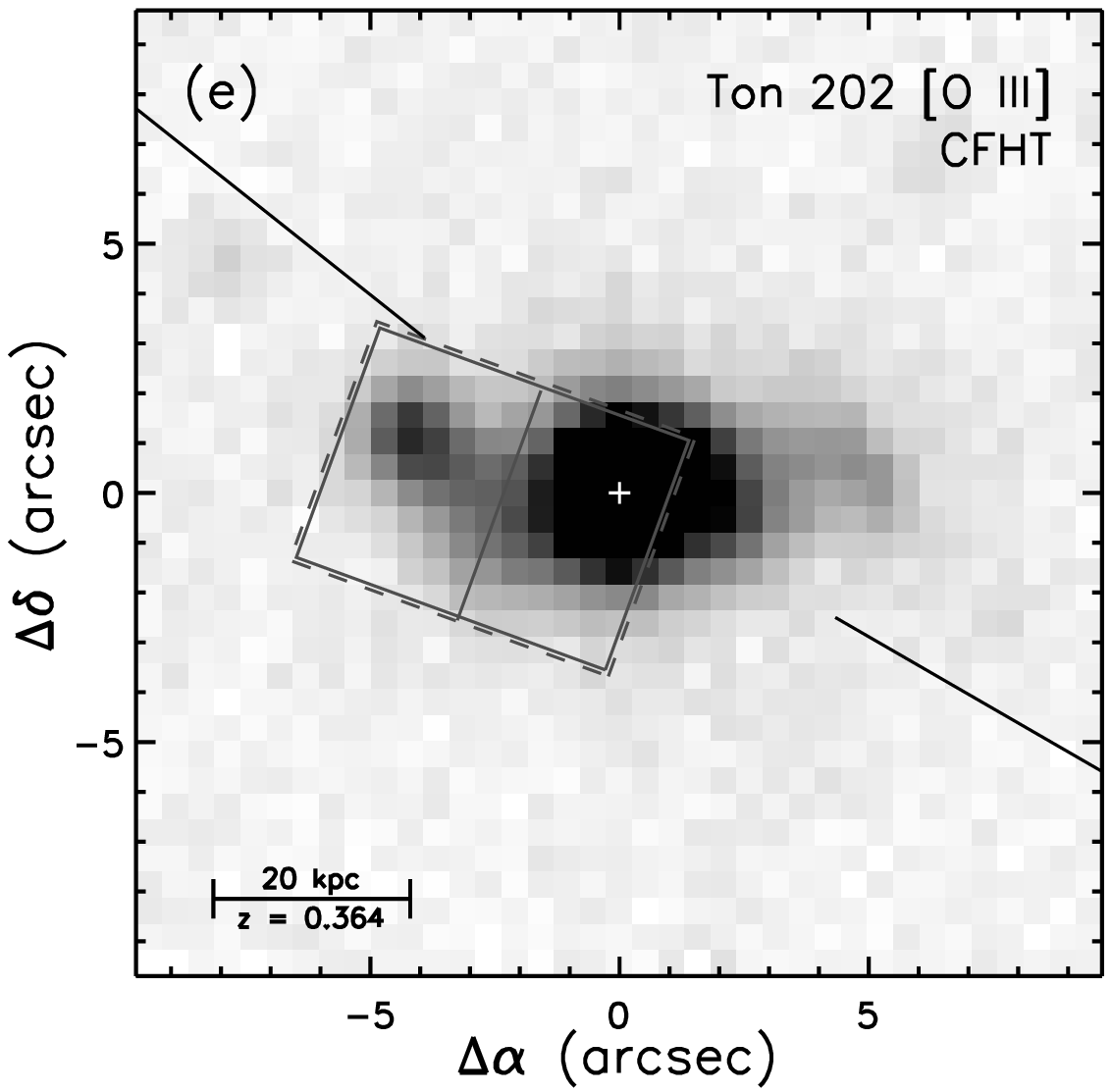}
\plotone{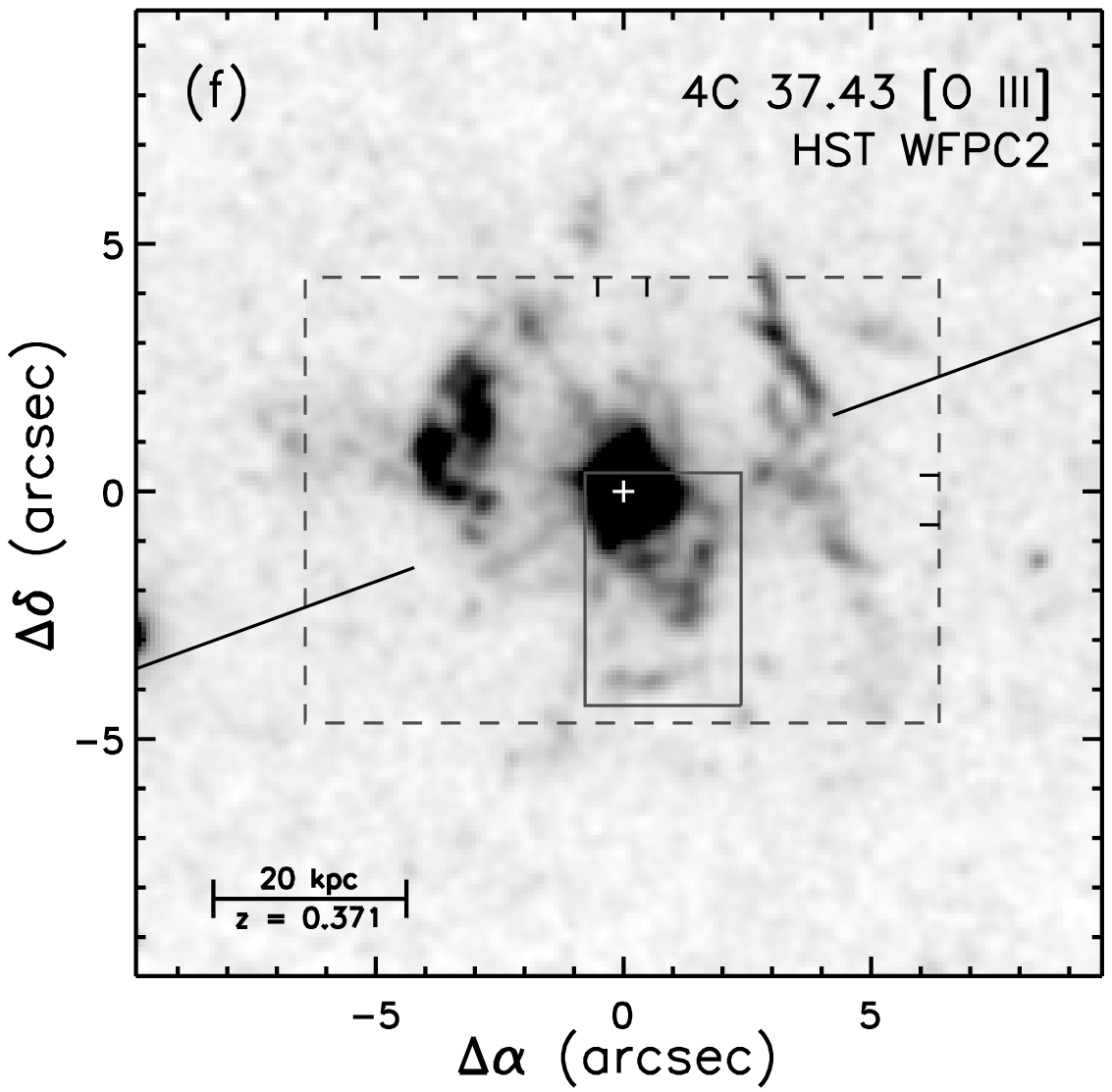}
\plotone{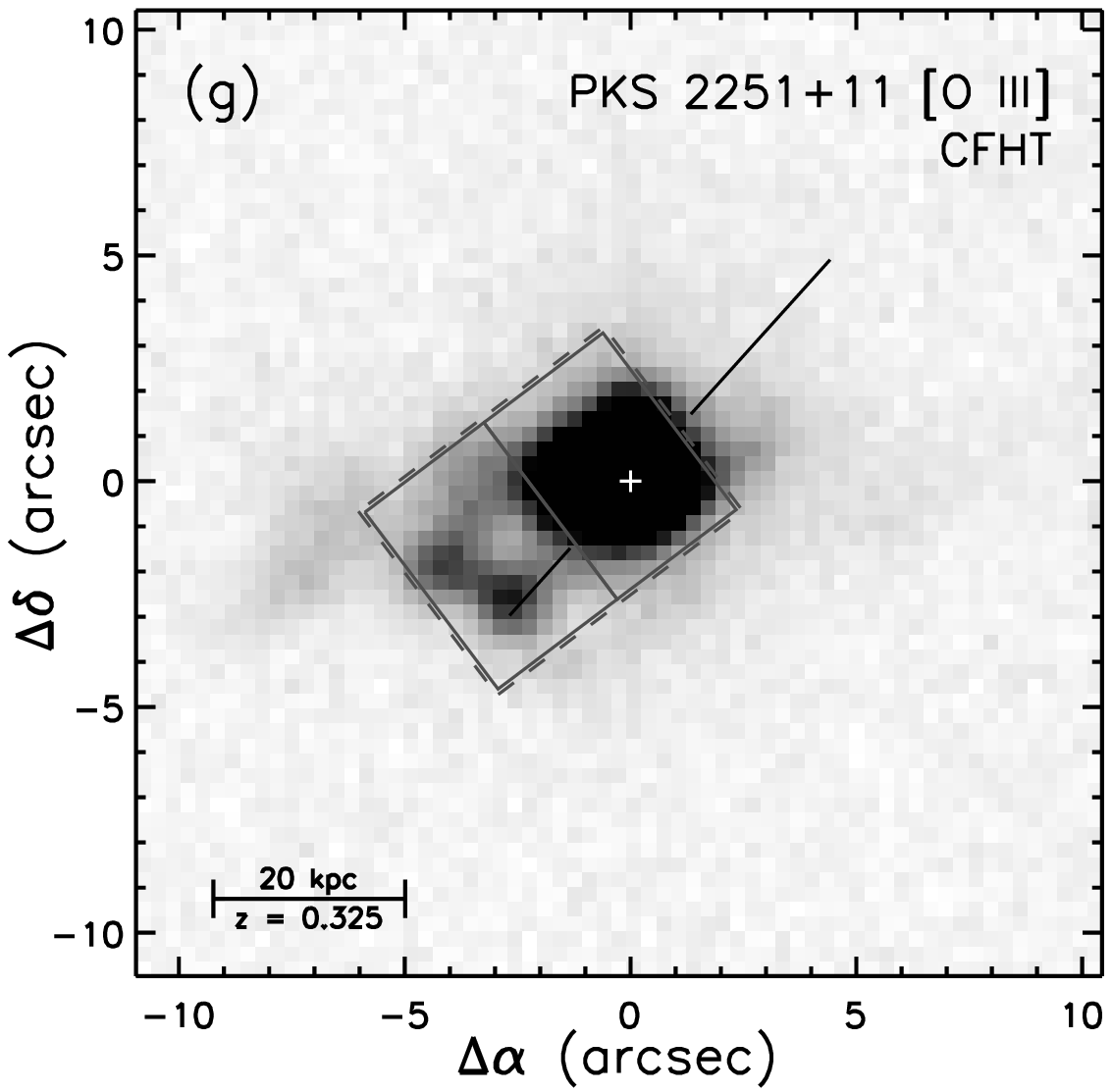}
\plotone{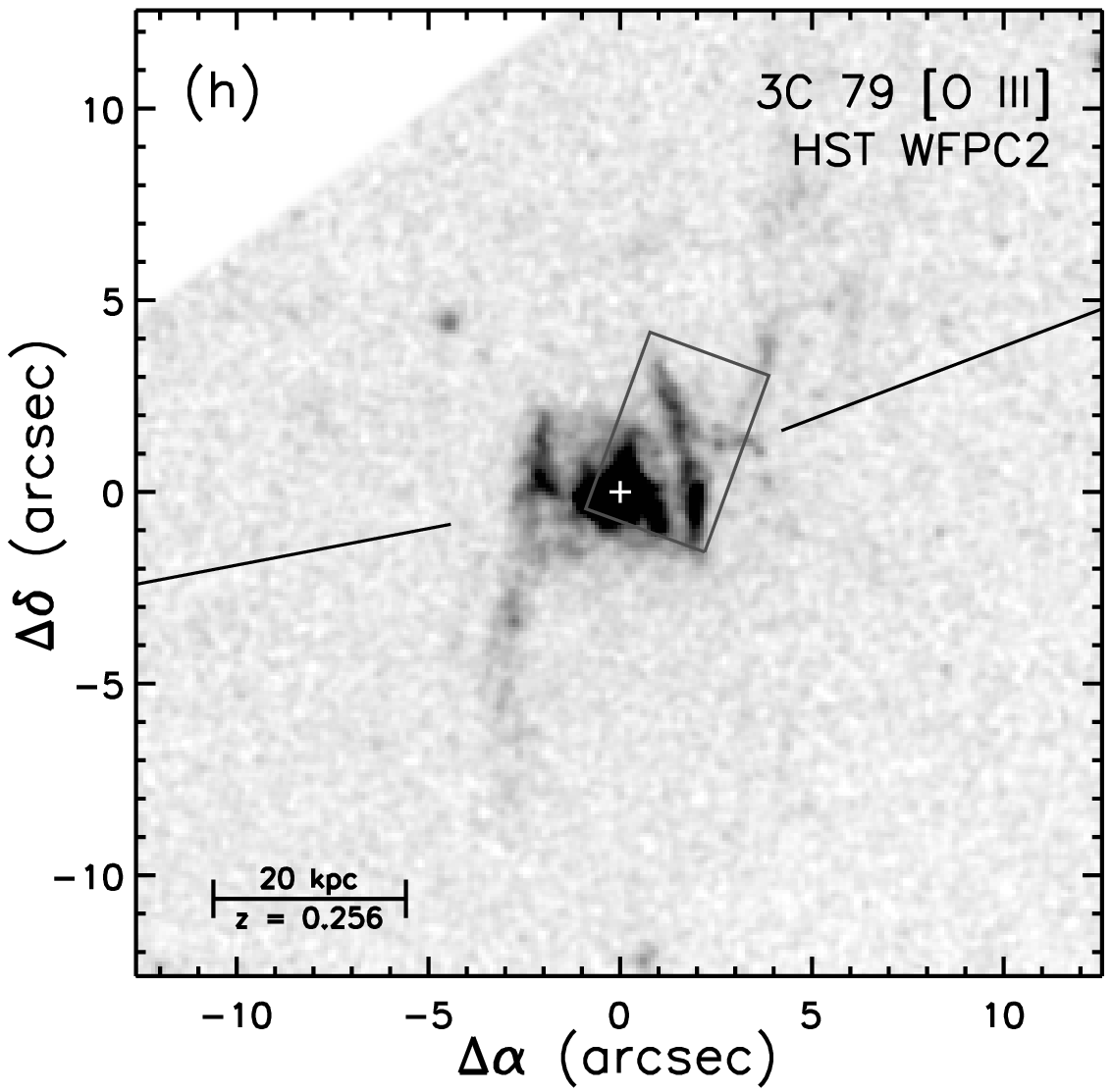}
\caption[Finding charts of GMOS observations]{
Narrow-band \othree\ $\lambda5007$ images of 3C\,48, Mrk\,1014, 3C\,249.1, Ton\,616, Ton\,202, 4C\,37.43, PKS\,2251+11, and 3C\,79 (panels $a$ to $h$). Overlaid are the fields of the GMOS IFUR observations ({\it solid boxes}) and those of the GMOS IFU2 observations ({\it long-dashed rectangles}). \hst\ linear-ramp-filter images are used if available. The radio jet direction is described by the solid lines for all of the objects except for 3C\,48 and Mrk\,1014, where the radio sources are less than 2\arcsec\ across. All images have been scaled to show the same physical size (150 kpc on a side) at the redshift of the object which is labeled under the scale bar. The images of 3C\,48 and Ton\,616 have been allowed to wrap around to show high surface brightness peaks. In the image of 4C\,37.43, the long-dashed rectangle shows the mosaicked field of four IFU2 pointings, the overlapping regions of which are shown by the tick marks on the inner edge.}
\label{introfig:obs}
\end{figure*}

The ultimate goal of this and our previous papers \citep{Fu06,Sto07,Fu07a,Fu07b,Fu08} has been to address these fundamental questions. The recent availability of integral-field units (IFUs) on large telescopes has made a dramatic improvement in our ability to do so. Most of the previous investigations on quasar EELRs have relied on either narrow-band imaging of the entire structure or slit spectroscopy of only a small fraction of the whole extended emission. IFUs allow spectroscopy in a two-dimensional area, combining the advantages of the two traditional observational methods. The earliest integral field spectrograph used in studying quasar EELRs was TIGER \citep{Bac95}, which was a lenslet-based instrument mounted on the 3.6-m Canada France Hawaii Telescope (CFHT). It offered a $7\arcsec\times7\arcsec$ field of view (FOV) with a spatial sampling of 0\farcs4. \citet{Dur94} observed three EELR quasars in the \citet{Sto87} sample with TIGER (all of which are included in our sample). Due to the limited spectral coverage of the instrument, their spectra included only the range from H$\beta$ $\lambda4861$ to \othree\ $\lambda5007$. The main result of this study was that the velocity fields of EELRs were rather chaotic. TIGER was soon superseded by ARGUS \citep{Van95}, which was also hosted by CFHT but had a ``lenslet+fiber" design. ARGUS covered a $\sim8\farcs5\times12\farcs4$ hexagonal FOV with a sampling of 0\farcs4. Thanks to its optical-fiber design, ARGUS offered a wider wavelength coverage than did TIGER, allowing \otwo\ $\lambda3727$ and \othree\ $\lambda5007$ to be observed simultaneously. \citet{Cra97,Cra00} performed ARGUS integral field spectroscopy of seven radio-loud quasars that were known to possess spatially extended emission-line gas, among which three are in the \citet{Sto87} sample and two are in common with our sample. Besides confirming the chaotic velocity fields, \citeauthor{Cra97} found that all of the extended nebulae had very similar \otwo/\othree\ ratios independent of location.
These pioneering studies using IFUs on CFHT were nevertheless restricted in signal-to-noise ratio (S/N) and in resolution because of the size of the telescope.

\section{Observations and Data Reduction}\label{obssec}

\begin{deluxetable*}{lcccccccc}
\tablewidth{0pt}
\tabletypesize{\footnotesize}
\tablecaption{Targets of Gemini GMOS Integral Field Spectrocopy \label{introtab:targets}}
\tablehead{ 
\colhead{Name} & \colhead{Designation} &
\colhead{$z'$} & \colhead{$z$} & \colhead{$M_B$} &
\colhead{$L_{\rm \lambda4861}$} & \colhead{\lnothree} & \colhead{\leothree} &
\colhead{$\alpha_{\nu}$} \\
\colhead{(1)} & \colhead{(2)} & \colhead{(3)} & \colhead{(4)} &
\colhead{(5)} & \colhead{(6)} & \colhead{(7)} &
\colhead{(8)} & \colhead{(9)}
}
\startdata
        3C 48&  0137$+$3309& 0.367&0.36926& -24.59&  41.29&  43.22&  42.75&   0.82\\
     Mrk 1014&  0159$+$0023& 0.163&0.16323& -23.14&  41.08&  42.88&  41.81&   RQ  \\
     3C 249.1&  1104$+$7658& 0.313&0.31172& -25.18&  41.71&  43.43&  42.82&   0.84\\
      Ton 616&  1225$+$2458& 0.268&0.26808& -23.17&  40.84&  42.94&  42.59&   0.80\\
      Ton 202&  1427$+$2632& 0.366&0.36367& -25.39&  41.37&  43.12&  42.32&   0.62\\
     4C 37.43&  1514$+$3650& 0.370&0.37120& -24.09&  41.45&  43.28&  42.97&   0.80\\
PKS 2251$+$11&  2254$+$1136& 0.325&0.32538& -24.88&  41.36&  42.94&  42.47&   0.73\\
        3C 79&  0310$+$1705& 0.256&0.25632&\nodata&\nodata&  42.21&  42.89&   0.87
\enddata
\tablecomments{
Col. (1): Common name. 
Col. (2): J2000.0 designation.
Col. (3): Redshift from literature.
Col. (4): Redshift measured from our GMOS/IFU data from the \othree\
$\lambda\lambda4959,5007$ lines.
Col. (5): Absolute $B$ magnitude ($k$-correction applied).
Col. (6): Continuum luminosity at H$\beta$ in logarithmic (erg s$^{-1}$ \AA$^{-1}$).
Col. (7): Nuclear \othree\ $\lambda5007$ luminosigy in logarithmic (\ergs).
Col. (8): Extended \othree\ $\lambda5007$ luminosity in logarithmic (\ergs).
Col. (9): Radio spectral index ($f_{\nu} \propto \nu^{-\alpha_{\nu}}$)
between 400 and 2700 MHz.
Data in Cols. (3) \& (5) are taken from \citet{Ver06}. Cols. (6-9) are
based on \citet{Sto87}. All luminosities and magnitudes in this table
have been converted to our concordance cosmology.  
3C\,48 is a CSS radio-loud quasar with a single-sided jet within $\sim$1\arcsec\ to the north
\citep{Fen05}; Mrk\,1014 is a radio-quiet quasar but shows two radio knots $\sim$1\arcsec\ to either side of the nucleus \citep{Lei06a,Ben02} and has a steep radio spectrum ($\alpha_{\nu} \sim 0.9$); 3C\,79 is an FR-II radio galaxy. The rest of the objects are all FR-II radio-loud quasars.
}
\end{deluxetable*}

\begin{deluxetable*}{lccccccccc}
\tablewidth{0pt}
\tabletypesize{\scriptsize}
\tablecaption{Gemini GMOS Integral Field Spectroscopy Observing Log
\label{introtab:allifuobs}}
\tablehead{
\colhead{Slit}&\colhead{Grating}&\colhead{CW (\AA)}&\colhead{Lines}&\colhead{P.A.}&\colhead{Field}&\colhead{Exposure (s)}&\colhead{Airmass}&\colhead{Seeing}&\colhead{UT Date} \\ 
\colhead{(1)}&\colhead{(2)}&\colhead{(3)}&\colhead{(4)}&\colhead{(5)}&\colhead{(6)}&\colhead{(7)}&\colhead{(8)}&\colhead{(9)}&\colhead{(10)}
}
\startdata
\multicolumn{10}{c}{3C 48 ($z$ = 0.369)}   \nl
\hline
IFUR&B600   &5750&\nev$-$\oiii &30$^{\circ}$ &on      &180$0\times$5   &1.201  &0\farcs45  &20030923\\ 
IFUR&B600   &5680&$-$          &90$^{\circ}$ &off     &270$0\times$3   &1.468  &0\farcs72  &20070819\\ 
IFU2&R400   &9000&\oi$-$\sii  &90$^{\circ}$ &on      &240$0\times$1   &1.094  &0\farcs46  &20070905\\ 
\hline
\multicolumn{10}{c}{Mrk 1014 ($z$ = 0.163)}   \nl
\hline
IFUR&R400   &6250&\hb$-$\sii   &90$^{\circ}$ &on      &240$0\times$1   &1.114  &0\farcs54  &20070813\\ 
$-$ &$-$    &$-$ &$-$          &$-$          &off     &240$0\times$1   &1.070  &$-$        &$-$     \\ 
IFUR&B600   &4890&\nev$-$\oiii &90$^{\circ}$ &on      &240$0\times$1   &1.245  &0\farcs49  &20070923\\ 
$-$ &$-$    &$-$ &$-$          &$-$          &off     &240$0\times$1   &1.403  &0\farcs73  &20070924\\ 
$-$ &$-$    &$-$ &$-$          &$-$          &on      &240$0\times$1   &1.430  &0\farcs91  &20071007\\ 
$-$ &$-$    &$-$ &$-$          &$-$          &off     &240$0\times$1   &1.618  &0\farcs50  &20071012\\ 
\hline
\multicolumn{10}{c}{3C 249.1 ($z$ = 0.312)}   \nl
\hline
IFUR&B600   &5800&\nev$-$\oiii &130$^{\circ}$&on      &282$5\times$2   &1.841  &1\farcs15  &20050407\\ 
$-$ &$-$    &$-$ &$-$          &$-$          &on      &282$5\times$1   &1.838  &1\farcs03  &20050408\\ 
IFUR&R400   &7170&\hb$-$\sii   &130$^{\circ}$&on      &157$5\times$1   &1.840  &1\farcs03  &20050408\\ 
\hline
\multicolumn{10}{c}{Ton 616 ($z$ = 0.268)}   \nl
\hline
IFUR&R400   &6570&\hb$-$\sii   &60$^{\circ}$ &off     &240$0\times$3   &1.207  &0\farcs57  &20070612\\ 
IFUR&R600   &7250&\hb$-$\sii   &60$^{\circ}$ &on      &180$0\times$3   &1.757  &0\farcs74  &20070717\\ 
IFU2&B600   &6332&\hb$-$\oiii  &0$^{\circ}$  &on      &240$0\times$1   &1.230  &0\farcs90  &20070703\\ 
\hline
\multicolumn{10}{c}{Ton 202 ($z$ = 0.364)}   \nl
\hline
IFUR&B600   &5680&\nev$-$\oiii &340$^{\circ}$&on      &210$0\times$3   &1.108  &0\farcs82  &20070414\\ 
$-$ &$-$    &$-$ &$-$          &$-$          &off     &210$0\times$3   &1.197  &$-$        &$-$     \\ 
IFU2&R600   &9107&\oi$-$\sii  &340$^{\circ}$&on      &180$0\times$2   &1.069  &0\farcs58  &20070709\\ 
\hline
\multicolumn{10}{c}{4C 37.43 ($z$ = 0.371)}   \nl
\hline
IFUR&B600   &6412&\nev$-$\oiii &0$^{\circ}  $&on      &240$0\times$5   &1.071  &0\farcs59  &20060524\\ 
IFU2&R831   &6585&\hb$-$\oiii  &0$^{\circ}  $&on      &720$\times1$2   &1.059  &0\farcs49  &20060523\\ 
\hline
\multicolumn{10}{c}{PKS 2251$+$11 ($z$ = 0.325)}   \nl
\hline
IFUR&B600   &5510&\nev$-$\oiii &217$^{\circ}$&on      &270$0\times$3   &1.064  &0\farcs39  &20070811\\ 
$-$ &$-$    &$-$ &$-$          &$-$          &off     &270$0\times$2   &1.031  &$-$        &$-$     \\ 
$-$ &$-$    &$-$ &$-$          &$-$          &off     &270$0\times$1   &1.059  &0\farcs33  &20070812\\ 
IFU2&R400   &8810&\oi$-$\sii  &217$^{\circ}$&on      &240$0\times$1   &1.224  &0\farcs39  &20070905\\ 
\hline
\multicolumn{10}{c}{3C 79 ($z$ = 0.256)}   \nl
\hline
IFUR&B600   &5500&\nev$-$\oiii &160$^{\circ}$&on      &288$0\times$3   &1.253  &0\farcs52  &20061221
\enddata
\tablecomments{
Col. (1): IFU configuration (IFUR: half-field one-slit; IFU2: full-field two-slit).
Col. (2): Grating.
Col. (3): Central wavelength.
Col. (4): Emission lines enclosed
(Abbreviations refer to \nev\,$\lambda3426$, \oiii\,$\lambda5007$, \oi\,$\lambda6300$,
and \sii\,$\lambda6731$).
Col. (5): Position angle.
Col. (6): Pointing position (``on-nucleus'' or ``off-nucleus'', refer to
Fig.~\ref{introfig:obs}), if there are multiple pointings.
Col. (7): Exposure time.
Col. (8): Mean airmass during the exposure.
Col. (9): Seeing measured from acquisition images. 
Col. (10): Observation date (yyyymmdd).
``$-$" means ``the same as above".
}
\end{deluxetable*} 

The newest optical integral field spectrograph on Mauna Kea is the IFU \citep{All02} of the Gemini Multiobject Spectrograph \citep[GMOS;][]{Hoo04}, installed on the Gemini North telescope. The IFU uses a ``lenslet+fiber" design similar to that of ARGUS and is the first such instrument on an 8-m telescope. When used in the full-field mode, it covers a fully filled contiguous field of $7\arcsec\times5\arcsec$ with 1000 hexagonal lenslets 0\farcs2 in diameter. A ``sky" field of $5\arcsec\times3\farcs5$ $1'$ from the science field is observed simultaneously by another 500 lenslets for background subtraction. Optical fibers behind the lenslet arrays reformat the focal plane into two parallel pseudo-slits that pass the light into the rest of the spectrograph. To avoid spectral overlapping between the two pseudo-slits, broad-band filters are often used to limit the spectral range. One can increase the spectral coverage to that of long-slit spectroscopy by sacrificing half of the FOV (\ie\ masking off one of the slits). 
This instrument provides sufficient spatial and spectral resolution to allow constructing unprecedentedly high-resolution velocity fields and resolving the density-sensitive \otwo\ doublet at redshifts beyond $\sim$0.3. In addition, the wide wavelength coverage in combination with the large aperture of Gemini permit detailed spatially resolved modeling of emission-line spectra accounting for weak but important lines such as [Ne\,{\sc v}] $\lambda3426$, \hetwo\ $\lambda4686$ and \othree\ $\lambda4363$ that were unexplored by previous IFU observations. 

The narrow-band \othree\ imaging survey of \citet{Sto87} published \othree\ $\lambda5007$ luminosities and morphologies for all cases of detected extended emission around 47 quasars with $z \leq 0.45$, $\delta > -25^{\circ}$, and $M_V < -23.4$\footnote{This absolute $V$ magnitude cut was calculated in their original cosmology ($H_0=75$ km s$^{-1}$ Mpc$^{-1}$, $\Omega_m=0$, and $\Omega_\Lambda = 0$) without $k$-correction. In our chosen cosmology and accounting for $k$-correction, all the 47 objects have $M_B < -22.4$ (see Fig.~\ref{introfig:corr}$c$).}. We have performed GMOS integral field spectroscopy of 7 of the 11 quasars that show luminous structures of extended emission with \leothree\ $> 5\times10^{41}$ \ergs---3C\,249.1, 4C\,37.43, 3C\,48, Mrk\,1014, Ton\,616, Ton\,202, and PKS\,2251+11. We also have obtained comparable observations of a radio galaxy, 3C\,79, which has a similarly luminous EELR. The properties of these objects have been listed in Table~\ref{introtab:targets}. The literature redshifts ($z'$) and absolute {\it B} magnitudes ($M_B$) in this table are compiled from \citet{Ver06}. 
The more accurate redshifts ($z$) are measured from the \othree\,$\lambda\lambda4959,5007$ lines in nuclear spectra
extracted from our GMOS/IFU data. The radial velocity maps presented 
in \S~\ref{resultssec}
have been calculated relative to these assumed systemic redshifts using the
relativistic Doppler formula \citep{Sto07}. In addition to the
rest-frame correction, the velocity dispersion ($\sigma$ = FWHM/2.355) 
maps have also been
corrected for the instrumental resolution. The typical spectral
resolutions of the B600 and the R400 gratings are respectively about
2.1 \AA\ and 2.7 \AA\ FWHM, as measured from night sky lines.
The last four columns give the continuum luminosity at H$\beta$, nuclear \othree\ $\lambda5007$ luminosity, extended \othree\ luminosity within an annulus of inner radius 11.2 kpc and outer radius 44.9 kpc centered on the nucleus, and radio spectral index between 400 and 2700 MHz. 
Results for the quasars 3C\,249.1 and 4C\,37.43, and for the radio galaxy 3C\,79, have been published in \citet{Fu06,Fu07b,Fu08}. In this paper we present results for the last five objects. The observations were performed in queue mode in 2007 between April and October. Various slit/grating/central wavelength combinations were used to avoid chip gaps for objects with different redshifts. Generally the single-slit half-field mode (IFUR) was used to cover the wider wavelength range from \nev\,$\lambda3426$ to \othree\,$\lambda5007$ (blue spectra), and the two-slit full-field mode (IFU2) was used to cover the narrower wavelength range between \oone\ $\lambda6300$ and \stwo\,$\lambda6731$ (red spectra). A detailed observing log can be found in Table~\ref{introtab:allifuobs}. Figure~\ref{introfig:obs} shows the areas covered by the 3\farcs5$\times$5\arcsec\ FOV of the IFUR mode and the 7\arcsec$\times$5\arcsec\ FOV of the IFU2 mode. Most of the observations were performed under excellent seeing (FWHM $<$ 0\farcs6). Spectrophotometric standard stars were observed with the same instrument settings as that of the science observations but often on different nights for relative flux calibration.

The data were reduced using the Gemini IRAF package (ver.\ 1.8), following the same procedure as described by \citet{Fu07b}. After correction for differential atmospheric refraction, the quasars were removed from the data cubes using the same Richardson-Lucy method as in \citet{Fu07b} with the STSDAS procedure $cplucy$ and point-spread functions (PSFs) generated in emission-line$-$free regions in the same data cube. 

To increase the S/N, we extracted spectra of distinct clouds with circular apertures of a radius between 0\farcs5 to 0\farcs7, depending on the seeing. We have concentrated on clouds at least 1\farcs5 away from the nucleus to avoid high levels of contamination from quasar light to the emission lines. The line-of-sight Galactic reddening was corrected using a standard Galactic reddening law \citep{Car89} with extinction values from \citet{Sch98}. Intrinsic reddening due to dust in front of or associated with the cloud was determined from the H$\alpha$/H$\beta$ or H$\gamma$/H$\beta$ ratio, assuming theoretical ratios of H$\alpha$/H$\beta$ = 3.1 and H$\gamma$/H$\beta$ = 0.468 \citep{Ost89}. Whenever H$\alpha$ and H$\beta$ were observed simultaneously, the H$\alpha$/H$\beta$ ratio was used to determine the intrinsic reddening; and in other cases, the H$\gamma$/H$\beta$ ratio was used. The emission-line fluxes were dereddened using the same reddening law and are tabulated in Table~\ref{tab:3c48lineratio}.
When there are overlaps between blue and red spectra, which is the case for Mrk\,1014, the spectra were normalized at the overlapping region according to the \othree\,$\lambda\lambda4959,5007$ flux and combined before dereddening. For 3C\,48, Ton\,202 and PKS\,2251+11, as there are no overlaps, the dereddened line intensities from the red spectra were divided by a constant so that H$\alpha$/H$\beta$ = 3.1. 
In contrast to the case of the intrinsically weaker extended emission in Seyfert galaxies, we do not here have to deal with the issue of correcting for stellar Balmer absorption in measuring the Balmer emission lines to determine the internal extinction. Most of the EELRs we are dealing with are projected onto regions with no evidence for stellar contamination detectible in our observations. Even for those in the inner regions, where we can detect the underlying host galaxy, the emission so strongly dominates the stellar continuum that absorption corrections are negligible.

\section{Results}\label{resultssec}

\subsection{3C\,48}\label{sec:3c48}

\begin{figure*}[!t]
\plotone{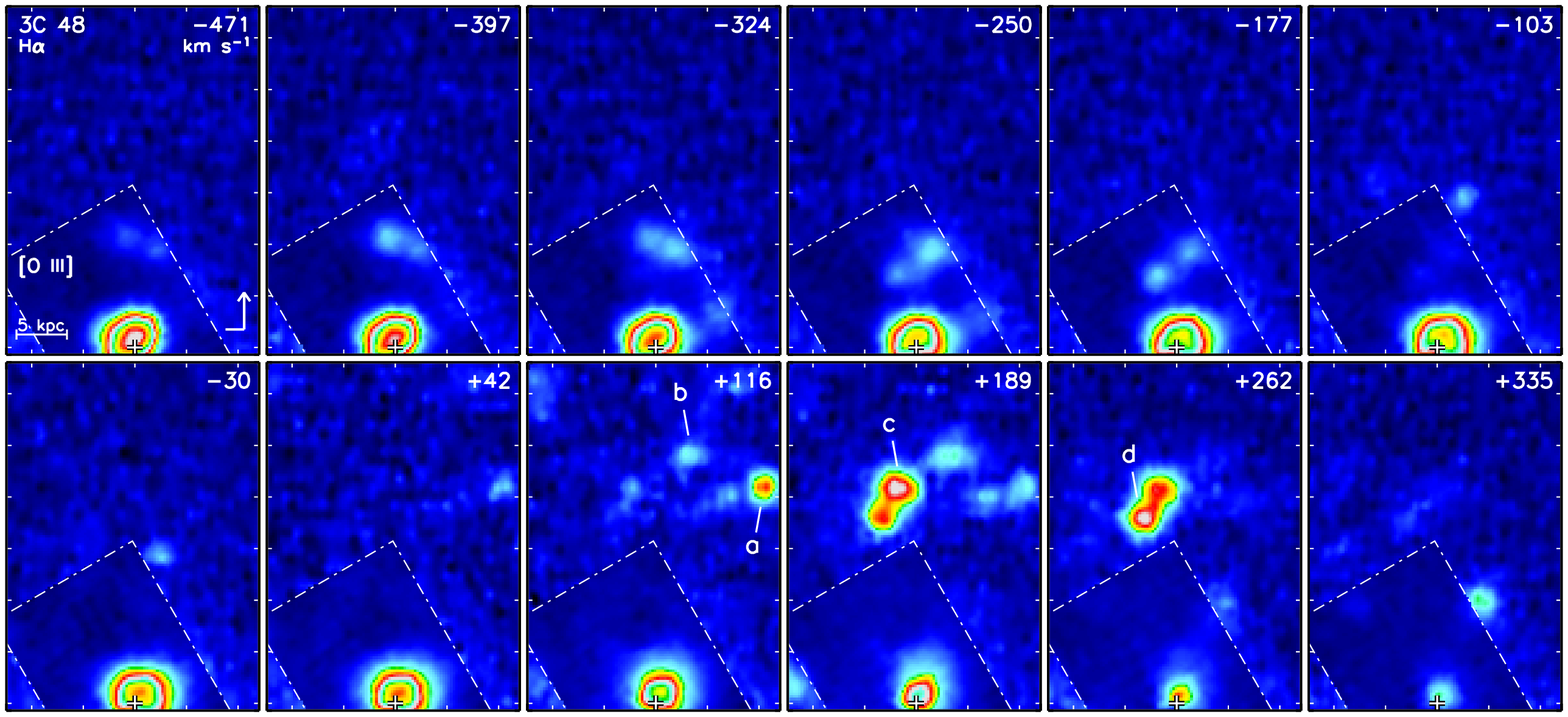} \vskip 0.1in
\plotone{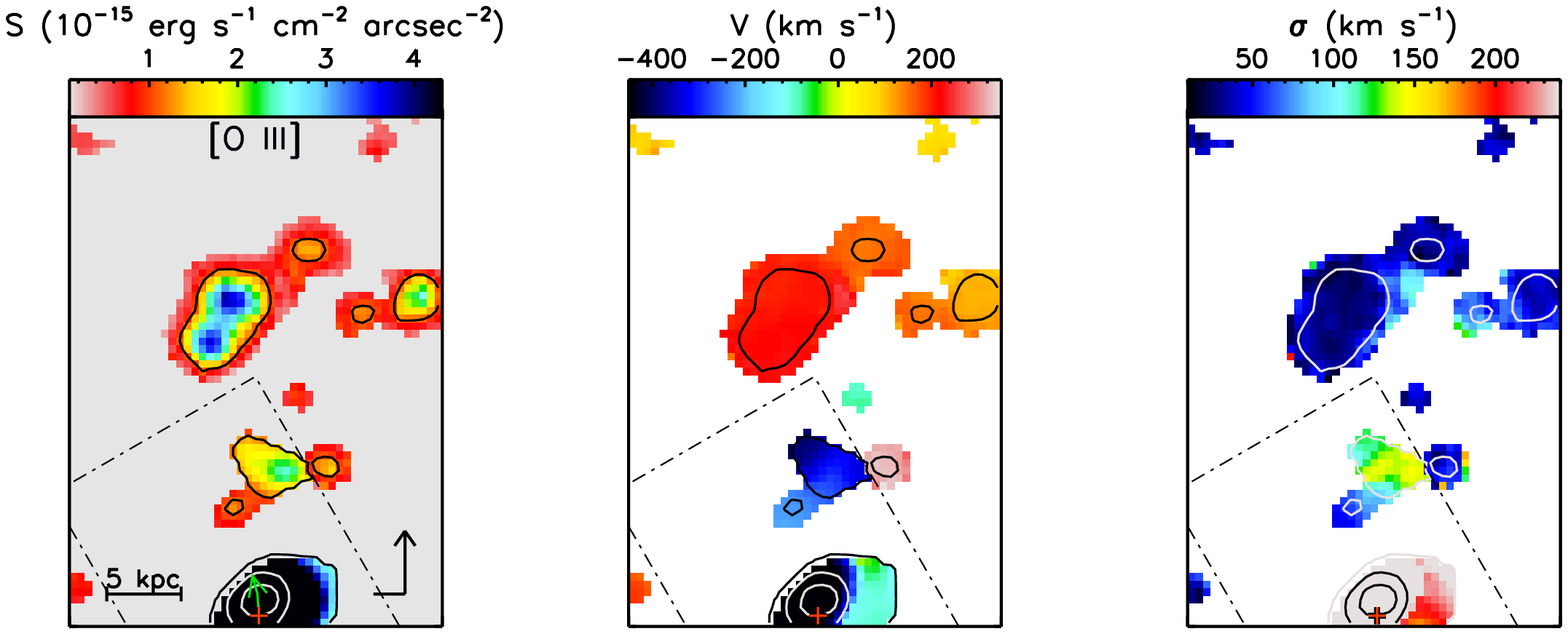}
\caption[Channel map and velocity field of the EELR of 3C\,48]{Channel map and velocity field of the EELR of 3C\,48. The plus signs indicate the position of the quasar before it was removed from the data cube. North and east are marked by the directional arrow. Scale bars of 5 kpc in length at the quasar redshift are shown. The channel maps are shown above the velocity field which covers the same area. The central velocities, in \kms, relative to that of the nuclear NLR (refer to Table~\ref{introtab:targets}) are indicated in the top right corner of each panel. The tick marks on the axes are spaced by 1\arcsec. Regions where full spectra have been extracted are labeled. The three columns of the velocity field are line intensity ($S$), radial velocity ($V$), and velocity dispersion ($\sigma$) maps. The anomalous cloud 0\farcs25 north of the nucleus has $V \simeq$ $-$500 \kms\ and $\sigma \simeq$ 390 \kms, and therefore exceeds the displayed range of the maps. Contours are from the intensity map, where the green arrow emanating from the nucleus delineates the compact radio jet of 3C\,48 \citep[the arrow ends at the component N2a in][]{Fen05}. The data outside of the dash-dot rectangle are from the H$\alpha$ line in the IFU2 data cube, which has been scaled and interpolated to match both the flux and the velocities of the \othree\,$\lambda5007$ line in the IFUR cube inside the rectangle.}
\label{fig:3c48}
\end{figure*}

3C\,48 ($z = 0.369$, 1\arcsec\ = 5.11 kpc) was not only the first quasar to be identified \citep{San61,Gre63}, but was also the first quasar to be associated with a luminous EELR \citep{Wam75}. It is a compact steep-spectrum (CSS) radio-loud quasar with a one-sided jet appearing disrupted within $\sim$0\farcs7 (3.6 kpc) to the north \citep{Wil91,Fen05}. 

The host galaxy of 3C\,48 is unusually luminous among low-redshift quasars \citep{Kri73}.  Strong Balmer absorption lines were detected long ago in the host galaxy, indicating the existence of a relatively young stellar population \citep{Bor82}. With deep spatially-resolved long-slit spectroscopy and a stellar population decomposition technique, \citet{Can00a} determined the ages for the most recent major star formation in 32 distinct regions
within 50 kpc of the nucleus, and they detected ongoing star formation and post-starburst populations less than 115 Myr old  in 26 of the 32 regions. Consistent with the spectroscopy results, the {\it IRAS} FIR luminosity and colors of 3C\,48 look quite similar to those of ULIRGs \citep{Sto91}; virtually all of the latter appear to be merger- or interaction-induced starbursts \citep{San96}. 
The optical morphology of the host galaxy also indicates an ongoing merger.
A tidal tail comprising an old stellar population is seen extending to the northwest \citep{Can00a}, and there is a bright continuum peak 1\arcsec\ northeast of the nucleus \citep[3C\,48A;][]{Sto91} comprising stars with a luminosity-weighted age of 140 Myr. 3C\,48A seems to be the distorted nucleus of the merging partner, and its stellar age may indicate star formation triggered during the previous close encounter \citep{Sto07}.

3C\,48 was first imaged in \othree\ through a narrow-band filter by \citet{Sto87}. Like the radio jet, the bulk of the emission-line gas is distributed to the north of the quasar nucleus (see Fig.~\ref{introfig:obs}$a$ for a \hst\ \othree\ linear-ramp filter [LRF] image, \citealt{Kir99}). \citet{Wam75} noticed that the \othree\,$\lambda5007$ line profile is unusually broad---26 \AA\ FWHM in rest-frame (\ie\ $\sim$ 1560 \kms). They also found strong \fetwo\ emission between H$\gamma$ and H$\beta$. Subsequent higher resolution spectra showed that the \othree\ line is  double-peaked \citep{Cha99,Can00a}: in addition to the usual nuclear NLR at the same redshift as the broad H$\beta$ line, there is an anomalous high-velocity narrow-line component 0\farcs25 north of the nucleus, precisely in the initial direction of the radio jet, but within the region where the jet appears relatively unperturbed \citep[][]{Sto07}. \citet{Cha99} also presented an \othree\,$\lambda5007$ velocity map covering a large fraction of the EELR, and they found no clear trends in the velocity field except for an apparent rotation within 1\arcsec\ to the nucleus.

In addition to the IFUR data cube that was used to study the high-velocity anomalous cloud and the 3C\,48A (both are within 1\arcsec\ from the nucleus), we obtained another IFUR data cube covering the EELR clouds further north and an IFU2 data cube covering the whole EELR (Fig.~\ref{introfig:obs}$a$ and Table~\ref{introtab:allifuobs}). As the IFU2 observation has the most complete spatial coverage and an excellent resolution, ideally we should construct a global velocity field from the IFU2 data cube alone using the H$\alpha$ line, which is the strongest line covered by the cube. However, the quasar was placed very close to the edge of the IFU2 FOV, making it difficult to remove the quasar scattered light. Therefore, we built a hybrid data cube by combining the \othree\,$\lambda5007$ region in the on-nucleus IFUR cube with the H$\alpha$ region in the IFU2 cube. First, the quasar was carefully removed using the Lucy-Richardson technique from the IFUR cube. Second, we shifted and interpolated the IFU2 H$\alpha$ cube to accurately match the velocities of the IFUR \othree\ cube, then we scaled the former by a constant to match the fluxes of the clouds 2\arcsec\ north of the quasar, which were covered by both cubes, to those of the IFUR cube. Finally, we rotated the IFUR cube and inserted it into the IFU2 cube in the overlapping area of the two. Figure~\ref{fig:3c48} shows the channel maps and the velocity field from this hybrid data cube. Our velocity maps are in generally good agreement with the previous integral field spectroscopy of \citet{Cha99}\footnote{Note that their relative velocities were calculated with respect to an arbitrary systemic velocity at $z$ = 0.368.}, but they offer a much finer spatial and velocity resolution. The highest relative velocities were observed in the anomalous cloud 0\farcs25 to the north and the group of 3 clouds $\sim$ 2\arcsec\ also to the north. The three-cloud group is also seen in the original IFU2 data cube in H$\alpha$ at the same relative velocities but with much lower S/N. The high-velocity cloud near the nucleus is also the only region where high velocity dispersions ($\sigma \gg$ 100 \kms) are seen.

We extracted a nuclear spectrum from the on-nucleus IFUR cube with an 0\farcs5-radius aperture centered on the quasar, from which we measured the redshifts of the high-velocity cloud and the NLR at systemic velocity. We found that (1) the high-velocity and systemic components were at redshifts $z$ = 0.36674$\pm$0.00006 and $z$ = 0.36926$\pm$0.00002; therefore there is a velocity difference of $-501\pm$16 \kms, and (2) their velocity dispersions ($\sigma$) are 392$\pm$21 and 148$\pm$5 \kms. The above nominal values and their errors were respectively the mean values and half of the differences measured from the \othree\,$\lambda5007$ line and the \othree\,$\lambda4959$ line. There are slight differences between our results and those of \citet{Sto07}, because we corrected for the atmospheric {\it B}-band absorption with a standard star spectrum taken on the same night. 

The \otwo\,$\lambda\lambda3726,3729$ doublet is well-resolved in the two brightest clouds ($c$ and $d$). Since their blue spectra come from the off-nucleus IFUR data cube which was taken during a period of poorer seeing ($\sim$ 0\farcs7), the two clouds were not resolved as clearly as in the IFU2 data cube. So, we combined their spectra and measured that \otwo\,$\lambda3726/\lambda3729$ = 0.946$\pm$0.036 and \othree\,$(\lambda4959+\lambda5007)/\lambda4363$ = 51$\pm$4, which implies an electron density ($N_e$) of 360$\pm$50 \cc\ and an electron temperature ($T_e$) of 17400$\pm$700 K (STSDAS task $temden$).

\begin{figure*}[!t]
\plotone{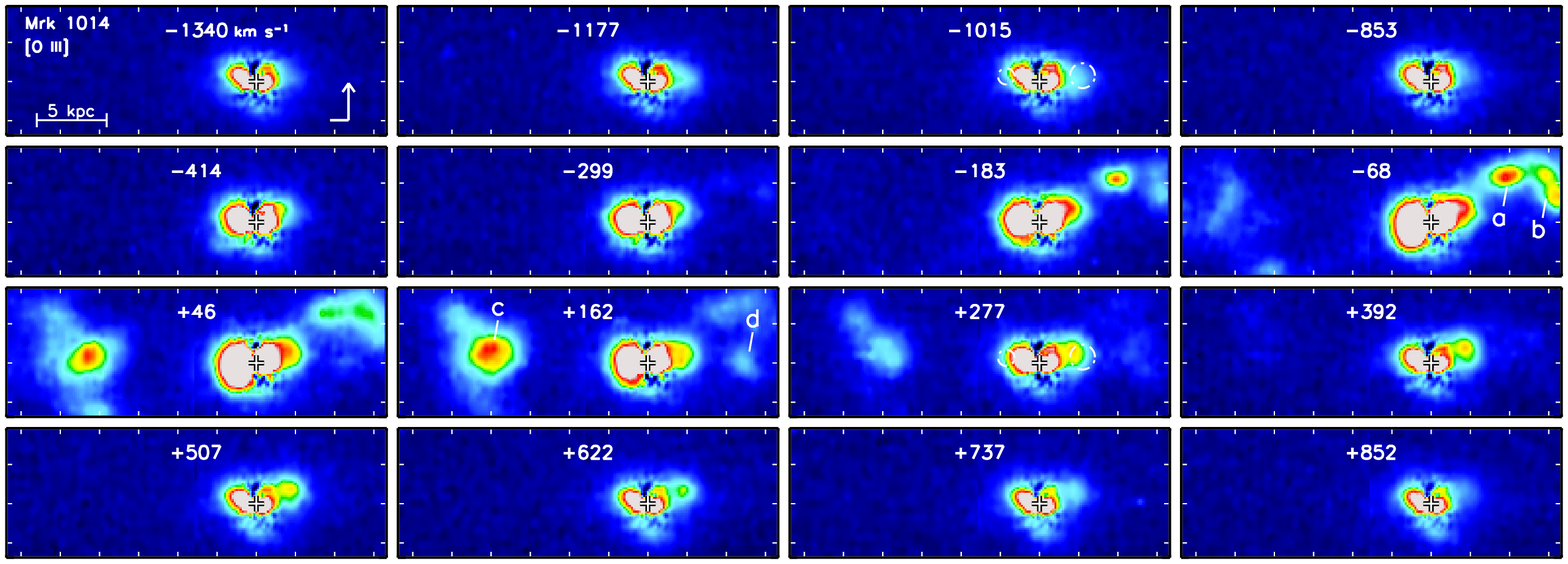} \vskip 0.1in
\plotone{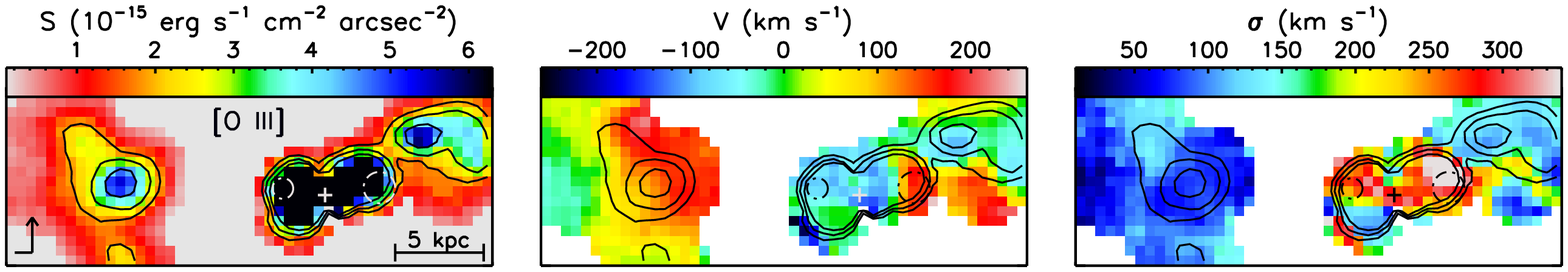}
\caption[Channel map and velocity field of the EELR of Mrk\,1014]{Channel map and velocity field of the EELR of Mrk\,1014. Keys are the same as in Fig.~\ref{fig:3c48}. Note that (1) there is a large gap in velocity between the first row and the second row of the channel map, and (2) the emission-line cloud appeared 1\arcsec\ to the west of the nucleus in panels at $-$1177 and $-$1105 \kms\ (This cloud is not shown in the velocity field, as it spatially overlaps with a more luminous cloud at $\sim$150 \kms). The dash-dot circles show the two off-nuclear radio components---they approximate the 0.06 mJy/beam contour in a VLA 8.4 GHz map \citep{Lei06a}. }
\label{fig:mrk1014}
\end{figure*}

\subsection{Mrk\,1014 (PG\,0157+001)}

Mrk\,1014 ($z$ = 0.163, 1\arcsec\ = 2.80 kpc) was discovered as an ultraviolet excess object in the Byurakan objective prism survey \citep{Mar77}. Subsequent spectroscopy observed a broad H$\beta$ line, confirming its nature as a low-redshift radio-quiet quasar \citep{Afa80} with an radio-loudness ($R$) parameter of 2.1 \citep{Kel89}. 
The low-luminosity radio source was resolved into a triple source at a resolution of 0\farcs36 by the Very Large Array (VLA)---two radio knots within 1\farcs1 (3.1 kpc) on either side of a central component that corresponds to the nucleus \citep{Lei06a,Mil93,Kuk98}. 
The host galaxy of Mrk\,1014 is very similar to that of 3C\,48. The FIR luminosity and colors qualify Mrk\,1014 as a ULIRG. It has a prominent spiral-like tidal tail extending to the northeast, the spectra of which showed stellar absorption features of a ``K+A" character \citep{Mac84}. Strong \fetwo\ emission is also present in the nuclear spectrum. Post-starburst populations of ages between 180 and 290 Myr were detected in 5 distinct regions in the host galaxy, which, together with its morphology, suggest that Mrk\,1014 is a result of a merger of two disk galaxies of comparable size \citep{Can00b}.

Mrk\,1014 is the only radio-quiet quasar with a luminous EELR in the \citet{Sto87} sample (Fig.~\ref{introfig:corr}$a$). The morphology of the EELR, although on a much larger scale, resembles that of the radio emission (Fig.~\ref{introfig:obs}$b$). There is evidence for strong jet-cloud interactions within the extent of the radio components ($<$ 1\farcs1 or 3.1 kpc). By subtracting an \hst\ narrow-band line-free image off an \othree\,$\lambda5007$ LRF image, \citet{Ben02} identified a filamentary emission-line structure 0\farcs9 to the west of the nucleus, which appeared coincident with the western radio knot 1\farcs1 from the central radio component at a P.A. of $-$82.2$^{\circ}$. Near the same region, \citet{Can00b} detected a weak high-velocity cloud at $-$1350 \kms\ as well as a strong emission-line cloud at $-$200 \kms\ (both velocities were calculated with respect to the underlying stellar features). This high-velocity cloud was confirmed by \citet{Lei06b}, who also noticed that there was a broad ($\sigma \sim$ 590 \kms) blueshifted component in addition to the classical NLR in the nuclear \othree\ lines, which are reminiscent of the double-peaked \othree\ lines in 3C\,48. 

\begin{figure*}[!t]
\plotone{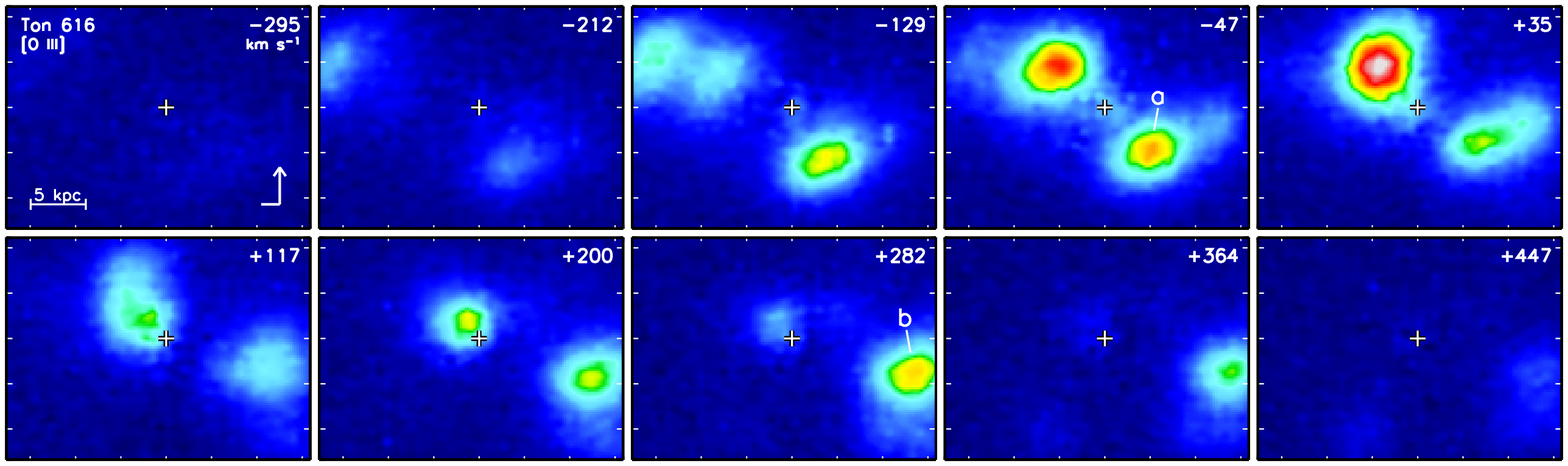} \vskip 0.1in
\plotone{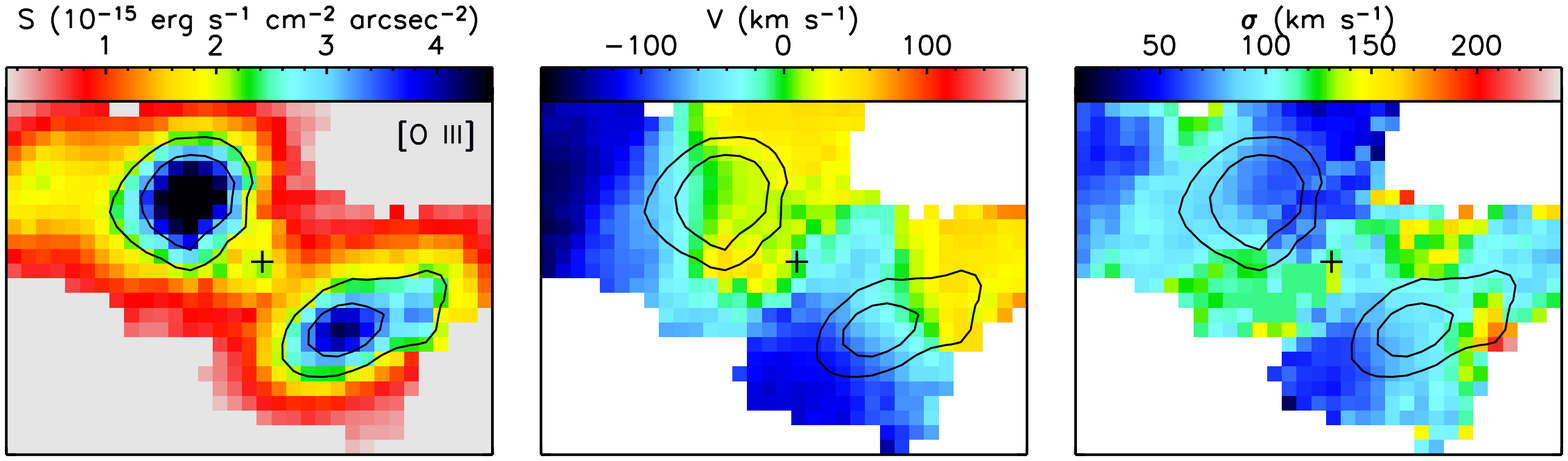}
\plotone{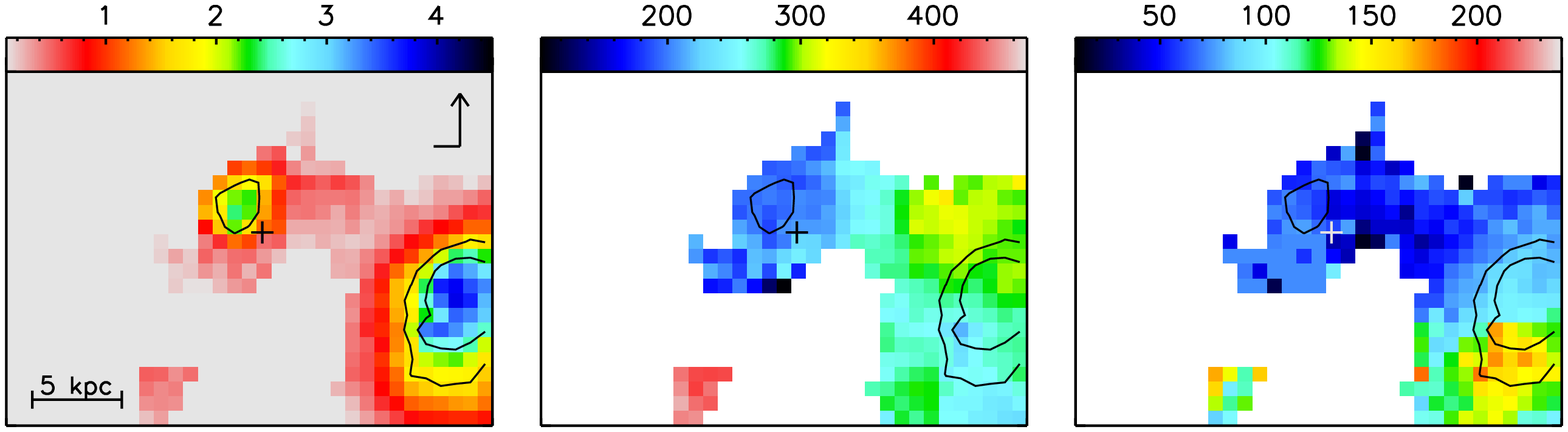}
\caption[Channel map and velocity field of the EELR of Ton\,616: central region]{Channel map and velocity field of the EELR of Ton\,616: central region. Keys are the same as in Fig.~\ref{fig:3c48}. To separate multiple clouds that are present along the same line of sight, the velocity field is shown in two velocity bands ($-$170 to 170 \kms\ and 105 to 470 \kms).}
\label{fig:ton616a}
\end{figure*}

When removing the quasar nucleus from the data cube with $plucy$, Mrk\,1014 had to be dealt with differently from the rest of the sample because the quasar is not the only non-negligable continuum source near the nucleus. \citet{Ben02} pointed out that the \hst\ line-free continuum image of Mrk\,1014 was clearly extended compared with that of a star. We also observed a significant {\it decrease} in the FWHM of the central source at the wavelengths of the broad H$\beta$ and H$\gamma$ lines in the data cube, implying the presence of a continuum source that has a spectrum different from that of the quasar. So, we decided to use the on-nucleus IFUR data cube of 3C\,48 to create the PSF because it was observed with a very similar instrument setting and under similar seeing. Due to the well known wavelength dependence of the spatial resolution, we had interpolated the 3C\,48 data cube to the same wavelength grid as that of the Mrk\,1014 cube before generating the PSF.

Figure~\ref{fig:mrk1014} shows the channel maps and the velocity field after removing the quasar nucleus. A peanut-shaped continuum source is evident at all velocities. It is unclear what the nature of the continuum source is from the current data, but it is possibly synchrotron radiation from the radio jets connecting the radio core and the knots. The high-velocity cloud 1\arcsec\ to the west is seen in the channel maps centered at $-$1177 and $-$1015 \kms and appears offset from the radio knot to the southeast. At almost the same location, another emission-line cloud is seen at lower velocities which appears to have a much higher velocity dispersion and to be offset from the radio knot to the northeast (see the panel at +277 \kms). Although rather disordered, the velocity field shows that most of the line-emitting clouds have velocities within 200 \kms\ from the systemic velocity. Another similarity to the EELR of 3C\,48 is that the only region of high velocity dispersions ($\sigma \gg$ 100 \kms) is the area occupied by the radio structure.

Although the \otwo\ doublet is unresolved as a result of the low redshift of Mrk\,1014, the \stwo\,$\lambda\lambda6716,6731$ doublet allows accurate measurements to be made. This is possible because at $z < 0.26$ the \stwo\ lines remain in the $i'$-band where CCD fringing is much less severe than in the $z'$-band. Unfortunately, the latter is the case for the other four quasars. The ratio of the \stwo\ doublet can be used to infer electron densities \citep{Ost89}. We measure \stwo\,$\lambda6716/\lambda6731$ ratios of 1.30$\pm$0.03, 1.29$\pm$0.03, 1.30$\pm$0.03, and 1.52$\pm$0.09 in regions $a$, $b$, $c$, and $d$, respectively. The results for $a$, $b$, and $c$ imply a density of $N_e$ = 121$\pm$30 \cc\ if $T_e$ = 10$^4$ K, or $N_e$ = 117$\pm$40 \cc\ if $T_e$ = 1.5$\times$10$^4$ K. The \stwo\ ratio of region $d$ is at the low density limit: $N_e <$ 80~\cc. 

The atmospheric {\it A}-band absorption affects the intensities of the \ntwo\,$\lambda6548$, H$\alpha$ and \ntwo\,$\lambda6583$ lines; therefore their intensities had to be corrected. The empirical absorption curve that we used to correct for the {\it A}-band absorption was derived from an observation of an A0V star (HIP\,13917) taken at a similar airmass immediately after the science exposures.

\subsection{Ton\,616 (4C\,25.40)}

\begin{figure*}[!t]
\plotone{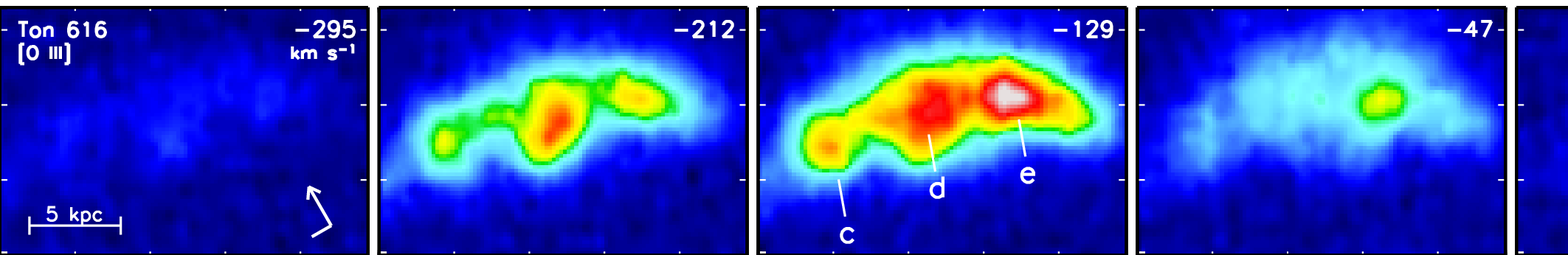} \vskip 0.1in
\plotone{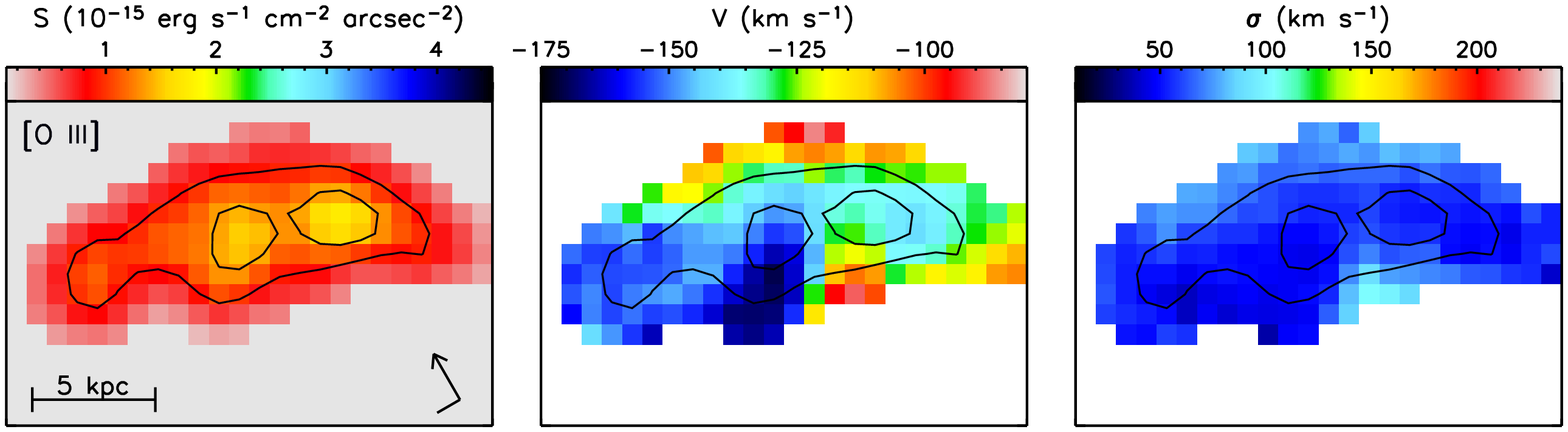}
\caption[Channel map and velocity field of the EELR of Ton\,616: northern arch]{Channel map and velocity field of the EELR of Ton\,616: the northern arch.}
\label{fig:ton616b}
\end{figure*}

Ton\,616 ($z$ = 0.268, 1\arcsec\ = 4.11 kpc) was first cataloged as a Tonantzintla blue star \citep{Iri57}. The bright radio source linked with Ton\,616 was found in the Fourth Cambridge radio survey \citep[4C;][]{Pil65}. Subsequent spectroscopy identified Ton\,616 as a quasar with a redshift of 0.268 \citep{Lyn68}. The radio emission has a classical FR II type morphology, showing hot spots $\sim$35\arcsec\ (144 kpc) to the northeast and southwest of a faint core at a P.A. of $30^{\circ}$ \citep{Gow84}. 

The host galaxy of Ton\,616 is an elliptical galaxy with an effective radius of $r_{1/2} \simeq$ 20 kpc\footnote{Converted to our chosen cosmology.}, as measured in a ground-based $K$-band image \citep{Dun93,Tay96}. No signs of disturbance have been reported in the literature. Unlike those of 3C\,48 and Mrk\,1014, the nuclear spectrum of Ton\,616 does not show strong \fetwo\ emission \citep[\eg][]{Bal75,Ste81}. 

The EELR of Ton\,616 consists of (1) an inversion-symmetric distribution of gas in the vicinity of the quasar nucleus, (2) a detached arch 7\farcs5 to the north, (3) and some fainter clouds scattered in the south and southwest \citep{Sto87}. Our integral field spectroscopy completely covered the first two components (Fig.~\ref{introfig:obs}$d$). The velocity field of the central region and the northern arch are shown in Fig.~\ref{fig:ton616a} and \ref{fig:ton616b}, respectively. Partly due to the mediocre seeing (0\farcs9) of the IFU2 observations, line blending (\ie\ multiple kinematic components present in the \othree\,$\lambda5007$ line of a single spectrum) occurs quite often. Thus, we display the velocity field in two velocity bands. The three clouds comprising the northern arch are essentially at the same velocity ($\sim -$160 \kms). Our results are in good agreement with the velocity map of \citet{Dur94} from TIGER integral field spectroscopy and the rather coarse map of \citet{Boi94} from long-slit spectroscopy. The much higher spectral resolution of GMOS/IFU in comparison with TIGER allowed us to cleanly disjoin spatially overlapping but kinematically independent components in the two regions where \citeauthor{Dur94} noticed line widths significantly broader than the instrumental resolution.  

Since only the southwest part of the central IFU2 field was observed with a wide spectral coverage setting (\ie\ the on-nucleus IFUR observations), we could only extract spectra from two clouds ($a$ and $b$) for the emission-line analysis. Unfortunately, the observations meant to cover the blue spectra were never executed---we thus do not have intensity measurements for emission lines shortward of \hetwo\,$\lambda4686$. The data cube of the northern arch in fact covers a wavelength range from \otwo\,$\lambda3727$ to \stwo\,$\lambda6731$. But as a result of the low efficiency ($<$ 40 \%) of the R400/G5305 grating at $\lambda <$ 5500 \AA, all of the emission-lines at wavelengths shorter than \hetwo\,$\lambda4686$ are too noisy to be measured. 

Only in region $e$ is the \stwo\ doublet sufficiently strong relative to the noise to allow a meaningful measurement of its ratio: \stwo\,$\lambda6716/\lambda6731$ = 1.33$\pm$0.14, implying $N_e$ = 90$\pm$160 (80$\pm$200) \cc\ if $T_e$ = $10^4$ ($1.5\times10^4$) K. 

\subsection{Ton\,202 (B2\,1425+267)}

Ton\,202 ($z$ = 0.364, 1\arcsec\ = 5.07 kpc) was also one of the Tonantzintia blue stars \citep{Iri57}. The radio source was first listed in the B2 catalogue \citep{Col72}. Ton\,202 was first suspected to be a quasar by \citet{Gre66} based on an ambiguous detection of several broad emission lines, and this identification was subsequently confirmed and the object was assigned a redshift of 0.366 as measured from the broad H$\beta$ line \citep{Wee76}. The radio source has an FR II morphology with one radio hot spot 140\arcsec\ (709 kpc) to the northeast at a P.A. of 52$^{\circ}$ and another one 83\arcsec\ (420 kpc) to the southwest at a P.A. of $-$120$^{\circ}$ \citep{Rog86}. 

By modeling an \hst\ WFPC2 broad-band images of Ton\,202, \citet{Kir99} classified the host galaxy as an elliptical and measured an effective radius of r$_{1/2} \simeq$ 12 kpc. An arc-like feature to the west of the nucleus was seen in the \hst\ F555W image \citep{Kir99} and a CFHT $H$-band adaptive optics image \citep{Mar01}, suggesting mild galactic interactions. This feature was also detected by \citet{Sto83} in their line-free continuum image, but the feature is not seen in the \hst\ F814W image. It is clear from these images that the host galaxy of Ton\,202 is not experiencing as dramatic interactions as those of 3C\,48 and Mrk\,1014. Like that of Ton\,616, the nuclear spectrum of Ton\,202 does not show strong \fetwo\ emission \citep[\eg][]{Ste81,Bor85}.

\begin{figure*}[!t]
\plotone{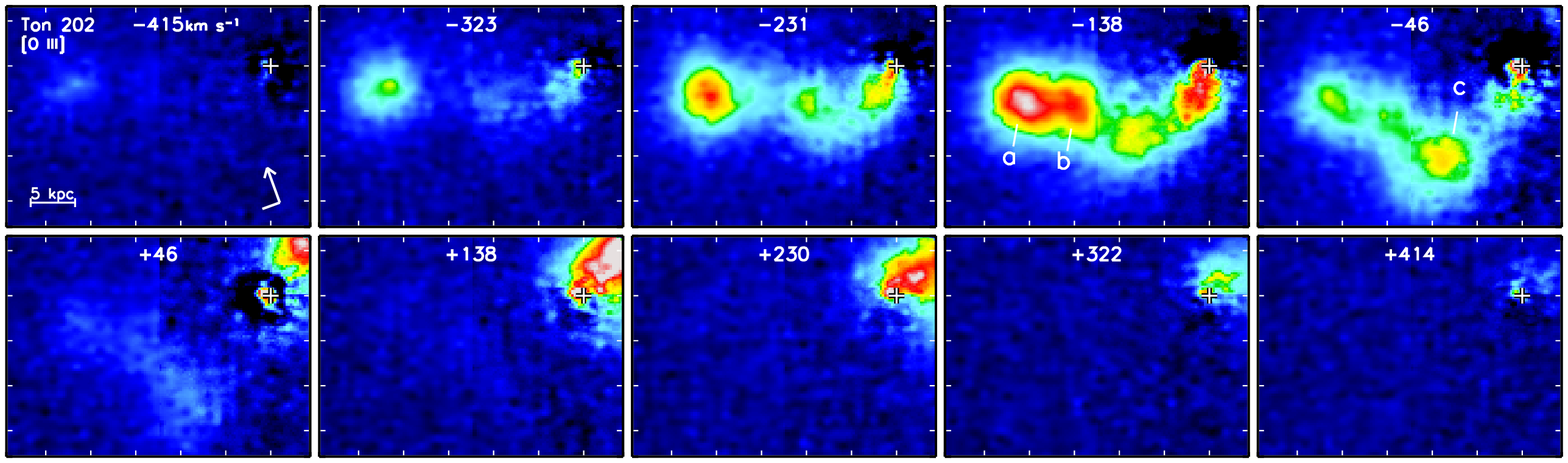} \vskip 0.1in
\plotone{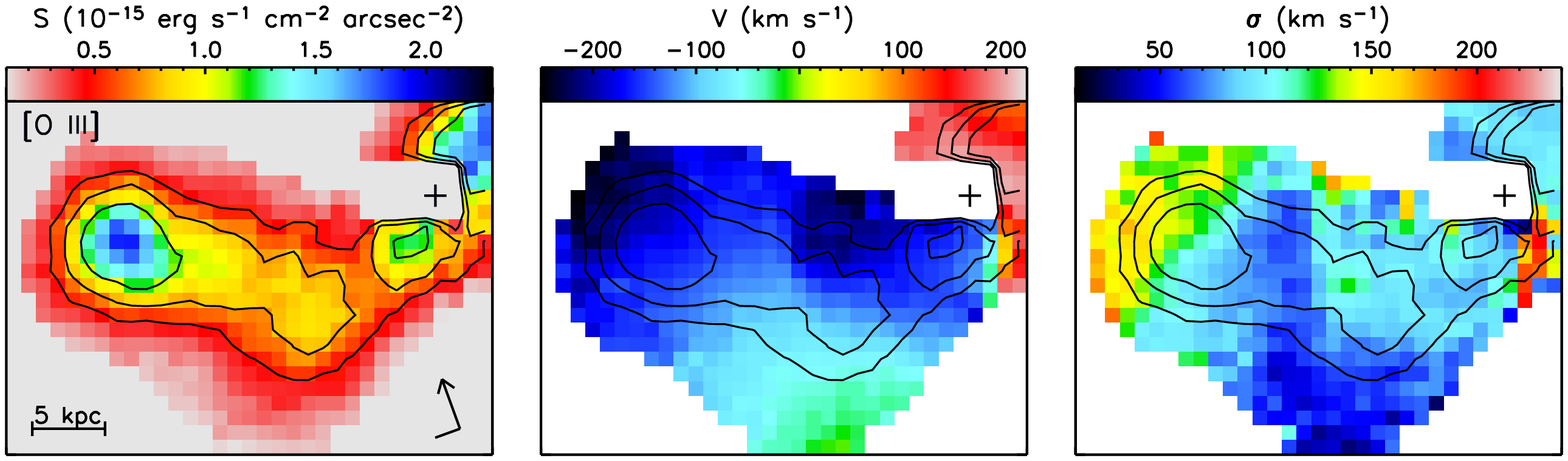}
\caption[Channel map and velocity field of the EELR of Ton\,202]{Channel map and velocity field of the EELR of Ton\,202. Keys are the same as in Fig.~\ref{fig:3c48}.}
\label{fig:ton202}
\end{figure*}

The EELR of Ton\,202 was discovered by \citet{Sto83} in their narrow-band \othree\ imaging survey. The nebulosity shows two impressive spiral-like arms extending over 40 kpc (Fig.~\ref{introfig:obs}$e$). Long-slit spectroscopy in the \othree\ showed that the brighter eastern arm was approaching at a velocity of $-$170 \kms\ while the fainter western arm is receding at $+$190 \kms\ with respect to the systemic velocity \citep{Boi94}. 
We present in Fig.~\ref{fig:ton202} the \othree\ channel maps and the velocity field of the brighter eastern arm from the mosaicked IFUR data cube. Similar trends were seen in H$\alpha$, from the IFU2 data cube (which only covers the H$\alpha$ $-$ \stwo\ region), although the data are much noisier. We measured a velocity range of $-$250 to $-$100 \kms\ in the brightest condensation $a$, consistent with the velocities reported by \citet{Boi94}. The entire velocity field is definitely more complex than pure rotation.

Because of the poor S/N at both the \otwo\ and the \stwo\ doublets, we were unable to measure  the electron density in this EELR.

\begin{figure*}[!t]
\plotone{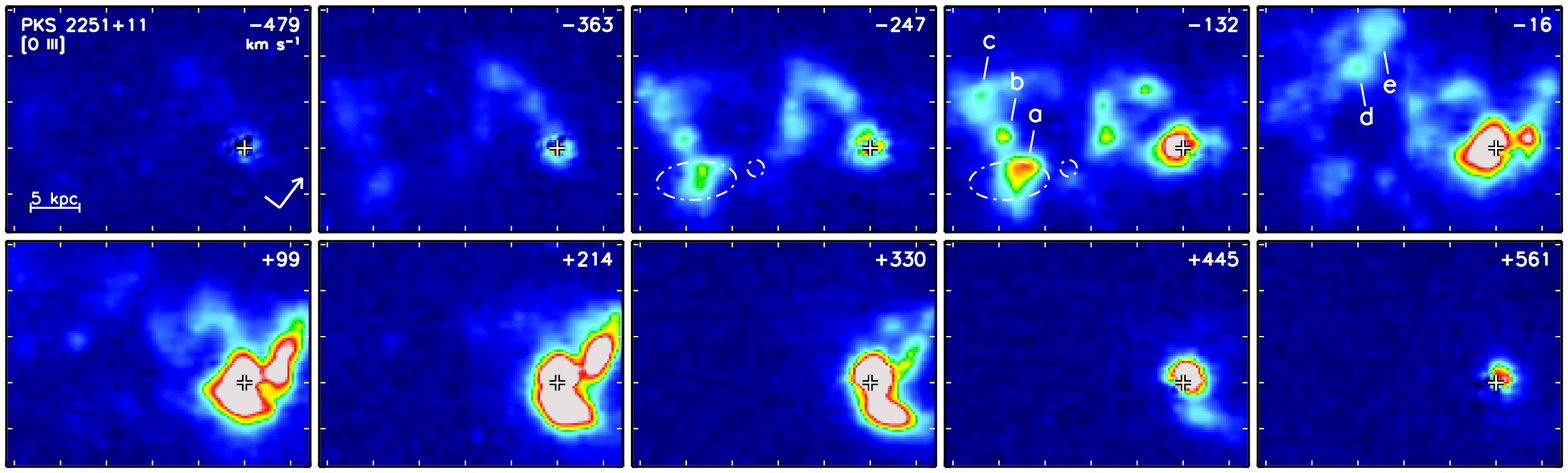} \vskip 0.1in
\plotone{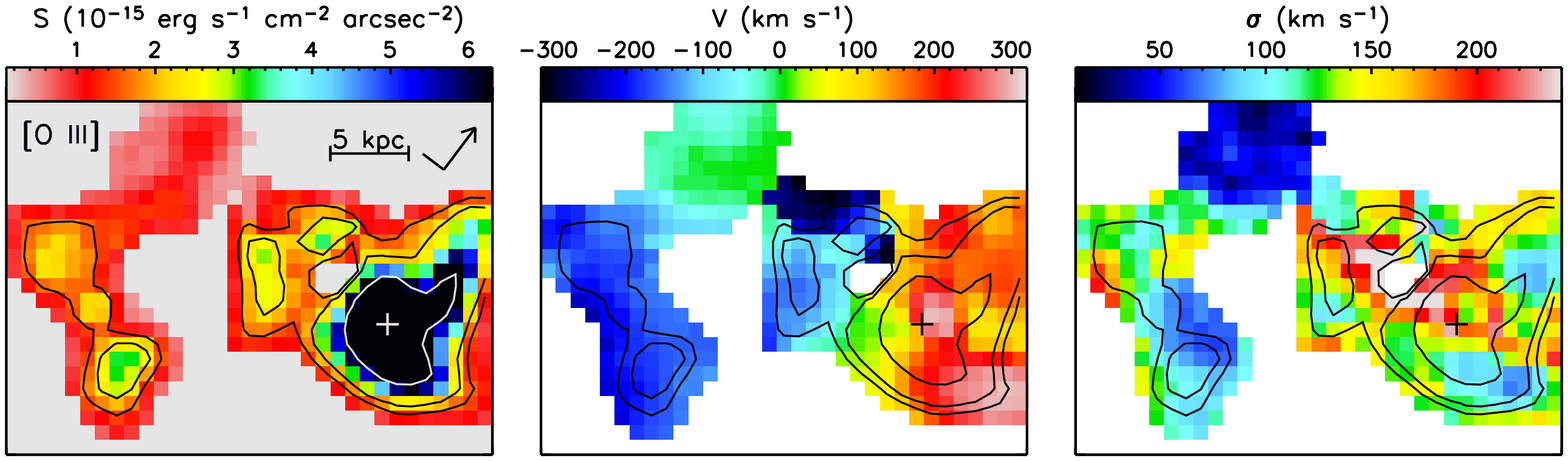}
\caption[Channel map and velocity field of the EELR of PKS\,2251+11]{Channel map and velocity field of the EELR of PKS\,2251+11. Keys are the same as in Fig.~\ref{fig:3c48}. The white dash dot ellipses in some panels describe the radio hot spots---they approximate the 12.7 mJy/beam contour in a VLA 4.86 GHz radio map \citep{Pri93}.}
\label{fig:pks2251}
\end{figure*}

\begin{figure*}[!t]
\epsscale{0.55}
\plotone{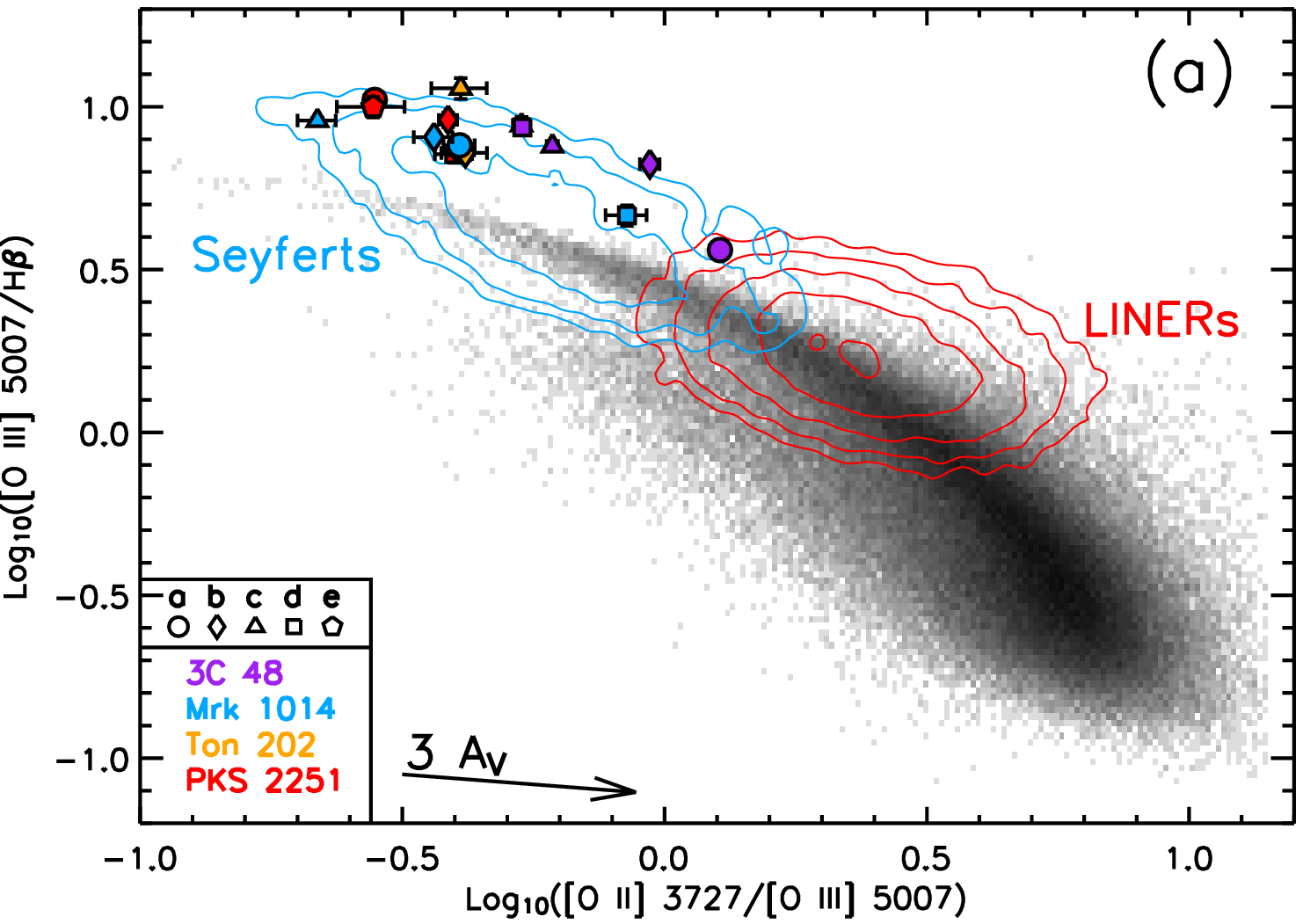}
\plotone{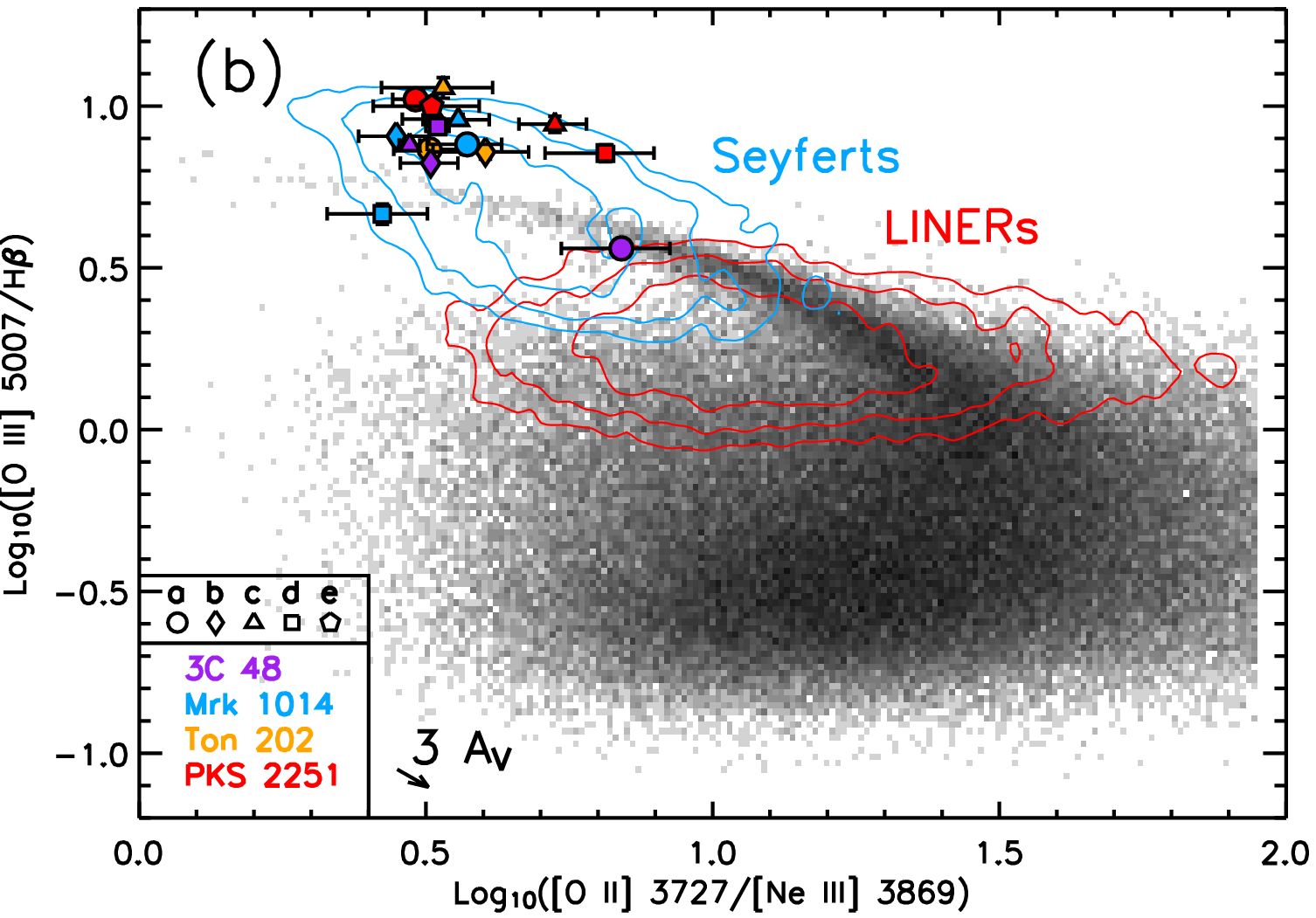}
\vskip 0.1in
\plotone{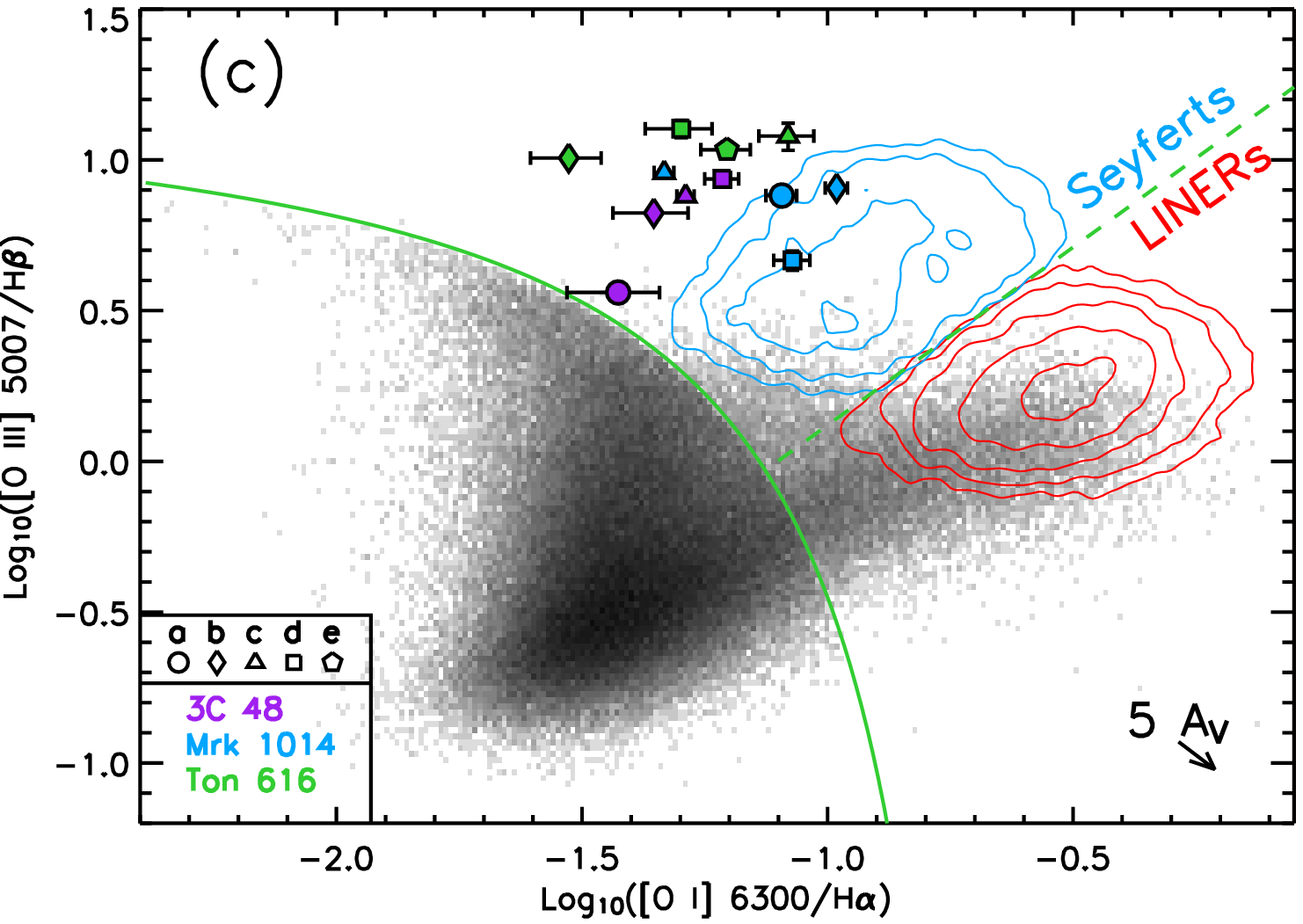}
\plotone{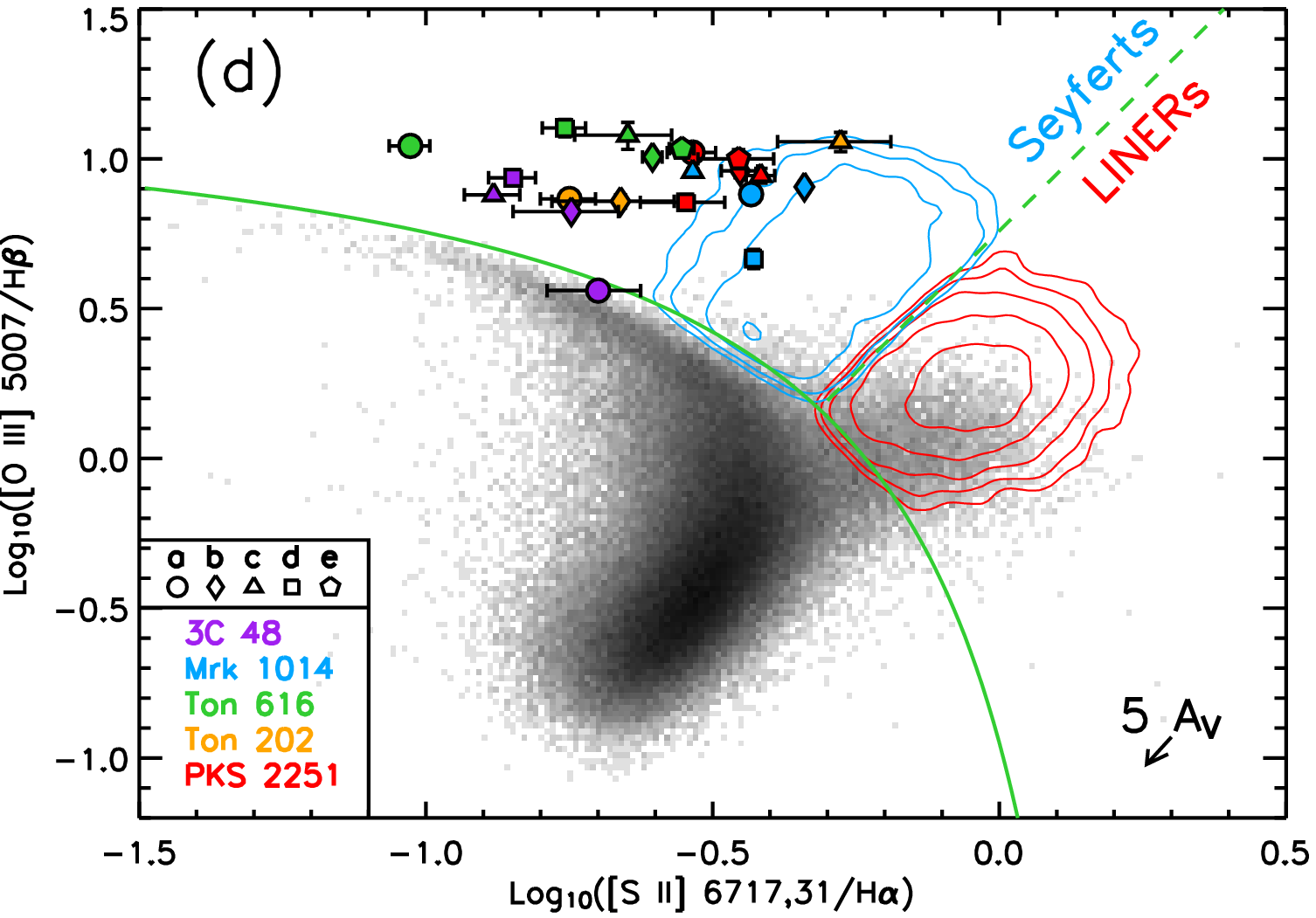}
\caption[Line ratios of quasar EELRs indicate photoionization by the central source]{
Line ratios of quasar EELRs indicate photoionization by an AGN-type ionizing spectrum. 
Log-scaled density distributions of SDSS DR4 emission-line galaxies are displayed in line-ratio diagrams: \othree\,$\lambda5007$/H$\beta$ vs.  
({\it a}) \otwo\,$\lambda\lambda3726,3729$/\othree\,$\lambda5007$,
({\it b}) \otwo\,$\lambda\lambda3726,3729$/[Ne\,{\sc iii}]\,$\lambda3869$,
({\it c}) \oone\,$\lambda6300$/H$\alpha$, 
and ({\it d}) \stwo\,$\lambda\lambda6717,6731$/H$\alpha$.
Measurements from various clouds in the EELRs of 3C\,48, Mrk\,1014, Ton\,616, Ton\,202, and PKS\,2251+11 are shown as color symbols with error bars. The symbol colors and shapes follow the coding system in the legend (\eg\ 3C\,48-$a$ is indicated by the purple circle). 
Seyfert2s, LINERs and star-forming galaxies are well separated, by definition, in the last diagram. However, they appear quite blended in the first three diagrams. Therefore,
Seyfert2s are shown as blue contours, LINERs as red contours, and star-forming galaxies (including star-forming---AGN composite galaxies) as the background image.
Arrows are reddening vectors. 
3C\,48-$a$ is the only cloud that has a significant star-forming component (see also Fig.~\ref{fig:3c48n2}$a$).
}
\label{fig:3c48o2}
\end{figure*}

\begin{figure*}[!t]
\epsscale{0.55}
\plotone{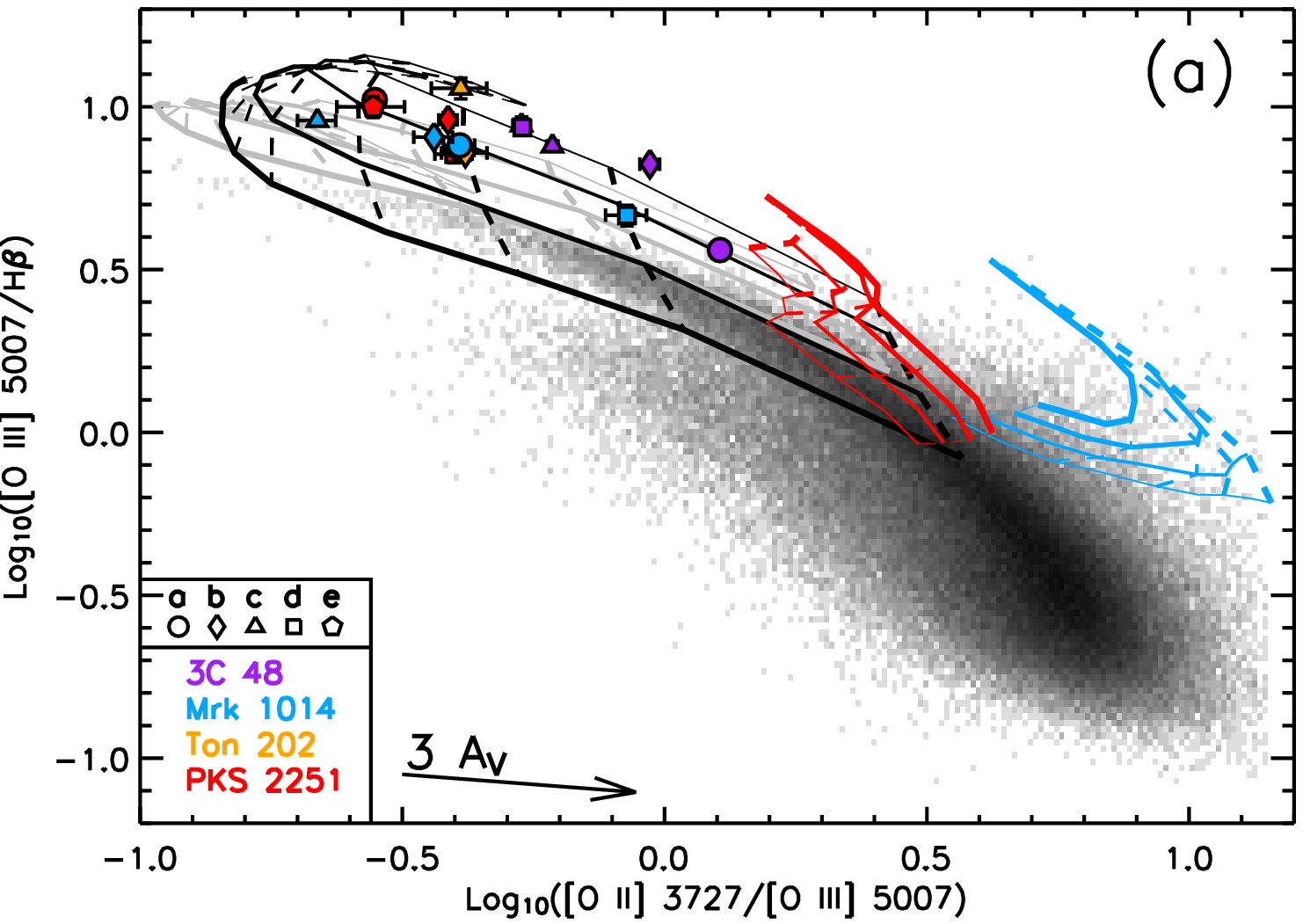}
\plotone{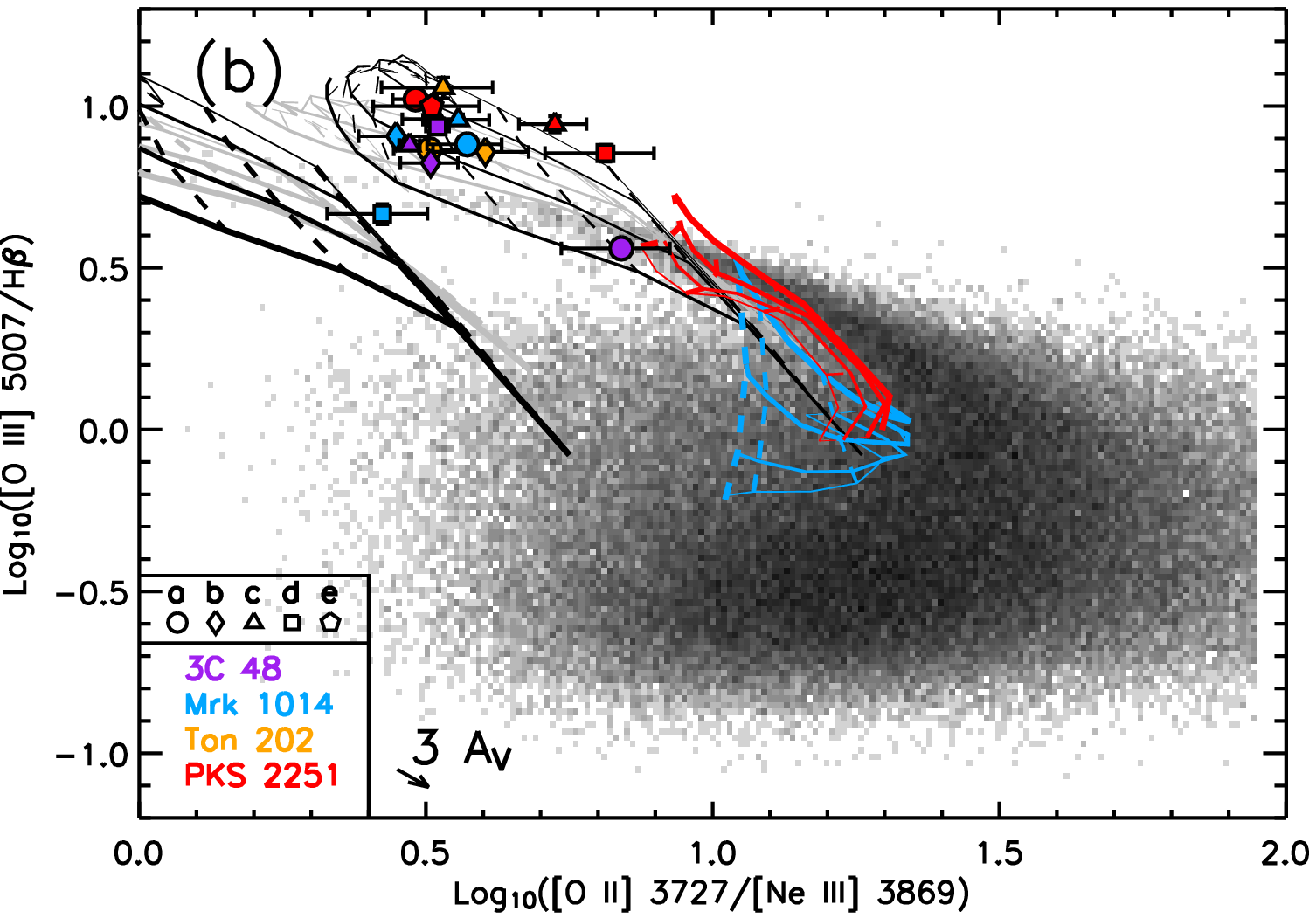}
\vskip 0.1in
\plotone{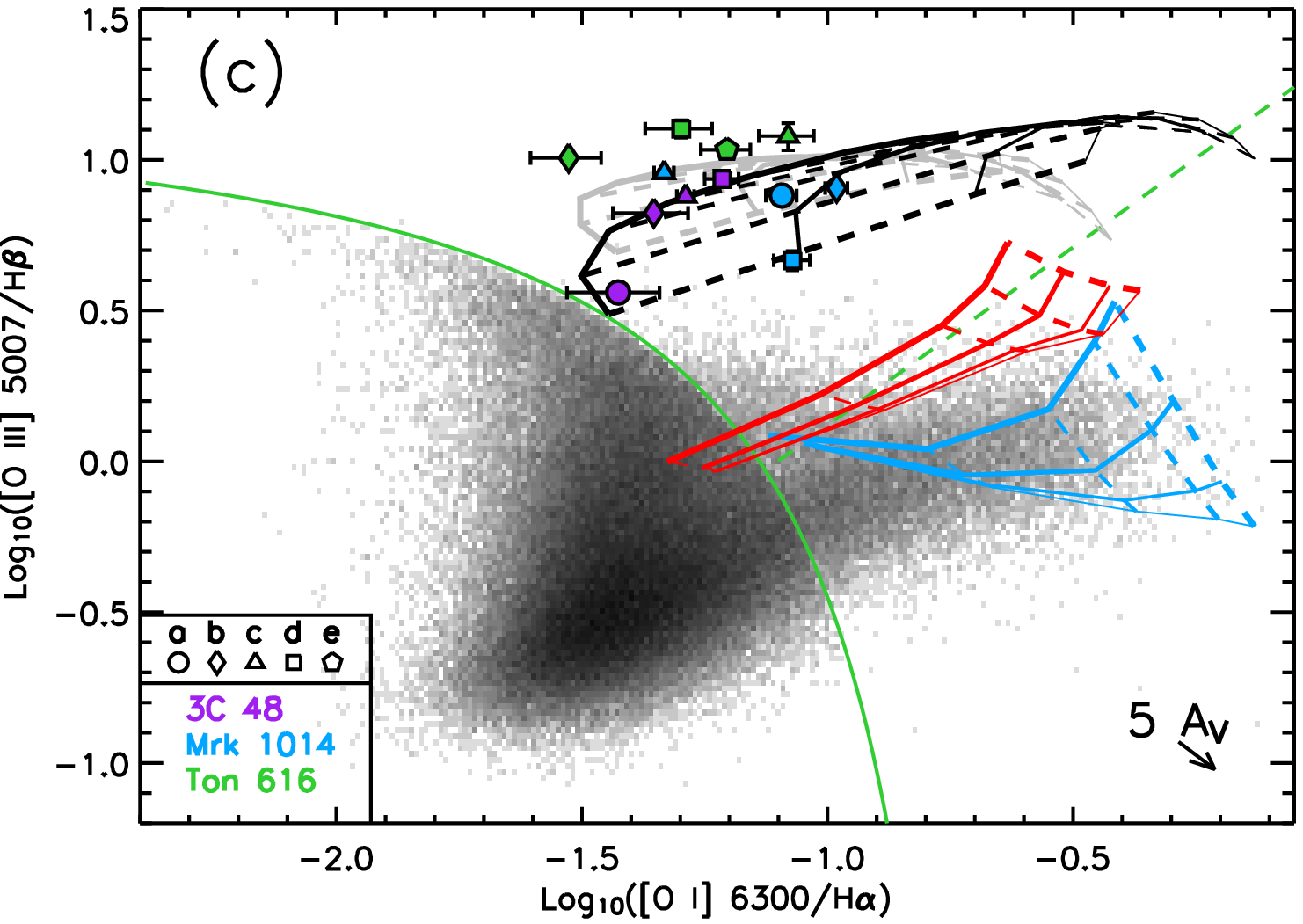}
\plotone{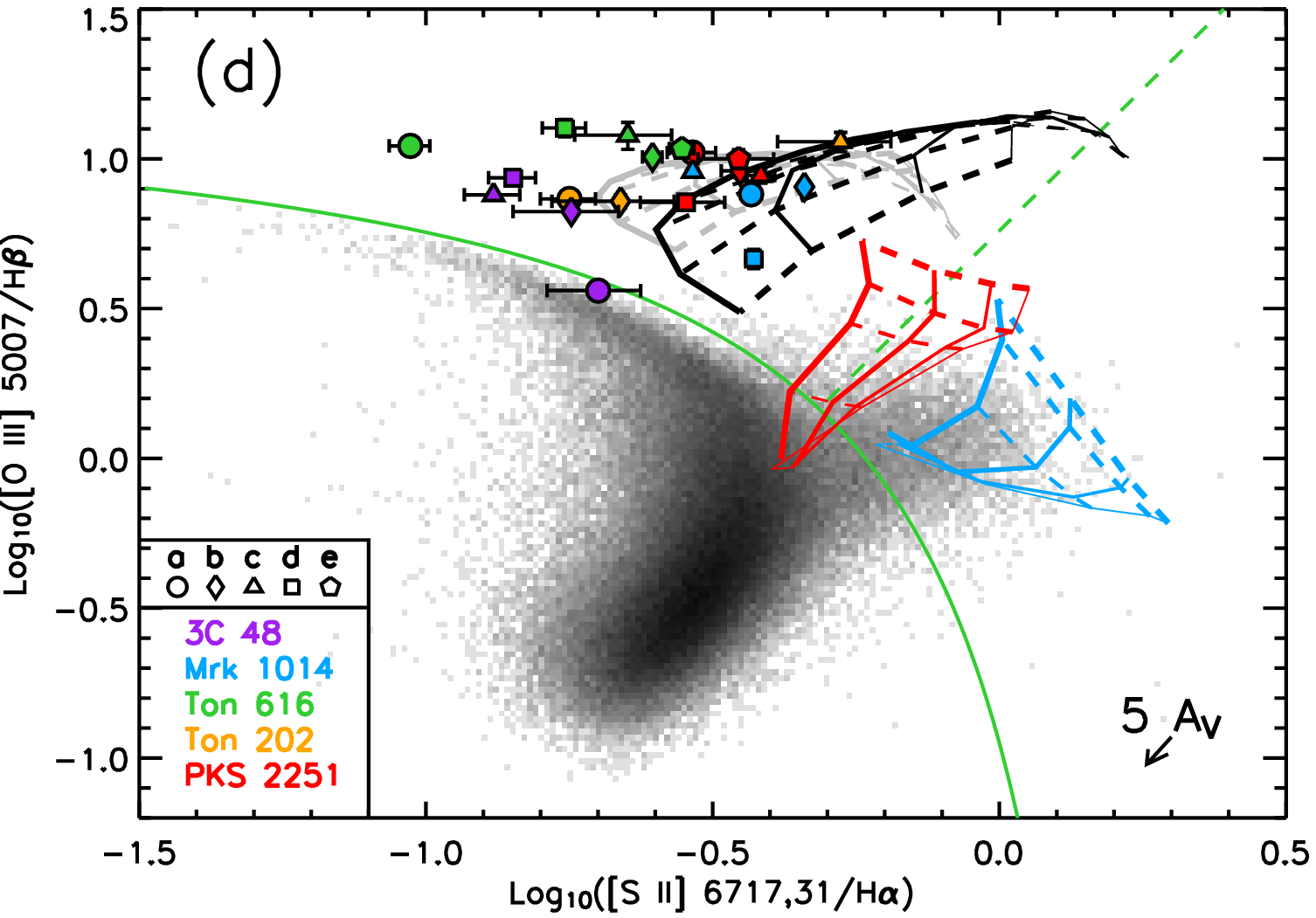}
\caption[Line ratios of quasar EELRs indicate photoionization by the central source]{
The same line-ratio diagrams as in Fig.~\ref{fig:3c48o2}, although ionization model grids have been plotted in replacement of SDSS Seyferts and LINERs. The dusty radiation-pressure dominated photoionization \citep{Gro04a} model grids for gas-phase metallicities of 1/3 and 2/3 \zsun\ are shown in gray and black, respectively. Both grids assume a hydrogen density near the ionization front of 1000 \cc, and they cover the same range of ionization parameters ($0.0 \geq {\rm log} U \geq -3.0$, dashed lines with increasing thickness) and the same four power-law indices for the ionizing continuum ($F_{\nu} \propto \nu^{\alpha}$; $\alpha = -1.2, -1.4, -1.7$, and $-2.0$; solid lines with increasing thickness). The shock-only ({\it blue}) and ``shock + precursor" models \citep[{\it red};][]{Dop96} show the same range of shock velocities ($200 \leq V_S \leq 400$ \kms; dashed lines with increasing thickness) and four magnetic parameters ($B/n^{1/2} = 0.5, 1, 2$, and 4 $\mu$G cm$^{3/2}$; solid lines with increasing thickness).
In panel $b$, the original photoionization model grids do not match the data. We applied a simple correction to the model grids by shifting them to the right by 0.51 in log, which is motivated by a recent update on the O/Ne abundance ratio [log (O/Ne) = 1.12 \citep{Glo07} as opposed to the old value of 0.61 that was used in \citet{Gro04a}]. The girds after the correction are shown in thiner black and grey lines.
}
\label{fig:3c48o2_grid}
\end{figure*}

\subsection{PKS\,2251+11}

PKS\,2251+11 ($z$ = 0.325, 1\arcsec\ = 4.70 kpc) was discovered in the Parkes radio sky survey \citep{Day66}. Soon the optical counterpart was identified \citep{Bol66}, and the object was confirmed to be a quasar from its ultraviolet excess \citep{Kin67a} and a redshift was determined \citep{Kin67b}. The extended radio emission is an FR II source with a P.A. of 138$^{\circ}$ and an total extent of only 11\arcsec\ (52 kpc), and the southeast lobe appears much closer to the core than the northwestern lobe \citep{Pri93,Mil93}. Optical broad-band imaging revealed that PKS\,2251+11 is hosted by an elliptical galaxy with no signs of disturbance \citep{Hut92}. A moderate level of \fetwo\ emission is present in the nuclear spectrum of the quasar \citep[\eg][]{Hut90}.

\citet{Sto87} noticed that there was a clear correspondence between the southeast part of the EELR and the radio structure based on a lower resolution VLA map \citep{Hin83}. One of the two bright emission-line clouds $\sim$ 4\arcsec\ southeast of the quasar appears to be coincident with a hot spot in the southeast radio lobe (Fig.~\ref{introfig:obs}$g$). \citet{Hut90} caught a glimpse of the chaotic velocity structure of the EELR by taking off-nucleus spectra along three slit positions. More detailed velocity maps became available as the object was investigated by integral field spectroscopy \citep{Dur94,Cra00}. A bipolar velocity structure was explicitly pointed out by \citet{Cra00}---redshifted clouds are distributed in an envelope in the vicinity of the nucleus, while more extended clouds are mostly blueshifted. It was also realized that a physical connection between the southeast cloud and the radio jet based on their spatial correspondence might be problematic, since the line width of the emission-line cloud is rather narrow ($\sigma < $ 170 \kms). 

Our GMOS data at 0\farcs4 resolution are presented as channel maps and two-dimensional velocity fields in Fig.~\ref{fig:pks2251}. The results are in general agreement with previous observations \citep{Hut90,Dur94,Cra00}. In a couple of the channel maps, we have overlaid the radio structure on top of the optical emission-line clouds. This comparison shows that the southeast cloud ($a$) that is alleged to coincide with the radio peak is actually a bit off-centered with respect to the hot spot. In agreement with \citet{Cra00}, no increase in velocity dispersion is seen in cloud $a$ relative to the rest of the EELR, indicating that the connection between the radio plasma and the warm ionized gas is not a physical one (\ie\ it is most likely a projection effect). This conclusion is reinforced by the emission-line spectra---the relative line fluxes of region $a$ is almost identical to those of the other four regions ($b - e$), none of which are on the path of the radio jet (Table~\ref{tab:3c48lineratio}). 

As with 3C\,48, the \otwo\ doublet is well resolved in PKS\,2251+11. But sufficient S/N is only reached in regions $a$ and $b$. We measured \otwo\,$\lambda3726/\lambda3729$ ratios of 0.97$\pm$0.08 and 0.80$\pm$0.07 for $a$ and $b$, respectively, which imply $N_e$ = 320$\pm$100 (380$\pm$120) and 130$\pm$70 (150$\pm$80) \cc\ if $T_e$ = 10$^4$ (1.5$\times$10$^4$) K. 

\section{Quasar Photoinionization and Gas Metallicity}\label{sec:3c48metal}

\begin{figure*}[!t]
\epsscale{0.55}
\plotone{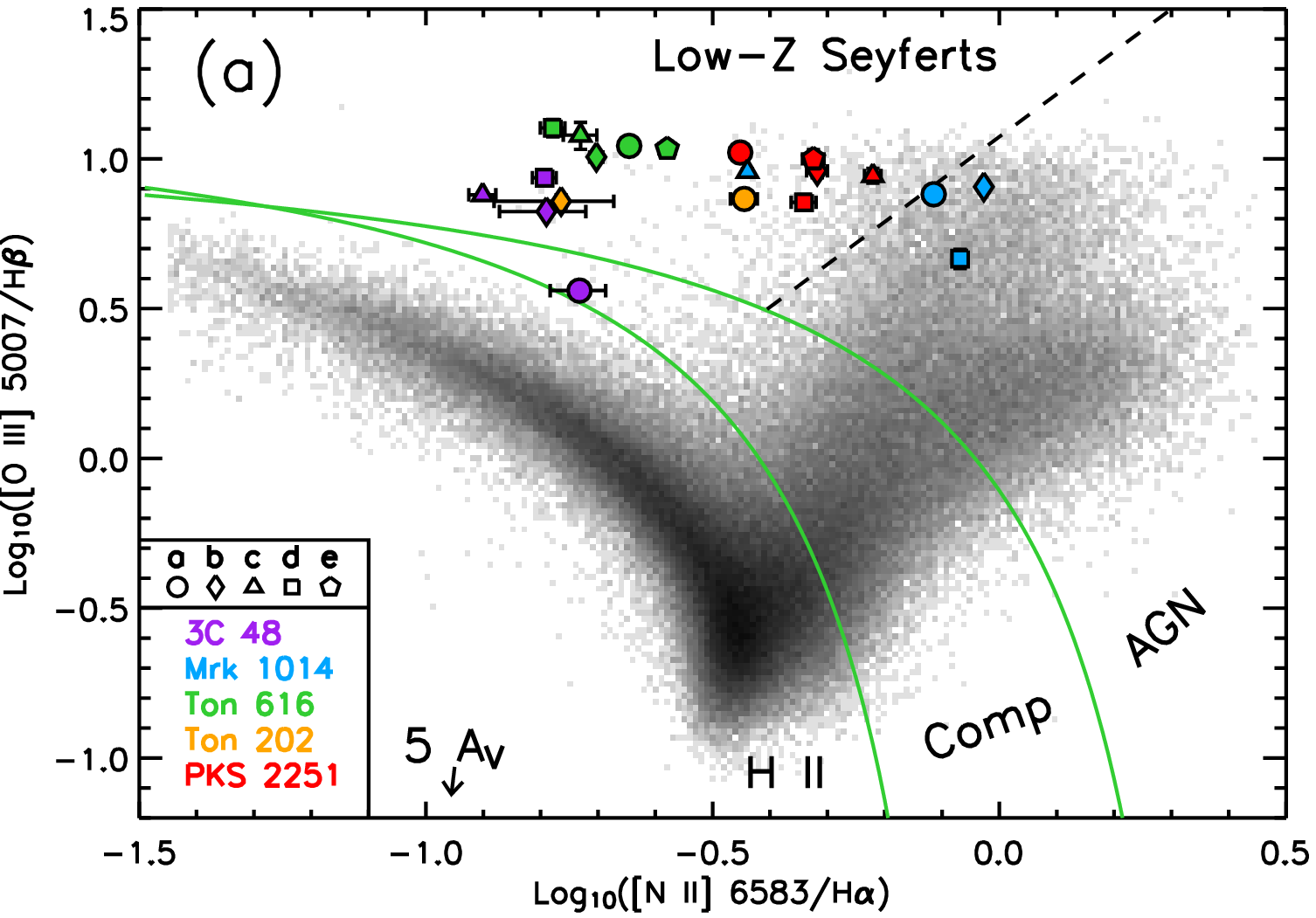}
\plotone{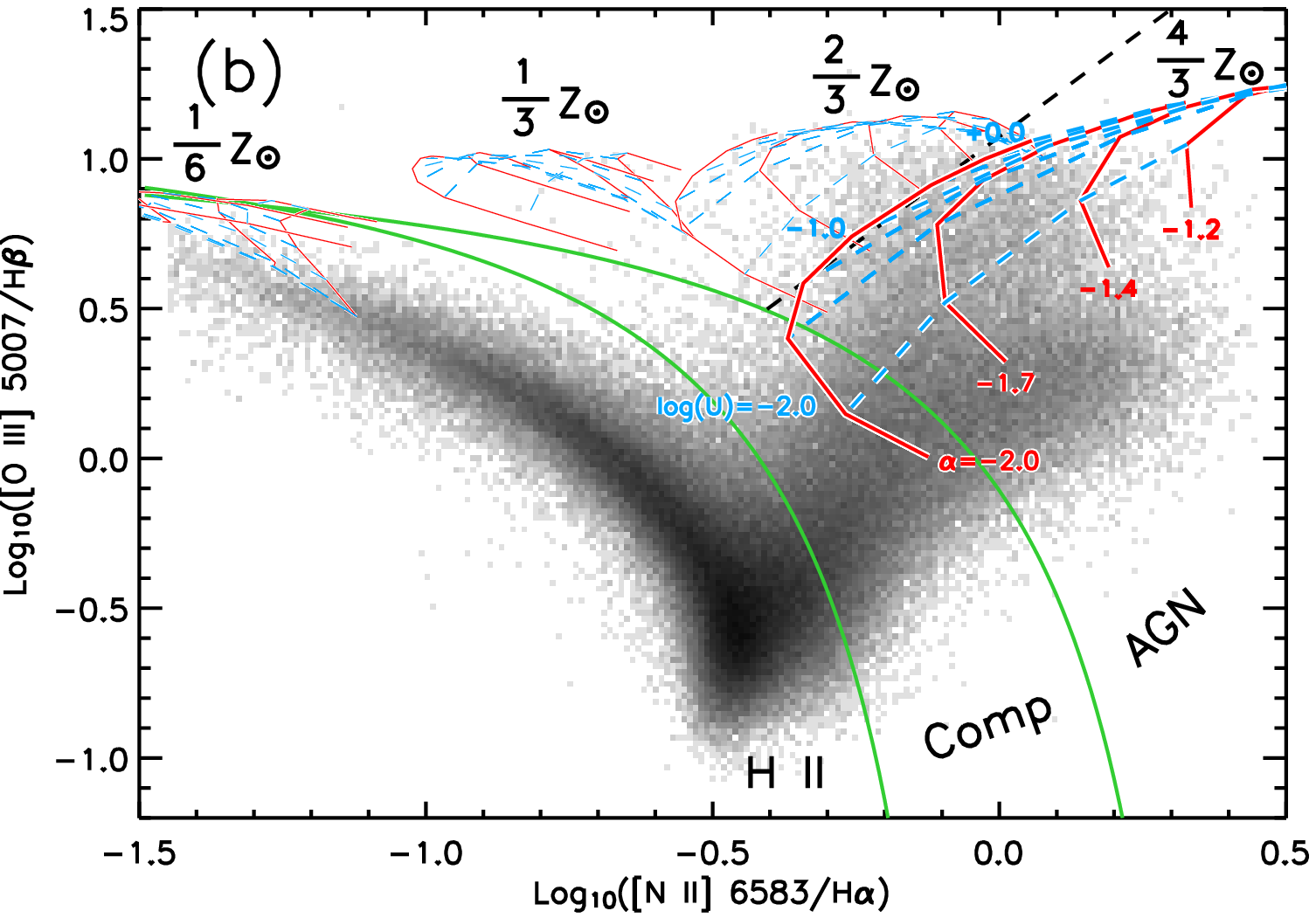}
\caption[Quasar EELRs are mostly more metal-poor than the nebulae in typical AGN]{Quasar EELRs are mostly more metal-poor than the nebulae in typical AGN, as indicated by the metallicity-sensitive diagram of \othree\,$\lambda5007$/H$\beta$ vs.\ \ntwo\,$\lambda6583$/H$\alpha$. The background image shows the density distribution of the SDSS DR4 emission-line galaxies in log scale. Objects above the upper green curve are objects dominated by AGN (As in Fig.~\ref{fig:3c48o2}$d$, LINERs are concentrated in the lower denser branch and Seyferts in the upper branch; \citealt{Kew06}), below the lower curve are star-forming galaxies \citep{Kau03a}, and AGN/star-forming composite galaxies are in between. In panel ({\it a}) we show our measurements from the EELRs; while in panel ({\it b}) we show photoionization model grids \citep{Gro04a} of four gas-phase metallicities from 0.2 to 1.3 \zsun.  For each metallicity, model predications are given for a range of ionization parameters ($-2.3 \leq$ log($U$) $\leq$ 0) and four power law indices representing the quasar ionizing continuum ($F_{\nu} \propto \nu^{\alpha}$; $\alpha = -1.2, -1.4, -1.7$, and $-2.0$). Most Seyferts can be fit quite well with the super-solar metallicity models. The top region bordered by the black dashed line and the upper green curve is where the low-metallicity Seyferts are located \citep{Gro06}.
} \label{fig:3c48n2}
\end{figure*}

In Figure~\ref{fig:3c48o2} we compare ratios of strong emission lines of the quasar EELRs with those of the emission-line galaxies in the Sloan Digital Sky Survey Data Release 4 (SDSS DR4). The emission-line fluxes, after correction for stellar absorption and foreground Galactic extinction, are publicly available for SDSS emission-line galaxies\footnote{http://www.mpa-garching.mpg.de/SDSS/DR4/; and see \citet{Tre04} for a description of the data.}. Following the same dereddening procedure as in \citet{Fu08} and the classification schemes of \citet{Kew06}, we classified the SDSS galaxies into Seyferts, LINERs, star-forming galaxies (H II), and AGN/star-forming composites. As can be seen in the BPT diagrams \citep{Bal81} shown in Fig.~\ref{fig:3c48o2}, the quasar EELRs show similar line-ratios to those of the Seyfert galaxies, but they are clearly distinguishable from star-forming and composite galaxies. This result, in combination with the inadequacy of shock and ``shock + precursor" models, as we demonstrated in Fig.~\ref{fig:3c48o2_grid}, indicate that the EELRs are most definitely photoionized by an AGN-type ionizing spectrum. The only exception is the region 3C\,48-$a$, which is below the AGN/star-forming dividing line in Fig.~\ref{fig:3c48o2}$d$ and Fig.~\ref{fig:3c48n2}$a$. 3C\,48-$a$ corresponds to a region between C5 and C6 as defined in the long-slit observations of \citet{Can00a}. C5 and C6 have roughly half of their stellar continuum coming from a young stellar population of only $\sim$9 Myr old. We believe that the low \othree/H$\beta$ ratio of 3C\,48-$a$ is due to a significant ionizing flux from the early-type stars in the region, \ie\ 3C\,48-$a$ is ionized not only by the central quasar $>$ 26 kpc away but also by the local massive stars. 

Although we have compared the results of the dusty radiation-pressure dominated
photoionization model \citep{Gro04a} with our data in Fig.~\ref{fig:3c48o2_grid}, we do not mean to imply that EELRs are actually dusty ionized surface layers of dense molecular clouds. We have
demonstrated previously \citep{Fu07b} that a two-phase photoionization model works equally well in reproducing the line-ratios. The dense ionization-bound clouds can easily be regenerated
through turbulent shocks that should be prevalent in EELRs, given the observed
supersonic velocity dispersions. Furthermore, it is difficult to understand how
dense molecular clouds could be present at large galactic radii, in the absence of any detectible underlying stellar continuum \citep{Sto02}. 

Some of the EELR clouds show noticeably weaker \stwo/H$\alpha$ ratios than those predicted by the photoionization models (Fig.~\ref{fig:3c48o2_grid}$d$). \stwo\ (like \oone) depends strongly upon the density structure of the cloud, since it is mostly emitted by the partially ionized region behind the ionization front. Therefore, the observed inconsistency probably implies that the density structure of EELR clouds is different from a radiation-pressure dominated NLR cloud, offering further evidence against the dense molecular cloud model. Because of the uncertainties in modeling the density structure of the NLR cloud, \citet{Gro06} have discounted \stwo\ and \oone\ ratios as reliable metallicity indicators. Instead, the authors concluded that ``line ratios involving \ntwo\ ratios are the most robust metallicity indicators in galaxies where the primary source of ionization is from the active nucleus,'' mainly because of the secondary production of the element. On the other hand, by comparing two power-law photoionization models assuming drastically different density structures, we have shown that the \ntwo\ line ratios are insensitive to the density profile but are almost entirely dependent on the nitrogen abundance \citep{Fu07b}.
 
In Figure~\ref{fig:3c48n2} we estimate the metallicity of the gas in the EELRs with the metallicity-sensitive \othree\,$\lambda5007$/H$\beta$ vs.\ \ntwo\,$\lambda6583$/H$\alpha$ diagram. We have converted the modeled total metallicities to gas-phase metallicities using the solar abundances defined by \citet{And89}, in order to be consistent both with previous studies on the metallicity of quasar BLRs and with metallicity results from stellar populations.  
One solar metallicity in \citet{Gro04a} corresponds to 1/3 \zsun\ in our figures, since approximately half of the metals are assumed to be depleted onto dust in their models, and their assumed solar metallicity is about 0.2 dex\footnote{The C, N, O, and Fe abundances in the solar abundance set assumed by \citet{Gro04a} are all $\sim$0.2 dex lower than those in \citet{And89}. Hence, the mass fraction of metals (i.e. $Z$) is about 1.5 times lower --- \zsun\ = 0.013 --- than that of \citet{And89}.} lower than that of \citet{And89}. In any case, in our comparisons with other AGN, it is the {\it relative} metallicities that are important, not the absolute values in solar terms.

In agreement with our previous results \citep[their Fig.~12]{Fu08}, all of the EELRs, except that of the only radio-quiet quasar, Mrk\,1014, are consistent with photoionization models with sub-solar metallicities (using the \citealt{And89} scale), which are significantly lower than the metallicities deduced from typical AGN in the SDSS that span a similar redshift range. For Mrk\,1014, metal-rich gas is detected in the western part of the EELR (regions $a$, $b$, and $d$), while the eastern cloud ($c$) shows a low metallicity similar to that of the other EELRs. Despite the precaution that \oone\ and \stwo\ ratios may not be reliable metallicity indicators, we note that the EELRs also lie clearly to the left side of the AGN branch in the BPT \oone and \stwo\ diagrams (Fig.~\ref{fig:3c48o2}$c$ and $d$), and they roughly overlap with the low-metallicity type-2 Seyferts that were selected in SDSS by \citet{Gro06}.

Recall that EELR quasars also distinguish themselves from other quasars
by their BLR metallicities \citep{Fu07a}---the former show
much lower metallicities ($Z \lesssim 0.6$ \zsun) than
the latter ($>$ 1 \zsun). The BLR metallicities are estimated from broad
permitted UV line ratios, such as \nfive\,$\lambda$1240/\cfour\,$\lambda$1549 and \nfive/\hetwo\,$\lambda$1640, based on the locally optimally emitting cloud model \citep{Bal95}.  The BLR result is therefore based on completely different
data and photoionization models than is the EELR metallicity, yet both indicate a low metallicity, if we adopt the most straightforward
interpretation for the low \ntwo/H$\alpha$ ratios. Therefore, although
neither the BLR model nor the NLR/EELR model is perfect, a lower gas
metallicity of EELR quasars most consistently explains the totality of
the data.

\section{Summary and Conclusions}\label{chap:sum}

Here we bring together our results from IFU spectroscopy of the EELRs around 7 quasars and 1 FR II radio galaxy, as have been presented in this and previous papers  \citep{Fu06,Fu07b,Fu08,Sto07}. We summarize these results for several key properties and interpret them in terms of a tentative ``toy model'' for the formation of quasar EELRs.

\subsection{Physical Properties of Extended Emission-Line Regions}

\subsubsection{Kinematics}

The velocity structure of the ionized gas is locally ordered but globally disordered in all of the 7 quasars, consistent with the results of previous work \citep{Dur94,Cra00,Sto02}. We are not aware of any simple dynamical models that can fit the data. The only velocity map that resembles a rotating disk is that of the radio galaxy 3C\,79; however, even in this case, this is true only when the fainter clouds are ignored.

The radial velocity maps have shown that most of the extended nebular emission is confined within 300 \kms\ of the quasar's systemic velocity, which is defined by the nuclear NLR. Nevertheless, extremely high velocities are observed in fainter or presumably less massive clouds in 5 of the 7 quasars---3C\,249.1 ($+550$ \kms), 4C\,37.43 ($-600$ \kms), Mrk\,1014 ($-1100$ \kms), 3C\,48 ($-500$ \kms), and Ton\,616 ($+440$ \kms). We believe such high velocities are unlikely to be solely gravitational in origin.

The velocity dispersion maps look quite quiescent with typical values between 30 and 100 \kms. Three of the 7 EELRs show clouds with velocity dispersions significantly greater than 100 \kms: in 3C\,249.1 there is a cloud to the SE of the nucleus with $\sigma \sim 310$ \kms; in Mrk\,1014 there are two clouds to the E and W of the nucleus with $\sigma \sim 280$ \kms; in 3C\,48 the high velocity cloud 0\farcs25 north of the nucleus shows a dispersion of $\sim$390 \kms. In all of the three cases, there seems to be a clear interaction between a radio jet and the cloud: the 3C\,249.1 cloud is aligned with the radio jet to the SE, the Mrk\,1014 clouds coincide with the two radio knots on either side of the core, and the high-velocity cloud in 3C\,48 is near the base of the 0\farcs7 radio jet. However, no such kinematic signatures were seen in PKS\,2251+11 where a positional correspondence between a radio hot spot and one of the brightest EELR cloud has long been noticed. Thus this is most likely a chance alignment; this conclusion is also supported by the lack of evidence of shocks from line intensity ratios.

\subsubsection{Ionization Mechanism}

There is now little doubt that EELRs are photoionized by the central quasars. We reached this conclusion because photoionization by a power-law continuum source is the only model that fits the data. 

Shock ionization and the ``shock + precursor" model cannot predict the correct line ratios; this result is especially clear in the \othree/H$\beta$ vs. \otwo/H$\beta$ diagram \citep[][their Fig.~5$a$]{Fu07b}. The high \othree/H$\beta$ ratios of the EELRs can only be reproduced by the ``shock + precursor" model when the shock speed exceeds 450 \kms, which is inconsistent with the low velocity dispersions measured in the clouds (typically 30 $< \sigma <$ 100 \kms). Further, the model predicts an \otwo/H$\beta$ ratio $\sim2\times$ higher than the observed values. 

Photoionization by star-forming regions has also been ruled out as a dominant ionization mechanism. For instance, the EELRs lie firmly above the maximum star-forming curve in the BPT diagrams of \othree/H$\beta$ vs. \stwo/H$\alpha$ and \othree/H$\beta$ vs. \ntwo/H$\alpha$ (\eg\ Figs.~\ref{fig:3c48o2} and \ref{fig:3c48n2}). The cloud 3C\,48-$a$ represents the only case where in addition to the quasar young stars may also have contributed significantly to the ionization (\S~\ref{sec:3c48metal}). 

\subsubsection{Metallicity}

EELRs consist of clouds with low metallicities ($Z \lesssim$ 2/3 \zsun, on the \citealt{And89} scale) as implied by their low \ntwo/H$\alpha$ line ratios. These clouds are significantly more metal-poor than the gas in similar environments, \eg\ the NLRs of type-2 Seyfert galaxies and radio galaxies at similar redshifts. Power-law photoionization models with low metallicities can correctly predict fluxes of optical lines other than \ntwo\ as well (see Fig.~\ref{fig:3c48o2_grid}). Four of the seven quasars in our sample were observed by the \hst\ FOS in the UV (3C\,249.1, Ton\,202, 4C\,37.43, and PKS\,2251+11), both the \nfive\ $\lambda1240$/\cfour\ $\lambda1549$ and \nfive/\hetwo\ $\lambda1640$ ratios of their BLRs indicate a metallicity less than 0.5 \zsun \citep{Fu07a}, remarkably consistent with the metallicity derived from the extended gas in a completely different density regime. 

The only radio-quiet quasar, Mrk\,1014, is also the only one where extended metal-rich gas is detected. It seems that the gas in this quasar has not been well mixed, as a bright EELR cloud to the east shows a low metallicity that is similar to that of the other EELRs in the sample. We suspect the clouds to the west have been enriched by recent star-forming activities. Evidence for a recent starburst in the host galaxy of Mrk\,1014 was reported by \citet{Can00b}.

\subsection{Star Formation in the Host Galaxies of EELR Quasars}

The host galaxies of a couple of our EELR quasars have been studied in great detail by \citet{Can00a,Can00b}. 3C\,48 and Mrk\,1014 were included in their sample because they fall into an intermediate region between quasars and ULIRGs in a FIR color-color diagram \citep{Lip94}, suggesting significant star-forming components in their host galaxies. Deep imaging of their host galaxies has shown clear evidence that these are cases of major mergers in their final stages. Deep spectroscopy revealed young stellar populations, confirming that FIR colors are indicative of star formation. In 3C\,48, \citeauthor{Can00a} found vigorous ongoing star formation and post-starburst populations with ages up to 114 Myr. In Mrk\,1014, only post-starburst populations with ages between 180 and 290 Myr were seen. 
Notice that 3C\,48 and Mrk\,1014 are the only two objects in our sample that do not have classical FR II double-lobe morphologies. 

Similar studies do not exist for the rest of the quasars in our sample. Nevertheless, an additional two EELR quasars have been detected at 60 $\mu$m by {\it IRAS} and one by ISO---3C\,249.1, 4C\,37.43, and 3C\,323.1. Although solid detections are only available at 25 $\mu$m and 60 $\mu$m, their large $\alpha(60,25)$ spectral indices place them firmly above the transition region defined by \citet{Can01}. The FIR colors\footnote{When calculating these spectral indices, photometric data from both {\it IRAS} and ISO have been considered. At 25 $\mu$m and 60 $\mu$m the priority list is {\it IRAS} solid detection $>$ ISO solid detection $>$ {\it IRAS} upper limits, and at 100 $\mu$m only {\it IRAS} upper limits were used.} are $\alpha(60,25) = -0.35, -0.23, -0.05$ and $\alpha(100,60) > -0.63, -2.10, -2.24$ for 3C\,249.1, 4C\,37.43, and 3C\,323.1, respectively. If ULIRG-unlike FIR colors are indicative of no recent star formation, then we would expect that the host galaxies of these EELR quasars will be devoid of young stellar populations less than a few Myr old. An indirect proof of this conjecture comes from our study of the host galaxy of 3C\,79 \citep{Fu08}.

Given that 3C\,79 is a type-2 EELR quasar with FR II morphology, the unique geometry of the radio galaxy offers a much clearer view of the host galaxy of an EELR quasar. We found that the host galaxy consists of an intermediate-age (1.3 Gyr) stellar population (4\% by mass) superimposed on a 10 Gyr old population, \ie\ there has been no recent star formation. 

In summary, the host galaxies of EELR quasars can have a large range of star formation histories. For  Mrk\,1014 and 3C\,48, stars have been vigorously forming in the past few hundred Myr; but for 3C\,79, and, apparently, for at least many of the FR II quasars in our sample, significant star formation has occurred only in the remote past ($\gtrsim 1$ Gyr ago).

\subsection{A Toy Model of Quasar Extended Emission-Line Regions}\label{sumsec:toy}

A scenario for producing quasar EELRs must be able to account for all of the following compelling results: (1) the correlation between extended nebular emission and radio spectral index (\S~\ref{introsec:corr}), (2) the correlation between extended nebular emission and quasar BLR metallicity \citep{Fu07a}, (3) the lack of detailed morphological correspondence between extended nebula and the host galaxy or the radio structure, (4) the chaotic velocity distribution but relatively calm velocity dispersions of the extended nebulae, (5) the frequent detection of extremely high velocity clouds ($V >$ 400 \kms), and (6) the large masses of the EELRs ($M \sim 10^9$--$10^{10}$ \msun; \citealt{Fu06,Fu07b,Fu08,Sto07}).

As mentioned in \S~\ref{introsec:orig}, there were three remaining at least somewhat plausible origins for the extended gas, and we now examine each one of them with the current set of observational results:
\begin{enumerate}
\item Cold accretion of intergalactic gas through Mpc-scale filaments. Although this model can easily explain the low metallicity gas comprising the EELRs, it is difficult for it to explain the quasar activity and the similarly low metallicity of the BLRs. Cold accretion by itself is unlikely to activate the central black hole as there is no likely mechanism to provide the drastic reduction of the angular momentum of the accreted gas that would be required to force a significant amount of it to within the accretion radius of the black hole.

\item Tidal debris from a galactic merger. It is difficult for this model to explain the clouds with velocities higher than reasonable for gravitational kinematics ($\gtrsim 400$ \kms), which have been detected in 5 of the 7 quasars, often at a considerable distance from the quasar and in locations not in the path of a radio jet (3C\,249.1, Ton\,616, 4C\,37.43). In addition, it is difficult to see how this tidal picture would give a situation where {\it no} luminous EELRs have been found associated with flat-spectrum core-dominated quasars, and those associated with radio-quiet QSOs are relatively rare and of generally lower luminosity than those found around steep-spectrum radio sources. Nevertheless, most of the quasars with luminous EELRs do show independent evidence for recent strong interactions or mergers \citep{Sto87}, and this evidence has to be taken into account in any reasonable physical model. But it is significant that we never see any indication of an underlying stellar component with a distribution similar to that of a luminous EELR.

\item Remnants of galactic superwinds driven by either the quasar itself or a coeval starburst. The starburst superwind picture has been found problematic both in 3C\,249.1 on the ground of the derived momentum injection rate \citep{Fu06} and in 3C\,79 based on the lack of young stellar populations \citep{Fu08}. In addition, starburst-driven superwinds reflect mass ejection from type-II supernovae, and thus they are expected to be metal enriched (it takes only $\sim$0.1 Gyr to reach a gas metallicity of 1 \zsun\ from the primordial abundances; \citealt{Fri98}). This contradicts the dominance of metal-poor gas in the environments of EELR quasars. It would be especially difficult to explain the metal-poor gas in the quasar BLRs, which are within 0.1 pc of the central black holes. Finally, a starburst-driven superwind cannot explain the fact that luminous EELRs are found virtually exclusively around steep-spectrum radio-loud quasars.  On the other hand, EELRs as remnants of superwinds driven by quasars themselves do not suffer from these difficulties. 
\end{enumerate}

We clearly need to combine aspects of the two last points into our over-all picture.  Both strong interactions or mergers, on the one hand, and the presence of a steep-spectrum radio source, on the other, each in general appear to be necessary, but not sufficient, ingredients for the production of a luminous EELR. There is considerable evidence that the merging companion supplies the gas \citep{Fu07a}, but that the distribution of the ionized gas is controlled by a massive outflow \citep{Sto02,Sto07,Fu06,Fu07b,Fu08}.
We suggest the following overall picture for the formation and the fate of the EELRs around quasars with large FR II radio jets: A large, late-type galaxy with a mass of a few $\times10^{10}$  M$_{\odot}$ of low-metallicity gas has recently merged with the gas-poor quasar host galaxy, which has a $\sim10^9$ M$_{\odot}$ black hole at its center. Low metallicity gas on the outskirts of the merging companion is driven to its center and is completely mixed with any higher metallicity gas there well before the final coalescence \citep[\eg][]{Kew06b}. The merger triggers the current episode of quasar activity, including the production of FR II radio jets. The initiation of the jets also produces a wide-solid-angle blast wave that sweeps most of the gas from the encounter out of the galaxy. This gas is immediately photoionized by UV radiation from the quasar when the blast wave clears the dust cocoon enshrouding the central engine. Turbulent shocks produce high-density ($\sim400$ cm$^{-3}$) filaments or sheets in the otherwise low-density ($\sim1$ cm$^{-3}$) ionized medium which is in hydrodynamical equilibrium with a hot ($T \sim 10^7$ K) diffuse medium, maintaining a low ionization state. The quasar will eventually shut itself off due to starvation, and some high velocity clouds will escape into the inter-galactic medium, but most of the warm gas will stay and recombine to neutral hydrogen (HI) $\sim10^4$ yr after the quasar shuts down. The chaotic kinematics will be washed away as the gas settles into a rotating disk in a crossing time of $\sim$0.5 Gyr. But before it settles down, it might look much like the HI gas in the nearby early-type galaxy NGC\,1023 \citep{Mor06}. The dynamically settled gas could probably account for a significant fraction of the dispersed HI in giant early-type galaxies.

Note that, in 3C\,48, we may be witnessing an early phase in one of these episodes (\S~\ref{sec:3c48}; \citealt{Sto07}).
The high-velocity gas currently concentrated near the base of the radio jet constitutes a wide-solid-angle outflow with velocities ranging up to at least 800 \kms\ and a total mass of ionized gas of at least $\sim10^9$ \msun, and quite likely much more.

This picture naturally explains the EELR---radio morphology correlation, since the superwind has been assumed to be associated with the formation of the jets. But the reason for the EELR---low gas metallicity correlation is much less clear. 
We can speculate that it may have something to do with the lower radiative coupling such gas would have to the quasar radiation field, allowing more efficient accretion. Specifically, a higher metallicity will lower the accretion rate of material towards the center, because both the higher opacity of the gas and larger amount of dust will couple the gas more efficiently to the radiation field of the quasar. Such a lowered accretion rate may delay the spin up of the black hole, consequently delaying the formation of the radio jet, assuming that the jet production is determined by the black hole spin \citep{Bla90}. Hence, most of the gas may have time to form stars before the jet is launched, leaving an insufficient amount of gas to form an EELR. It is interesting to note that related considerations have been invoked recently to explain the link between low metallicity and long duration $\gamma$-ray bursts \citep{Fru06}.

Whether this particular explanation for the low metallicities in the BLRs of EELR quasars is correct or not, the correlation itself clearly suggests {\it some} physical mechanism by which the injection of a large amount of low-metallicity gas into a system with a super-massive black hole can result in a large-scale, quasar-powered superwind. This sort of situation is likely to have been relatively common in the early universe, when mergers were more frequent and gas-rich galaxies more numerous. This form of feedback might therefore have been of some importance in restricting the growth of the most massive galaxies at early epochs.

\acknowledgments
We thank the Gemini North staff for carrying out the GMOS IFU observations. We also thank the anonymous referee for a careful reading of the manuscript and for useful comments that helped improve the presentation. This research has been partially supported by NSF grant AST~03-07335. The authors recognize the very significant cultural role that the summit of Mauna Kea has within the indigenous Hawaiian community, and we are grateful to have had the opportunity to conduct observations from it.


\clearpage 

\clearpage
\begin{landscape}

\begin{deluxetable}{lccccccccccccc}
\rotate
\setlength{\tabcolsep}{.05cm}
\tablewidth{0pt}
\tabletypesize{\scriptsize}
\tablecaption{Line Ratios of Quasar Emission-Line Clouds Relative to H$\beta$ \label{tab:3c48lineratio}}
\tablehead{ \colhead{ID} & \colhead{$A_V$}
& \colhead{[Ne\,{\sc v}]\,$\lambda3426$}
& \colhead{[O\,{\sc ii}]\,$\lambda3727$}
& \colhead{[Ne\,{\sc iii}]\,$\lambda3869$}
& \colhead{[O\,{\sc iii}]\,$\lambda4363$}
& \colhead{He\,{\sc ii}\,$\lambda4686$} & \colhead{H$\beta$}
& \colhead{[O\,{\sc iii}]\,$\lambda5007$}
& \colhead{[O\,{\sc i}]\,$\lambda6300$} & \colhead{H$\alpha$}
& \colhead{[N\,{\sc ii}]\,$\lambda6583$}
& \colhead{[S\,{\sc ii}]\,$\lambda6716$}
& \colhead{[S\,{\sc ii}]\,$\lambda6731$}
\\
\colhead{(1)} & \colhead{(2)} & \colhead{(3)} & \colhead{(4)} & \colhead{(5)} &
\colhead{(6)} & \colhead{(7)} & \colhead{(8)} & \colhead{(9)} & \colhead{(10)} &
\colhead{(11)} & \colhead{(12)} & \colhead{(13)} &
\colhead{(14)}
}
\startdata
\multicolumn{14}{c}{3C 48 ($z$ = 0.369)}   \nl
\hline
$a$&  (2.80)&       \nodata & $4.63\pm0.19$ & $0.67\pm0.14$ &       \nodata &       \nodata & $1.00\pm0.04$ &\phn $3.63\pm0.04$ & $0.12\pm0.02$ & $3.10\pm0.08$ & $0.57\pm0.06$ & $0.35\pm0.06$ &       \nodata  \\
$b$&  (2.80)&       \nodata & $6.25\pm0.27$ & $1.94\pm0.21$ &       \nodata & $0.42\pm0.03$ & $1.00\pm0.03$ &\phn $6.67\pm0.02$ & $0.14\pm0.02$ & $3.10\pm0.10$ & $0.50\pm0.08$ & $0.31\pm0.06$ &       \nodata  \\
$c$&   2.80 & $2.80\pm0.18$ & $4.63\pm0.11$ & $1.56\pm0.05$ & $0.21\pm0.02$ & $0.43\pm0.01$ & $1.00\pm0.01$ &\phn $7.58\pm0.01$ & $0.16\pm0.01$ & $3.10\pm0.03$ & $0.39\pm0.02$ & $0.23\pm0.03$ &       \nodata  \\
$d$&  (2.80)& $1.54\pm0.17$ & $4.63\pm0.13$ & $1.40\pm0.05$ & $0.21\pm0.02$ & $0.42\pm0.01$ & $1.00\pm0.02$ &\phn $8.64\pm0.01$ & $0.19\pm0.01$ & $3.10\pm0.03$ & $0.50\pm0.02$ & $0.25\pm0.02$ &       \nodata  \\
\hline
\multicolumn{14}{c}{Mrk 1014 ($z$ = 0.163)}   \nl
\hline
$a$&   0.61 &       \nodata & $3.10\pm0.21$ & $0.83\pm0.11$ &       \nodata & $0.46\pm0.03$ & $1.00\pm0.04$ &\phn $7.61\pm0.03$ & $0.25\pm0.02$ & $3.10\pm0.02$ & $2.38\pm0.02$ & $0.65\pm0.01$ & $0.50\pm0.01$  \\
$b$&   0.34 &       \nodata & $2.93\pm0.25$ & $1.04\pm0.12$ &       \nodata & $0.38\pm0.03$ & $1.00\pm0.04$ &\phn $8.06\pm0.04$ & $0.32\pm0.02$ & $3.10\pm0.02$ & $2.91\pm0.02$ & $0.80\pm0.01$ & $0.62\pm0.01$  \\
$c$&  (0.00)& $1.48\pm0.34$ & $1.98\pm0.17$ & $0.55\pm0.06$ &       \nodata & $0.47\pm0.02$ & $1.00\pm0.02$ &\phn $9.07\pm0.03$ & $0.13\pm0.01$ & $2.86\pm0.01$ & $1.04\pm0.01$ & $0.47\pm0.01$ & $0.36\pm0.01$  \\
$d$&   1.15 &       \nodata & $3.94\pm0.35$ & $1.48\pm0.26$ &       \nodata & $0.42\pm0.06$ & $1.00\pm0.07$ &\phn $4.64\pm0.06$ & $0.26\pm0.02$ & $3.10\pm0.02$ & $2.65\pm0.02$ & $0.70\pm0.03$ & $0.46\pm0.02$  \\
\hline
\multicolumn{14}{c}{Ton 616 ($z$ = 0.268)}   \nl
\hline
$a$&   0.81 &           $-$ &           $-$ &           $-$ &           $-$ & $0.56\pm0.03$ & $1.00\pm0.03$ &$11.05\pm0.03$ &       \nodata & $3.10\pm0.03$ & $0.70\pm0.02$ & $0.15\pm0.02$ & $0.14\pm0.02$  \\
$b$&   0.59 &           $-$ &           $-$ &           $-$ &           $-$ & $0.31\pm0.03$ & $1.00\pm0.03$ &$10.14\pm0.02$ & $0.09\pm0.02$ & $3.10\pm0.02$ & $0.61\pm0.02$ & $0.44\pm0.02$ &       \nodata  \\
$c$&   0.28 &           $-$ &       \nodata &       \nodata &       \nodata & $0.42\pm0.09$ & $1.00\pm0.10$ &$11.99\pm0.07$ & $0.26\pm0.03$ & $3.10\pm0.05$ & $0.58\pm0.04$ & $0.39\pm0.08$ &       \nodata  \\
$d$&   0.14 &           $-$ &       \nodata &       \nodata &       \nodata &       \nodata & $1.00\pm0.07$ &$12.68\pm0.05$ & $0.16\pm0.02$ & $3.10\pm0.03$ & $0.52\pm0.02$ & $0.31\pm0.03$ &       \nodata  \\
$e$&  (0.00)&           $-$ &       \nodata &       \nodata &       \nodata & $0.26\pm0.05$ & $1.00\pm0.06$ &$10.81\pm0.05$ & $0.19\pm0.02$ & $3.00\pm0.03$ & $0.79\pm0.02$ & $0.48\pm0.03$ & $0.36\pm0.03$  \\
\hline
\multicolumn{14}{c}{Ton 202 ($z$ = 0.364)}   \nl
\hline
$a$&  (2.30)& $1.34\pm0.20$ & $3.02\pm0.16$ & $0.94\pm0.12$ &       \nodata & $0.37\pm0.03$ & $1.00\pm0.03$ &\phn $7.35\pm0.02$ &       \nodata & $3.10\pm0.07$ & $1.11\pm0.06$ & $0.28\pm0.04$ & $0.27\pm0.04$  \\
$b$&  (2.30)& $1.59\pm0.29$ & $3.01\pm0.30$ & $0.75\pm0.12$ &       \nodata & $0.42\pm0.05$ & $1.00\pm0.05$ &\phn $7.22\pm0.04$ &       \nodata & $3.10\pm0.15$ & $0.53\pm0.12$ & $0.38\pm0.09$ &       \nodata  \\
$c$&  (2.30)& $2.28\pm0.56$ & $4.66\pm0.56$ & $1.38\pm0.25$ &       \nodata & $0.43\pm0.07$ & $1.00\pm0.07$ &$11.41\pm0.07$ &       \nodata & $3.10\pm0.22$ &       \nodata & $0.93\pm0.20$ &       \nodata  \\
\hline
\multicolumn{14}{c}{PKS 2251$+$11 ($z$ = 0.325)}   \nl
\hline
$a$&  (0.46)&       \nodata & $2.94\pm0.12$ & $0.97\pm0.07$ &       \nodata & $0.37\pm0.03$ & $1.00\pm0.04$ &$10.49\pm0.04$ &       \nodata & $3.10\pm0.05$ & $1.10\pm0.04$ & $0.51\pm0.05$ &       \nodata  \\
$b$&  (0.46)& $1.01\pm0.21$ & $3.52\pm0.14$ & $1.08\pm0.12$ &       \nodata &       \nodata & $1.00\pm0.05$ &\phn $9.12\pm0.04$ &       \nodata & $3.10\pm0.06$ & $1.49\pm0.05$ & $0.60\pm0.06$ & $0.49\pm0.05$  \\
$c$&  (0.46)&       \nodata & $4.69\pm0.17$ & $0.88\pm0.11$ &       \nodata & $0.21\pm0.05$ & $1.00\pm0.06$ &\phn $8.79\pm0.03$ &       \nodata & $3.10\pm0.06$ & $1.87\pm0.05$ & $0.70\pm0.05$ & $0.49\pm0.04$  \\
$d$&  (0.46)& $0.89\pm0.21$ & $2.84\pm0.23$ & $0.44\pm0.09$ &       \nodata & $0.26\pm0.05$ & $1.00\pm0.06$ &\phn $7.16\pm0.04$ &       \nodata & $3.10\pm0.07$ & $1.41\pm0.06$ & $0.50\pm0.08$ &       \nodata  \\
$e$&  (0.46)& $1.32\pm0.30$ & $2.78\pm0.41$ & $0.86\pm0.13$ &       \nodata &       \nodata & $1.00\pm0.07$ &\phn $9.99\pm0.06$ &       \nodata & $3.10\pm0.06$ & $1.47\pm0.06$ & $0.59\pm0.12$ & $0.50\pm0.11$ 
\enddata
\tablecomments{
Col. (1): Cloud name.
Col. (2): Intrinsic reddening. The values in parentheses are not
directly derived from the Balmer decrements of the corresponding clouds.
The $A_V$ values are measured from the H$\alpha$/H$\beta$ ratio for
Mrk\,1014 and Ton\,616, where the two lines were observed
simultaneously. For 3C\,48, Ton\,202 and PKS\,2251+11, they have been
fixed respectively to those derived from the H$\gamma$/H$\beta$ ratios
of clouds 3C\,48-$c$, Ton\,202-$a+b$, and PKS\,2251+11-$a+b+c$, because
the S/N ratio of an individual cloud at H$\gamma$ is normally
insufficient for a meaningful measurement of the Balmer decrement. $A_V$
with derived values slightly less than zero have been fixed to zero. 
Col. (3--14): Dereddened intensities of emission lines relative to
H$\beta$. The line-of-sight Galactic extinction has also been taken into
account. Undetected lines (S/N $<$ 4) have been omitted. 
}
\end{deluxetable}
\clearpage
\end{landscape}

\end{document}